\documentclass[10pt,onecolumn]{IEEEtran}
\usepackage{amssymb}
\usepackage{bbm}
\ifCLASSINFOpdf
  \usepackage[pdftex]{graphicx}
 \DeclareGraphicsExtensions{.pdf,.jpeg,.png}
\else
\fi
\usepackage[cmex10]{amsmath}
\usepackage{tikz}
\usepackage{tikz}
\usepackage{bigdelim}
\usepackage{graphicx}
\usepackage{epstopdf}
\DeclareGraphicsExtensions{.eps}

\newtheorem{thm}{Theorem}
\newtheorem{lem}{Lemma}

\newcommand{\qed}{\nobreak \ifvmode \relax \else
      \ifdim\lastskip<1.5em \hskip-\lastskip
      \hskip1.5em plus0em minus0.5em \fi \nobreak
      \vrule height0.75em width0.5em depth0.25em\fi}
\begin{document}
\baselineskip 8.0 mm
%
\title{Interference Alignment for the K-user Interference Channel with Imperfect CSI}
%
%
%
\author{\begin{normalsize}Milad Johnny and Mohammad Reza Aref
\\Information System and Security Lab (ISSL),\\Sharif Universiy of Technology, Tehran, Iran\\E-mail: Johnny@ee.sharif.edu, Aref@sharif.edu
\end{normalsize}
}
\maketitle
\begin{abstract}
In this paper we explore the information-theoretic aspects of interference alignment and its relation to channel state information (CSI). For the $K-$user interference channel using different changing patterns between different users, we propose several methods to align some parts of interferences and to increase what is achieved by time sharing method. For more practical case when all the channel links connected to the same destination have the same changing pattern, we find an upper-bound and analyze it for the large interference channel network. This result shows that when the size of the network increases, the upper-bound value goes to $\frac{\sqrt{K}}{2}$. For the fast fading channel when all the channels have the same changing pattern, we show that when the direct links have different characteristic functions (channel permutation or memory), in the absence of half part of CSI (cross links) at both transmitters and receivers, one can achieve $K/2$ degrees-of-freedom (DoF). Also by the converse proof we show that this is the minimum channel information to achieve maximum DoF of $\frac{K}{2}$. Throughout this work, this fact has been pinpointed to prove statements about more general partial state CSI and achievable DoF. In other words, for the 3-user fully connected interference channel we find out while $\frac{3}{2}$ lies in achievable DoF, we don't need to know half part of the CSI. Also, the result has been extended to a more general form for $K-$user interference channel and through the converse proof, its functionality on channel state is proved to be optimum.    
\end{abstract}


\IEEEpeerreviewmaketitle
\section{Introduction}
\IEEEPARstart{T}{he} interference channel is a channel with several pairs of input-output terminals, where each input communicates with its receiver through a common channel. The increasing demand for higher data rate in wireless networks motivates researchers to find solutions for channel rate constraints such as interference between the users. The application of wireless interference network is so essential that it must be evaluated by channel capacity and achievable transmission rate. Accordingly, using any method to reduce interference effects on the communication rate and improving bandwidth assignment for the users is an essential field of research in the wireless networks.

Time and frequency division medium access schemes, also known as orthogonal access schemes, divide the entire transmission signal duration and spectrum, respectively. Another approach to improve channel spectral efficiency and achieve higher data rate is to provide full cooperation either among the transmitters or among the receivers. For instance, the authors in \cite{shamai} and \cite{foschini}, employ full cooperation among transmitters to propose a signaling scheme that reduces the system to a single MIMO broadcast channel. In this case, cooperation can increase the capacity of the network. However, since the full cooperation among multiple network users involves joint processing and data sharing over separate nodes, it seems to be infeasible in practical scenarios to provide full cooperation among transmitters and receivers. Shannon in \cite{shannon2} initiated interference channels and his basic idea expanded further by Ahlswede \cite{Ahlswede} who gave fundamental inner and outer bounds. Carleial in \cite{Carleial} by using the basic idea of superposition coding introduced by Cover \cite{Cover} pays the way for considerable improvement for achievable rate region of interference channel. In \cite{Han}, Han and Kobayashi based on Carleial and Sato's work introduced a new achievable rate region for interference channel. Etkin, Tse and Wang in \cite{Etkin} found the capacity region of the two-user Gaussian interference channel within a single bit per second per hertz (bit/s/Hz) of the capacity for all values of the channel parameters. However, the problem of interference channel for simple configuration form of 2-user interference case is still open.
The capacity of an arbitrary interference network is an important and an open problem to information theorists. Therefore, parallel to the works related to finding exact channel capacity, scientists define a new mathematical intuition on networks capacity called degrees-of-freedom (DoF), or capacity prelog. In other words, DoF characterizes the network sum capacity as follows \cite{jafar}:
\begin{equation}
C=\mathrm{DoF} \log{\left(\mathrm{SNR}\right)+\mathrm{o}\left(\log{(\mathrm{SNR})}\right)}.
\end{equation}
DoF is well-suited for approximating capacity because it becomes increasingly accurate in high signal-to-noise ratio ($\mathrm{SNR}$) regime. In \cite{Maddah}, Maddah-Ali, Motahari and Khandani implicitly introduced the concept of interference alignment (IA) and showed its capability in achieving the full DoF for certain classes of the two-user $X$ channels.
Using IA method in \cite{jafar}, Cadambe and Jafar (C-J scheme) showed that, contrary to the popular belief, the $K$-user Gaussian interference channel with varying channel gains can achieve $\frac{K}{2}$ DoF which was proved to be the capacity upper bound in the high $\mathrm{SNR}$ regime. The assumption of channel gains knowledge, is unrealistic which limits the application of this important theoretical result in practice. This fact becomes seriously important when each transmitter needs to knowledge the perfect CSI of every link in the interference network, (whether it is linked to the transmitter or not). Many researchers believe that in the absence of CSI for many interference channels, DoF region collapses entirely to what is achieved by simple interference reduction methods such as time or frequency division multiple access (TDMA or FDMA). Several papers recently focused on this problem. In \cite{jafar} a scheme introduced  where all transmitters are assumed to have no knowledge about exact channel coefficient values but are aware of connectivity between different users. Authors in \cite{Nosrat} designed algorithms to perform IA given only local CSI. They provided examples of iterative algorithms that utilize the reciprocity of wireless networks to achieve IA with only local channel knowledge at each node. In \cite{Ayach}, authors provided an approximate $\mathrm{SINR}$ ratio expression for IA over MIMO channels with imperfect channel state information and transmit antenna correlation. In \cite{Ayach}, authors presented the average achievable rate under a given measurement error power, and \cite{Tresch} established bounds on the average achievable rate with Gaussian CSI errors. Also the results in \cite{Ayach}, \cite{Tresch} were deactivated and the average rates are operationally unachievable. In \cite{Guiazon} an achievable capacity lower bound for IA with imperfect CSI derived under the model that the CSI errors are bounded. There are some other basic ideas related to blind IA using staggered antenna switching and implicitly using channel changing pattern \cite{Guiazon}, \cite{Jafar3} and \cite{Gou}. Also in \cite{Khandani}, IA with delay CSIT has been considered. Since their work does not have the converse proof, whether they reach the optimum DoF or not, they find a solution to reach $\frac{4}{6 \ln{(2)}-1}, K\rightarrow \infty$ DoF for each user. In this paper our goal is to find such tools and ideas to take a forward step which to the best of our knowledge was not discussed before.
\begin{figure}
  \centering
  \begin{tikzpicture}
  \draw (0,0) -- (0,1.5) -- (3.5,1.5) -- (3.5,2) -- (4.5,1) -- (3.5,0) -- (3.5,0.5) -- (3,0.5) -- (3,-1) -- (3.25,-1) -- (2.5,-1.5) -- (1.75,-1) -- (2,-1) -- (2,0) -- (0,0);
  \node            (a) at (-1.5,0.75) {$H(X)\equiv Q$};
  \node            (b) at (7,1) {$H(X)-H(X|Y)\equiv \triangle U$};
  \node            (c) at (2.5,-2) {$H(X|Y)\equiv W$};
  \node            (d) at (1.5,0.75) {$Q=\triangle U+W$};
  \end{tikzpicture}
  \caption{In this structure ``$Q$" is the total energy injected to the system which is equivalent to the Shannon entropy of random variable ``$X$", some part of this energy is released to do work and in Shannon interpretation is equivalent to irreversible ambiguity which is injected to the system by noise. The term ``$\triangle U$" is what remains in the system and is equivalent to capacity of the Shannon peer-to-peer channel.}
\end{figure}
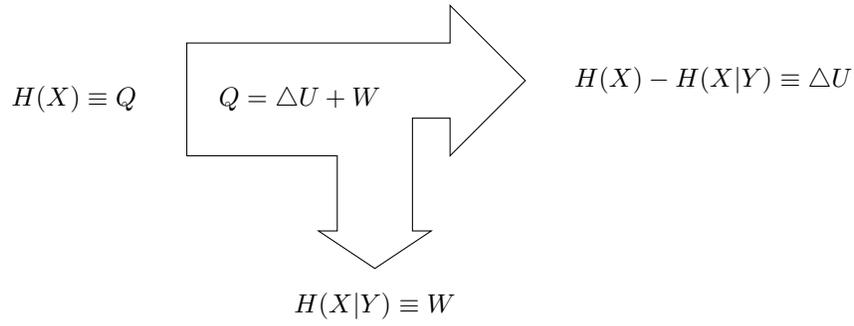
\subsection{The Motivation:}
In 1948, when Shannon published his paper ``A Mathematical Theory of Communication" \cite{shannon1}, information theory had been considered as a new field of study. But without any doubts, the starting point of Shannon idea was adopted from equations of Von-Neumann and Boltzmann in the field of statistical mechanics. Looking back at the history of information theory, we find that when scientists have some physical viewpoint on the problems of this field, they pave the way for most of them. For example, the Kullback-Leibler divergence has strong relation with conservation law and Gibbs' inequality in the jargon of physical scientists. In \cite{Merhav} Merhav, has gathered some of physical interceptions related to channel and source coding problems. In general form, every uncertainty in channel or source coding problems can be modeled with irreversible energy losing in an isolated system. But how can this fact help us in our specific problem? Consider a peer to peer communication channel, in this case transmitted signal entropy $H(X)$ can be modeled with a throughput thermal energy to the lossy channel by wasting rate of $H(X|Y)$ in the unit of time. In other words, every uncertainty in the channel coding problem can be modeled by losing entropy (see Figure 1). In the $K$-user IA with perfect CSI, we have two types of wanted and unwanted information flow rates. The information flows from transmitters to desired receivers as the first and wanted type and as well as to undesired receivers in the form of interference as the second and unwanted type. Some questions that may be raised here are as follows: Is it required to restrict the second type of the information flow? If so, what are the benefits? The important role of CSI on achievable DoF of $K$-user interference channel cannot be disregarded, but finding CSI is not an easy job. Any uncertainty about CSI may lead to drastically reducing DoF, but this uncertainty in physical interview can be modeled by losing entropy or wasting information flow in the communication scenario. Although the restriction of the second type of information flow is not needed, it is better for wasting information flow caused by imperfect CSI condition just to affect this type of information flow (leakage rate). Therefore, this fact not only solves some of our problems about imperfect CSI but also enables us somehow to create confidential communication between transmitters and their desired receivers, which is not in the main scope of this paper. This starting point gives us assurance about finding solution but unfortunately cannot shed any light on the details of this work. 
In this paper we want to prepare several tools to introduce some solutions for this problem. Considering a case where transmitters $\{1,\dots,K\}$ send their messages $\{M^{[1]},\dots,M^{[K]}\}$ during $n$ transmissions, we want to create some trade-off between decodable information received from interference paths $R_{L}^{[j]}=I\left( M^{[1]},\dots,M^{[k]},\dots,M^{[K]};Y^{[j]} \vert {\bf{\Theta'}} \right),~k\neq j$ at $j^{th}$ receiver and channel state information at both transmitters and receivers. In this relation ${\bf{\Theta'}}$ shows some parts of total channel state information. Indeed, for the specific cases in the absence of some part of channel state information and some limitation on the minimum time variation of channels, one can achieve more than one DoF for the well-known $K$-user time-varying interference channel problem. Throughout the paper, this fact is pinpointed to prove statements about more general partial CSI and achievable DoF.
\subsection{Particular Features of Our Solution}
In this paper we discuss a number of fundamental topics as follows:
\begin{itemize}
  \item In some cases of imperfect CSI, using interference alignment (IA) we can achieve DoF beyond what can be achieved with time or frequency sharing methods.
  \item We define ``channel mobility" and we show that the achievable DoF have direct relation with this term.
  \item We define ``channel changing pattern" and we show that having different changing patterns between transceivers can help us to achieve more than one DoF.
  \item In more practical scenarios when the channels with the same destination have the same changing patterns e.g. all the channels which are connected to the $j^{th}$ receiver, we propose a method to achieve more than one DoF. In this case we find out an upper-bound. This upper-bound asymptotically goes to $\frac{\sqrt{K}}{2}$, when the number of users goes to infinite.
  \item In the fast fading interference channel when half of the channel values among unintended transceivers are unknown, the $\frac{K}{2}$ DoF is also achievable.
  \item Using converse proof we show that knowledge of the half of channel values among unintended transceivers is essential to achieve maximum DoF of $\frac{K}{2}$. 
  \item We show that there has to be a trade-off between leakage rate and CSI uncertainty.
\end{itemize}
\subsection{Organization}
The rest of the paper is organized as follows. The next section describes the system model and the channel setup considering $K$ transmitters and receivers. In section II we introduce some definitions considering DoF, channel setup and both perfect and imperfect IA. Some examples are provided in section III and continued by finding much more efficient precoder designing. We found out a new DoF rate region when the perfect channel state information is not available at transmitters and there are different characteristic function between direct and interference channels. For more applied case when all the channel links ending in the same destination have the same changing pattern, we find an upper-bound. In section IV, we proposed a method to achieve $\frac{3}{2}$ DoF for the 3-user interference channel with partial unknown interference channel. This method is generalized to achieve $\frac{K}{2}$ DoF for the $K$-user interference channel in appendix. We showed that the proposed method of section IV for the $K$-user interference channel is optimum in the case of minimum channel knowledge. In section VI, we explore information theoretic interpenetration of our solution method and finally conclusions are presented in section VII. 
\subsection{Notation}
 Throughout the paper, boldface lower-case letters stand for vectors while upper case letters show matrices. ${\bf{A}}^\mathrm{H}$ shows Hermitian of matrix $\bf{A}$. $\mathrm{tr}\{\bf{A}\}$ is defined to be sum of elements on the main diagonal of $\bf{A}$. ${\bf{A}}^\mathrm{T}$ means transpose operation on $\bf{A}$. $\left \lfloor.\right \rfloor$ and $\left \lceil.\right \rceil$ represent floor and ceiling operations, respectively. Also, for the set $C$, $\lvert C \rvert$ shows the cardinality of the set $C$. 
\begin{figure}
  \centering
  \begin{tikzpicture}
  \draw (-4,6) circle (0.45cm);
  \node            (a) at (-4,6){$\text{TX}_1$};
  \draw (-4,4.5) circle (0.45cm);
  \node            (b) at (-4,4.5){$\text{TX}_2$};
  \node            (c) at (-4,3){$\vdots$} ;
  \draw (-4,1.5) circle (0.45cm);
  \node            (d) at (-4,1.5){$\text{TX}_K$};
  \draw (4,6) circle (0.45cm);
  \node            (a1) at (4,6){$\text{RX}_1$};
  \draw (4,4.5) circle (0.45cm);
  \node            (b1) at (4,4.5){$\text{RX}_2$};
  \node            (c1) at (4,3){$\vdots$} ;
  \draw (4,1.5) circle (0.45cm);
  \node            (d) at (4,1.5){$\text{RX}_K$};
  \node            (e) at (-3.55,6){};
  \node            (f) at (-3.55,4.5){};
  \node            (g) at (-3.55,1.5){};
  \node            (h) at (3.55,6){};
  \node            (m) at (3.55,4.5){};
  \node            (j) at (3.55,1.5){};
  \draw[->] (e) edge (h);
  \draw[->] (e) edge (m);
  \draw[->] (e) edge (j);
  \draw[->] (f) edge (h);
  \draw[->] (f) edge (m);
  \draw[->] (f) edge (j);
  \draw[->] (g) edge (h);
  \draw[->] (g) edge (m);
  \draw[->] (g) edge (j);
  \end{tikzpicture}
  \caption{Trivial form of interference channel in which through a channel there is a signaling path between each transmitter and receiver. In this figure we have two different types of channels. For $i^{th}$ user the first one is the channel in which the connection is generated between $\text{TX}_i$ and $\text{RX}_i$. The second one is the channel in which the connection is generated between undesired transceivers e.g. $\text{TX}_i$ and $\text{RX}_j$ where, $i \neq j$.}
\end{figure}
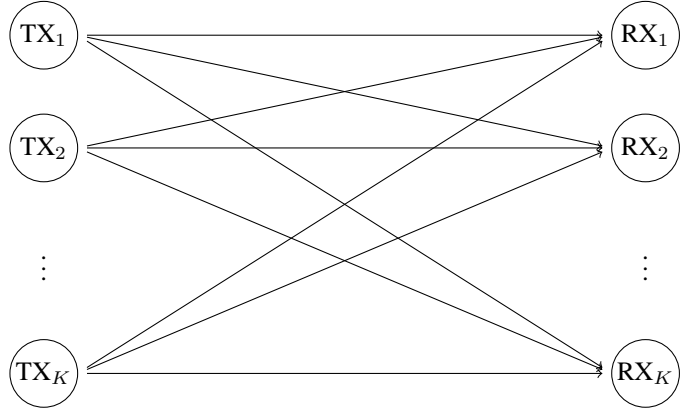
\section{Problem Formulation}
As illustrated in Figure 2, we consider the $K$-user Interference Channel in which there is a signaling path between different transmitters and receivers. This channel consists of $K$ transmitters ${\left \{\mathrm{TX}_k  \right \}_{k=1}^{K}}$ and $K$ receivers ${\left \{\mathrm{RX}_k  \right \}_{k=1}^{K}}$. Let a discrete interference channel be $K^{2}+2K$-tuple $\left({\bar{\bf{H}}}^{[11]},{\bar{\bf{H}}}^{[12]},...,{\bar{\bf{H}}}^{[KK]},\bar{\bf{X}}^{[1]},...,\bar{\bf{X} }^{[K]},\bar{\bf{Y}}^{[1]},...,\bar{\bf{Y}}^{[K]}\right)$, where $\left(\bar{\bf{X}}^{[1]},...,\bar{\bf{X}}^{[K]}\right)$ and $\left(\bar{\bf{Y}}^{[1]},...,\bar{\bf{Y}}^{[K]}\right)$ are $K$ finite inputs and outputs of the channel, respectively. The input of $\mathrm{TX}_k$ is represented by $\bar{\bf{X}}^{[k]}=[{X}_{1},....,{X}_{n}]^T$.
For a specific case where the thermal noise power is zero, ${\bar{\bf{H}}}^{[pq]}$ is a collection of such  $\mathrm{diag}\left(\left[{h_{1}^{[pq]},h_{2}^{[pq]},...,h_{n}^{[pq]}}\right]\right)$ matrices where they map $\bar{\bf{X}}^{[q]}$ to received signal at $\mathrm{RX}_p$ and represent the channel model. In all the sections of this paper, the channel coefficients are assumed to be generic, i.e., drawn from a continuous distribution, and the values of direct channels are assumed to be known to their receivers perfectly. The received signal at $\mathrm{RX}_p$ can be represented as follows:
\begin{equation}
\bar{\bf{Y}}^{[p]}=\sum_{\underset{q\neq p}{q=1}}^K \bar{{\bf H} }^{[pq]}\bar{\bf{X}}^{[q]}+F_p \left(\bar{{\bf H} }^{[pp]}\bar{\bf{X}}^{[p]}\right)+\bar{\bf{Z}}^{[p]},
\end{equation}
where, $\lim_{n \to \infty}{\frac{1}{n}\mathrm{tr}\{\bar{\bf{X}}^{[q]}(\bar{\bf{X}}^{[q]})^H\}}=\mathrm{SNR}$ and $\lim_{n \to \infty}\frac{1}{n}\mathrm{tr}\{{\bar{\bf{H}}}^{[pq]}({\bar{\bf{H}}}^{[pq]})^{\mathrm{H}}\}=1$. $\bar{\bf{Z}}^{[p]}$ is a $n\times 1$ column matrix which shows additive white Gaussian noise with zero mean and variance of one. Also, $F_p \left(.\right)$ shows linear transform function of direct channel at $p^{th}$ receiver. In other words, if ${\bf{x_1}}$ and ${\bf{x_2}}$ are two column matrices of size $n \times 1$, we have:
\begin{equation}
F_p(a{\bf{x_1}}+b{\bf{x_2}})=aF_p({\bf{x_1}})+bF_p({\bf{x_2}}),~a,b \in \mathbb{R}.
\end{equation}
The function $F_p(.)$ can be modeled with the matrix ${\bf{F}}_{p}$ which is an $n\times n$ full rank matrix (not just diagonal) which transfers $\bar{{\bf H} }^{[pp]}\bar{\bf{X}}^{[p]}$ to the form of ${\bf{F}}_{p}\bar{{\bf H} }^{[pp]}\bar{\bf{X}}^{[p]}$. Throughout this paper, we assume that each transmitter is unaware of the data being sent by other transmitters, i.e, there is no cooperation among different transmitters.\\
{\textit{Note:}} The $F_p(.)$ shows some channel properties such as channel permutation and memory.\\
{\textit{Note:}} The ${\bf{F}}_{p}$ is the full rank matrix and its elements are either 0 or nonzero.\\
{\textit{Note:}} At each transmission block, we have $\lim_{n \to \infty}\frac{1}{n}\mathrm{tr}\{{\bf{F}}_{p}{{\bf{F}}}_{p}^{\mathrm{H}}\}=1$.
\subsection{Degrees of Freedom} 
In the $K$-user interference channel the total power across all transmitters is $\rho$. The capacity region $\mathcal{\emph{C}}(\rho)$ of the $K$-user interference channel is set to be $\mathcal{\emph{R} }(\rho)=\left({R}_{1}(\rho),{R}_{2}(\rho),\dots,{R}_{K}(\rho)\right)$.
In the $K$-user interference channel, we define the degrees of freedom region as follows\cite{jafar}:
\begin{eqnarray}
\Bigg\{(d_1,d_2,\dots,d_K)\in \mathbb{R}_{+}^{K}:&&\forall (w_1,\dots,w_K) \in \mathbb{R}_{+}^{K},{w_1}{d_1}+\dots + {w_K}{d_K}\nonumber \\
&&\leq\lim_{\rho \to \infty} \sup\left [\underset{\mathcal{\emph{R}(\rho) \in \mathcal{C}(\rho) }}{\sup}\frac {(w_1{R}_{1}(\rho)+\dots+w_K{R}_{K}(\rho))}{\log(\rho)}\right ]\Bigg\}.
\end{eqnarray}
\subsection{Channel modeling related to transmission rate and coherence time of the channel}
Coherence time is the time duration over which the channel response is considered to be constant. If $T_t$ is the total transmission time duration and $T_s$ be the total time duration of each symbol, we have:
\begin{equation}
T_t=nT_s.
\end{equation}
In this work, we assume a block fading model in time, where channel states are constant for an average time duration of $T_c$. Therefore during transmission time consisted of $n$ time slots, if $T_s \leq T_c$ we have:
\begin{equation}
\label{channel1}
 {h}_{{1}}^{[pq]}=\dots={h}_{{c_{1}^{[pq]}-1}}^{[pq]}\neq {h }_{{c_{1}^{[pq]}}}^{[pq]}=\dots={h}_{{c_{2}^{[pq]}-1}}^{[pq]} \neq \dots \neq {h}_{{c_{R(p,q)}^{[pq]}}}^{[pq]}=\dots={h}_{{n}}^{[pq]}.
\end{equation}
Where, $c_{j}^{[pq]},~j\in \{1,\dots,R(p,q)\}$ shows the $j^{th}$ point of altering state of channel between $\mathrm{TX}_q$ and $\mathrm{RX}_{p}$. The value $c_{j}^{[pq]}-c_{j-1}^{[pq]}$, is a random variable with the mean of $\left \lfloor \frac{T_c}{T_s} \right \rfloor$.
All ${h}_{j}^{[pq]}$ are $i.i.d$ random variables with a specific distribution and are bounded between a nonzero and a finite maximum value. Since $\left \lfloor \frac{T_c}{T_s} \right \rfloor$ can be equal to 1, the assumption of coherence time of channel does not reduce the generality of our problem.\\ 
{\bf{Definition:}}~We define $\lim_{n \to \infty}\frac{R(p,q)}{n}$ as channel mobility rate, which shows how fast the channel beats between $\mathrm{TX}_q$ and $\mathrm{RX}_p$.\\
{\bf{Definition:}}~The set ${C}^{[pq]}_l$ is called the changing pattern of the channel between $\mathrm{TX}_q$ and $\mathrm{RX}_p$. This changing pattern probably can be chosen among $L^{[pq]}$ sets.\\
{\bf\emph{Remark:}}
There may be ambiguities among $L^{[pq]}$ sets of ${C}_{l}^{[pq]},p\neq q, l\in \left\{1,...,\lvert{L^{[pq]}}\rvert \right\}$.\\
{\bf{Definition:}}~$U^{[pq]} \subseteq \{1,\dots,n\}$ is the set in which the exact channel value between $\mathrm{TX}_q$ and $\mathrm{RX}_p$ during $\lbrace 1,2,\dots,n\rbrace$ time snapshots is not known.

In this paper, we consider the following assumptions:
\begin{enumerate}
\item The $\mathrm{RX}_p$ has knowledge of ${\bf{F}}_{p}{\bar{{\bf H} }}^{[pp]},p\in\left\{1,\dots,K\right\}$.
\item All transceivers know the values of ${h}^{[pq]}_{j},p,q\in\left\{1,\dots,K\right\}$, where $j \in \lbrace 1,2,\dots,n\rbrace-U^{[pq]}$.
\item In relation \eqref{channel1}, all transceivers know all sets of ${C}^{[pq]}_l=\left\{{c}_{1}^{[pq]_{l}},\dots,{c}_{R(p,q)}^{[pq]_{l}}\right\}$ where, ${c}_{1}^{[pq]_l}<\dots<{c}_{R(p,q)}^{[pq]_l}$.
\end{enumerate}
\subsection{Linear precoding, Perfect and imperfect IA}
IA is an elegant method to reduce the effects of some parts of the interference signals. Authors in \cite{jafar} showed that IA is the optimum scheme of enhancing DoF for each user in linear interference channels. It is based on designing precoding matrices $\bar{\bf{V}}^{[q]}$ with the size of ${n}\times{d_q}$ to encode transmitted information. Let $M^{[1]}=\{1,2,...,2^{nR_1}\},M^{[2]}=\{1,2,...,2^{nR_2}\},\dots, M^{[K]}=\{1,2,...,2^{nR_K}\}$ be the message sets of transmitters. These messages are encoded as $\bar{\bf{X}}^{[q]}=\bar{\bf{V}}^{[q]}\bar{\emph{X}}^{[q]}$ using the encoding function $e_{q}\left(M^{[q]},\bar{\emph{X}}^{[q]} |{\bar{\bf{H}}}^{[pq]} \right), p,q\in{1,\dots,K}$. The conditional term in the encoding function shows the designed precoder depends on the local view of transmitters from estimated interference channels. In other words, in the extended channel mode $\bar{\bf{X}}^{[q]}$ can be represented as follows:
\begin{equation}
\bar{\bf{X}}^{[q]}=\sum_{m=1}^{d_q}x_{m}^{[q]}{\bf{v}}_{m}^{[q]}.
\end{equation}  
For the perfect IA, encoding function should preserve the following conditions 
\begin{equation}
\label{align1}
\mathrm{span}\left({{\bar{\bf{H}}}^{[pq]}\bar{\bf{V}}^{[q]}}\right) \prec \mathrm{span}\left({{\bar{\bf{H}}}^{[p1]}\bar{\bf{V}}^{[1]}}\right), q\neq 1,p,~~
\footnote[2]{\begin{math}
\mathrm{span}\left({\bar A}\right) \prec \mathrm{span}\left({\bar B}\right)
\end{math} means that the span of the matrix
\begin{math}
 \bar A
\end{math}
 is the subset of the span of the matrix
\begin{math}
 \bar B
\end{math}.}
\end{equation}
and
\begin{equation}
\label{align2}
\bar{{\bf H}}^{[12]}\bar{\bf{V}}^{[2]}={\bar{\bf{H}}}^{[13]}\bar{\bf{V}}^{[3]}=\dots={\bar{\bf{H}}}^{[1K]}\bar{\bf{V}}^{[K]}.
\end{equation}
Moreover, relation \eqref{align2} can be generalized as follows:
\begin{equation}
\label{align3}
\mathrm{span}\left({\bar{\bf{H}}}^{[12]}\bar{\bf{V}}^{[2]}\right)=\mathrm{span}\left({\bar{\bf{H}}}^{[13]}\bar{\bf{V}}^{[3]}\right)=\dots=\mathrm{span}\left({\bar{\bf{H}}}^{[1K]}\bar{\bf{V}}^{[K]}\right) 
\end{equation}
The above conditions for the perfect IA can be degraded to the following conditions for imperfect IA:
\begin{equation}
{\bar{\bf{H}}}^{[pq]}\bar{\bf{V}}^{[q]} \prec {\bar{\bf{I}}^{[p]}},~~\sum_{p=1}^{K}{\dim(\bar{\bf{I}}^{[p]})}<(K-1)n,
\end{equation}
where, ${\bar{\bf{I}}^{[p]}}$ is the interference subspace at $p^{th}$ receiver. The messages are decoded at receivers using the decoding function $d_{p}\left({\bar{\bf{Y}}}^{[p]}|{\bar{\bf{H}}}^{[pq]}\right), p,q\in \lbrace 1,2,\dots,K \rbrace$ based on zero-forcing interference signal from received signal. Also, the desired signal subspace at $p^{th}$ receiver is shown by the ${\bar{\bf{D}}^{[p]}}$. For the perfect IA, the number of dimensions of the desired signal at each receiver can be limited as follows:
\begin{equation}
\dim{\left({\bar{\bf{D}}}^{[p]} \right)}=\frac{n}{2},
\end{equation}
and in the case of imperfect IA, we have:
\begin{equation}
\sum_{p=1}^{K}{\dim{\left({\bar{\bf{D}}^{[p]}}\right)}}>n.
\end{equation}
\textit{Note:} The total dimension of interference and desired signal at $\mathrm{RX}_p$ should be limited as follows:
\begin{equation}
\dim{\left( {\bar{\bf{D}}^{[p]}+{\bar{\bf{I}}^{[p]}}} \right)} \leq n.
\end{equation}
\section{Preliminary IA Examples and BIA DoF Rate Regions}
In this section a number of preliminary examples are explored to explain the main idea of IA with imperfect CSI. The interference alignment in multi-user scenario is based on beamforming over multiple symbol extensions of the time varying channel. The first example discuses the idea of BIA when the direct channels (e.g. $\bar{\bf{H}}^{[pp]}$) have more channel mobility than the cross ones (e.g. $\bar{\bf{H}}^{[pq]}, p \neq q$). Since in the most cases the assumption of $C^{[pp]} \neq C^{[pq]}$ is infeasible, in the second example the idea of BIA is extended to to more practical form of $C^{[pp]}=C^{[pq]}$. The assumption of $C^{[pp]}=C^{[pq]}$ shows that all channels which are ended to the same destination experience similar changing pattern. In the third example, the method of second example is improved. We continue this section by finding both achievability and converse proof for the case of $C^{[pp]}=C^{[pq]}$. At last, when the direct channels have more ``channel mobility" than cross ones we present an algorithm to achieve more than one DoF. Also, we show that achievable rate region for each transmitter has a direct relation to the ``channel mobility" parameter.\\
{\bf{Example 1:}} IA with imperfect CSI using channel mobility

Consider $K$-user interference channel where each transmission slot consisted of $n\in\{2,4,\dots\}$ time snapshots. We assume that $\bf{F}_p$ is an $n \times n$ identity matrix. Therefore, the received signal at $p^{th}$ receiver can be modeled as follows:
\begin{equation}
\label{received signal model}
\bar{\bf{Y} }^{[p]} =\sum_{q=1}^K \bar{{\bf H}}^{[pq]} \bar{\bf{X}}^{[q]} +\bar{\bf{Z}}^{[p]},
\end{equation}
where, $\bar{\bf{Y}}^{[p]} $ is an $n\times 1$ received matrix and transmitted signal represented by $\bar{\bf{X}}^{[q]} =\bar{\bf{V}}^{[q]} \bar{X}^{[q]}$. The $\bar{\bf{V}}^{[q]} $ is an $n\times\frac{n}{2}$ matrix and show precoder which is used by transmitter $q$. The ${\bar{X}}^{[q]}$ shows transmitted information and can be represented as follows:
\begin{equation}
 \bar{X}^{[q]}={\left[ x^{q}_1,\dots,x^{q}_{\frac{n}{2}} \right]}^{\mathrm{T}},~x^{q}_j \in \{x_1,x_2,\dots,x_{2^{n{R_q}}}\}. 
\end{equation}
Let, $\bar{{\bf H}}^{[pq]} , p\neq q$ is an $n\times n$ diagonal matrix with constant diagonal elements of $h^{[pq]}$\footnote[1]{$\bar{\bf H}^{[pq]} = \mathrm{diag} \left( {\left[ {h^{[pq]},\dots,h^{[pq]}} \right]} \right)$.}.
The $\bar{{\bf H} }^{[qq]} $ is an $n\times n$  diagonal matrix with random elements. In order to keep interference aligned in all the receivers we choose $\bar{\bf{V}}^{[1]}=\bar{\bf{V}}^{[2]}=\dots=\bar{\bf{V}}^{[K]}$. Since $\bar{{\bf H} }^{[pq]}$ is a diagonal matrix, $\bar{{\bf H} }^{[pq]} \bar{\bf{V}}^{[q]},p\neq q $ aligned with ${\bar{\bf{V}}^{[q]} }$. Similarly at all receivers all interferences arrive along ${\bar{\bf{V}}^{[q]} }$. In this case, the desired signal arrives along ${\bar{\bf{H}}}^{[pp]} \bar{\bf{V}}^{[p]} $ but almost surely due to random nature of ${\bar{\bf{H}}}^{[pp]}$ the space span by this vector is linearly independent of interference subspace. Therefore, the achievable DoF for $p^{th}$ user can be calculated as follows:
\begin{equation}
\frac{d_p}{n}=\frac{\mathrm{dim}\left({\bf{D}}^{[p]}\right)}{\mathrm{dim}\left({\bf{D}}^{[p]}+{\bf{I}}^{[p]}\right)}=\frac{1}{2}
\end{equation}
\subsection{BIA using different changing pattern between direct and interference channels:}
Consider $K-$user interference channel with unknown channel state information defined in section II. During signaling time this channel consists of $n$ time snapshots. The set ${C} ^{[pq]_l},~l=[1:L^{[pq]}]$, shows all the possible changing patterns of the channel between $\mathrm{TX}_q$ and $\mathrm{RX}_p$. In this case we assume there is no uncertainty about the direct channel changing pattern set ($L^{[pp]}=1$). Let, all the channel matrices e.g. ${\bar{\bf{H}}}^{[pp]} $ are available at $\mathrm{RX}_p$. We assume all the receivers know the value of its corresponding channel. In other words, for all the receivers, $U^{[pp]}=\varnothing,~p\in \{1,\dots,K\}$ and $U^{[pq]}=\{1,2,\dots,n\},~p\neq q.$
 In this subsection our goal is to generalize Example1 and find a solution for more general cases. Our objective is to design proper encoding $e_{q}\left(M^{[q]},\bar{\bf{X}}^{[q]} |{\bigcup_{l,p,q}{C}^{[pq]}_{l}}\right)$ and decoding $d_{q}\left(\bar{\bf{Y}}^{[q]} |{\bigcup_{l,p,q}{C}^{[pq]}_{l}},\bar{\bf{H}}^{[pp]}\right), p,q\in \{1,\dots,K\}$
functions in transmitters and receivers, respectively that satisfy perfect or imperfect IA conditions. For the $K$-user interference channel with channel definition of \eqref{channel1}, the received signal like \eqref{received signal model} is given by following relation:
\begin{equation}
\label{RX2}
\bar{\bf{Y}}^{[p]} ={\bar{\bf{H}}}^{[pp]}{\bf{F}}_{p} \bar{\bf{X}}^{[p]}+\sum_{q=1, q \neq p}^{K} {\bar{\bf{H}}}^{[pq]} \bar{\bf{X}}^{[q]}+\bar{\bf{Z}}^{[p]},
\end{equation}
where, $\bf{F}_{p}$ is an identity matrix. Let we define all the collection of the sets ${C}^{[pq]}_l$ by the set $C$ as follows:
\begin{equation}
{C}={\bigcup_{l,p,q}{C}^{[pq]}_{l}} {\text{where,}}~ l=[1:L^{[pq]}],~p,q \in \{1,\dots,K\}.
\end{equation}
The set ${C}$ can be represented as follows:
\begin{equation}
{C}=\lbrace c_1,c_2,\dots,c_{\sigma}\rbrace, ~\lvert{C}\rvert =\sigma.
\end{equation}
In order to analysis the direct and cross channels separately, the set ${C}$ contains two subsets of ${C}'$ and ${C}''$ in which:
\begin{equation}
\left\{\begin{matrix}
{C}'&=&\bigcup_{l,p,q}{C}^{[pq]}_{l},&~p\neq q \in \{1,\dots,K\},l=[1:L^{[pq]}]\\ 
{C}''&=&\bigcup_{l,p}{C}^{[pp]}_{l},&~p \in \{1,\dots,K\},~l=1
\end{matrix}\right.
\end{equation}
We assume ${{C}'}=\lbrace c'_1,\dots,c'_{\sigma'}\rbrace$ and  ${{C}''}=\lbrace c''_1,\dots,c''_{\sigma''}\rbrace$.
Now, we explore a lemma in order to analyze $\bar{{\bf H} }^{[pq]} $ with some basic matrices.
\begin{lem}
\it{The matrix $\bar{{\bf H} }^{[pq]} , p \neq q$ can be represented by $\sum_{j=1}^{\sigma '+1}\beta^{[pq]}_j{\bf{\bar{I}}}{{\bar{\bf{Q}}}}^j$ where, $\beta^{[pq]}_j{\bf{\bar{I}}}$ is a matrix with $\beta^{[pq]}_j$ diagonal elements and ${\bar{{\bf{Q}}}}=\mathrm{diag}\left(\left[{{q}_{1},\dots,{q}_{n}}\right]\right)$ where,
\begin{equation}
\left\{\begin{matrix}
{q}_{1}&=&\gamma_{1}~~~~~~~~~~~~~\\
{q}_r&=&{q}_{r-1}~\text{if}~r\notin {C'} \\
{q}_r&=&\gamma_{r}~~~~\text{if}~r\in {C'}
\end{matrix}\right.
\end{equation}
and $\gamma_r$ is a random number generated from arbitrary distribution.}
\end{lem}
\begin{IEEEproof}
The proof establish with finding such $\beta^{[pq]}_j, j=\left\{1,\dots,\sigma'+1\right\}$, that satisfy following linear equation:\\
\begin{equation}
\begin{bmatrix}
{q}_{1}^{1}& \dots& {q}_{1}^{j}& \dots& {q}_{1}^{\sigma'+1}\\
{q}_{c'_1}^{1}& \dots& {q}_{c'_1}^{j}& \dots& {q}_{c'_1}^{\sigma'+1}\\
\vdots& \dots& \vdots& \dots& \vdots\\
\\
{q}_{c'_{\sigma'}}^{1}& \dots& {q}_{c'_{\sigma'}}^{j}& \dots& {q}_{c'_{\sigma'}}^{\sigma'+1}\\
\end{bmatrix}
\begin{bmatrix}
\beta^{[pq]}_1\\
\vdots\\
\beta^{[pq]}_j\\
\vdots\\
\beta^{[pq]}_{\sigma '+1}\\
\end{bmatrix}=
\begin{bmatrix}
h^{[pq]}_1\\
h^{[pq]}_{c'_1}\\
h^{[pq]}_{c'_2}\\
\vdots\\
h^{[pq]}_{c'_{\sigma'}}\\
\end{bmatrix}.
\end{equation}
 Since in the above equation the left square matrix has random elements it is a full rank and invertible matrix with most probability. Therefore, this linear equation system has single unique solution for $\sigma'+1$ tuple $\big(\beta^{[pq]}_1,\dots,\beta^{[pq]}_j,\dots,\beta^{[pq]}_{\sigma'+1}\big)$, this completes the proof of this lemma.
\end{IEEEproof}
{\textit{Note:}} The direct result from lemma 1 is that, because $\bar{\bf{Q}}$ has the similar changing pattern to the matrix $\bar{{\bf H} }^{[pq]} , p \neq q$, the diagonal matrix $\left(\bar{\bf{H}}^{[pq]} \right)^{j},~j\geq 0$ can be represented by $\sum_{j=1}^{\sigma '+1}\beta^{[pq]}_j{\bf{\bar{I}}}({\bar{\bf{Q}} })^j$.\\
Let $n=2\varrho (\sigma'+1)$ and $\varrho$ be such a natural number that $n >> \max_{l,p,j}{\{c^{[pp]_l}_j-c^{[pp]_l}_{j-1}\}},~p\in\{1,\dots,K\},~l=1$ and $j\in\{2,\dots,R(p,p)\}$. Referring Lemma 1, since $\beta^{[pq]}_j\bar{\bf{I}}$ is a diagonal matrix with the diagonal elements of $\beta^{[pq]}_j$, the relation \eqref{RX2} can be rewritten by the following relation:
\begin{equation}
\label{RX2 signal}
\begin{aligned}
\bar{\bf{Y}}^{[p]} &={\bar{\bf{H}}}^{[pp]} \bar{\bf{V}}^{[p]} \bar{X}^{[p]} +\sum_{k=1, k \neq p}^{K}{\left(\sum_{j=1}^{\sigma'+1}(\bar{\bf{Q}} )^{j}\beta^{[pk]}_j{\bf{\bar{I}}}\right)\bar{\bf{V}}^{[k]}{\bar{X}}^{[k]}} \\
&={\bar{\bf{H}}}^{[pp]} \bar{\bf{V}}^{[p]} \bar{X}^{[p]} +\sum_{k=1, k \neq p}^{K}{\left(\sum_{j=1}^{\sigma'+1}(\bar{\bf{Q}} )^{j}\bar{\bf{V}}^{[k]} \beta^{[pk]}_j{\bf{\bar{I}}}\right){\bar{X}}^{[k]}} \\
\end{aligned}
\end{equation} 
where, $\bar{\bf{V}}^{[p]} \bar{X}^{[p]} $ shows transmitted signal of $\mathrm{TX}_p$. From the view point of $\mathrm{RX}_p$, $M^{[p]}$ is the message of $\mathrm{TX}_p$ in the transmission time which is consisted of $d_p$ independent streams $x^{(p)}_{s} , s=1,2,\dots,d_p$ along vector $\bar{\bf v }^{[p]}_{s} $. Therefore, we define $\bar{\bf X}^{[p]} $ as follows:
\begin{equation}
\bar{\bf X}^{[p]} =\sum_{s=1}^{N} x^{(p)}_{s} \bar{\bf v }^{[p]}_{s} =\bar{\bf{V}}^{[p]}\bar{X}^{[p]}.
\end{equation}
where, $\bar{\emph{X} }^{[p]} $, $\bar{\bf v }^{[p]}_{s} $ are $d_p\times 1$ and $n\times 1$ column matrices, respectively. 
In all the above relations $x^{(q)}_{s} \in\{x_1,\dots,x_{Q}\}$, $|M^{[q]}|=2^{n\log _2 Q}$. All the receivers decode the desired signal by zero-forcing the interference vectors. In order to align interference signals at each receiver it is sufficient to design such $\bar{\bf{V}}^{[q]} $ vectors that satisfy the following relations:
\begin{equation}
{{\mathrm{span}}} \left({{\bar{\bf{Q}}}^{j}\bar{\bf{V}}^{[q]}}\right)\subseteq {\mathrm{span}}\left(\bar{\bf{I}}^{[p]} \right),(p\neq q)~\text{and}~j\in\lbrace {1,2,\dots,L^{[pq]}} \rbrace,
\end{equation}
where, $\bar{\bf{I}}^{[p]}$ is the interference subspace at $\mathrm{RX}_p$ receiver. Let us choose $\frac{n}{2}$ column vectors of $\bar{\bf{I}}^{[p]} $ and $\frac{n}{2}$ column vectors of $\bar{\bf{V}}^{[k]} $ from the following set:
\begin{equation}
\label{S}
I=\left\lbrace {\left( \bar{\bf{Q}}^{\alpha}{\bar{\bf{\Gamma}}}^{j} \right){\bf{W}}: \forall {\alpha} \in \{1,\dots,\sigma'+1 \},\forall {j} \in \{1,\dots,\varrho\}}\right\rbrace ,
\end{equation}
where, ${\bf{W}}=\left[1~1 \dots1\right]^{\mathrm{T}}, n\times1$ column matrix and ${\bar{\bf{\Gamma}}}=\mathrm{diag}\left(\left[{\Gamma_{11},\dots,\Gamma_{nn}}\right]\right)$ is a random diagonal matrix with an arbitrarily distribution.\\
{\textit{Note:}} Both interference subspace and transmitted vectors choose their basic vectors from the similar set of $I$. 
\begin{thm}
\label{theorem_one} {In the $K-$user interference channel with designed precoders,  ${{\mathrm{span}}} ({\bar{\bf{H}}}^{[pk]} \bar{\bf{V}}^{[k]} )\subseteq {{\mathrm{span}}} \left({\bar{\bf{I}}^{[p]}}\right),(p\neq k)$.}
\end{thm}
\begin{IEEEproof}
The proof is prepared in appendix A.
\end{IEEEproof}
{\bf{Definition:}} We define the set $\mathcal{B}_{m}, m \in \mathbb{N}$ as follows:
\begin{equation}
\mathcal{B}_{m}=\left\lbrace i \vert i\in [c'_{m-1}:c'_{m}-1], i \in \mathbb{N} \right\rbrace 
\end{equation} 
{\bf{Definition:}} We define the set $C^{[pq]}_{\mathcal{B}_{m}}, m \in \mathbb{N}$ as follows:
\begin{equation}
C^{[pq]}_{\mathcal{B}_{m}}=\left\lbrace c_i \vert c_i \in \mathcal{B}_{m}, c_i \in C^{[pq]} \right\rbrace 
\end{equation} 
For the case of perfect or imperfect IA we need to find the number of desired and interference dimensions at all the receivers. We calculate the number of desired and interference dimensions in the following theorem.
\begin{thm}
\label{theorem_two}  {In the $K-$user interference channel with defined channel setup matrices and designed precoders, during $n=2\varrho(\sigma'+1)$ transmissions we can find $D_{k}= \min \left({\frac{n}{2}},\varrho \sum_{m} {\mathbbm{1}}{\left(\left\vert {C}^{[kk]}_{\mathcal{B}_{m}}-\bigcup_{p,q} {C}^{[pq]}_{\mathcal{B}_{m}}\right\vert > 0\right)}\right),~p\neq q$ free interference dimensions at $\mathrm{RX}_k$.}
\footnote[1]{The function ${\mathbbm{1}}(x>0)$ returns one if $x>0$ and returns zero if $x\leq 0$.}
\end{thm}
\begin{IEEEproof}
See Appendix B.
\end{IEEEproof}
From Theorem 2, it is obvious we can find $D_{k}$ number of vectors which are free from interference at each receiver. Therefore, the achievable DoF for $k^{th}$ user can be obtained from the following relation:
\begin{equation}
\frac{d_k}{n}=\frac{D_{k}}{n}.
\end{equation}    
{\textit{Note:}} It is clear that the above lower bound on the achievable DoF can be improved using time sharing. Therefore, we have:
\begin{equation}
\frac{1}{n}\sum_{k=1}^{K}{d_k} \geq \max{\left(1,\sum_{k=1}^{K}{\frac{D_k}{n}}\right)}.
\end{equation}       
Example1 shows that if the direct channels have different changing pattern from interference channels ({$\lvert {C}_{l=1}^{[pp]}\rvert \geq \lvert {C}_{l=1}^{[pq]} \rvert$}), we can achieve more than one DoF. This example generalized by Theorem 2 but unfortunately since in the most cases the assumption of $\lvert {C}_{l=1}^{[pp]}\rvert \geq \lvert {C}_{l=1}^{[pq]} \rvert$ is not applicable we cannot use this method in practical systems. Therefore, in the next example, without the aid of direct channel mobility we propose a method to align interference signals.\\ 
\begin{figure}
  \centering
  \begin{tikzpicture}
  \draw (-4,6) circle (0.45cm);
  \node            (a) at (-4,6){$\text{TX}_1$};
  \draw (-4,4.5) circle (0.45cm);
  \node            (b) at (-4,4.5){$\text{TX}_2$};
  \draw (-4,3) circle (0.45cm);
  \node            (c) at (-4,3){$\text{TX}_3$} ;
  \draw (-4,1.5) circle (0.45cm);
  \node            (d) at (-4,1.5){$\text{TX}_4$};
  \draw (4,6) circle (0.45cm);
  \node            (a1) at (4,6){$\text{RX}_1$};
  \draw (4,4.5) circle (0.45cm);
  \node            (b1) at (4,4.5){$\text{RX}_2$};
  \draw (4,3) circle (0.45cm);
  \node            (c1) at (4,3){$\text{RX}_3$};
  \draw (4,1.5) circle (0.45cm);
  \node            (d) at (4,1.5){$\text{RX}_4$};
  \node            (e) at (-3.55,6){};
  \node            (f) at (-3.55,4.5){};
  \node            (g) at (-3.55,3){};
  \node            (h) at (-3.55,1.5){};
  \node            (m) at (3.55,6){};
  \node            (j) at (3.55,4.5){};
  \node            (k) at (3.55,3){};
  \node            (l) at (3.55,1.5){};
  \node            (i1) at (7,6){${C}^{[1q]}_{1}=\lbrace 3,4,5,9,10\rbrace$};
  \node            (j1) at (7,4.5){${C}^{[2q]}_{1}=\lbrace 3,4,5,6\rbrace$};
  \node            (k1) at (7,3){${C}^{[3q]}_{1}=\lbrace 1,2,7,8,9,10\rbrace$};
  \node            (l1) at (7,1.5){${C}^{[4q]}_{1}=\lbrace 5,6,7,8,9,10\rbrace$};
  \draw[->] (e) edge (m);
  \draw[->] (e) edge (j);
  \draw[->] (e) edge (k);
  \draw[->] (e) edge (l);
  \draw[->] (f) edge (m);
  \draw[->] (f) edge (j);
  \draw[->] (f) edge (k);
  \draw[->] (f) edge (l);
  \draw[->] (g) edge (m);
  \draw[->] (g) edge (j);
  \draw[->] (g) edge (k);
  \draw[->] (g) edge (l);
  \draw[->] (h) edge (m);
  \draw[->] (h) edge (j);
  \draw[->] (h) edge (k);
  \draw[->] (h) edge (l);
  \end{tikzpicture}
  \caption{4-user interference channel in which there is a signaling path between each transmitters and receivers. All the channels which are ended to the same destination have the same changing patterns. For example the set ${C}^{[1q]}_{1}=\lbrace 3,4,5,9,10\rbrace, q \in \lbrace 1,2,3,4\rbrace$ shows that during 10-time snapshots all the channel matrices ${\bar{\bf{H}}}^{[11]} $, ${\bar{\bf{H}}}^{[12]} $, ${\bar{\bf{H}}}^{[13]} $ and ${\bar{\bf{H}}}^{[14]} $ experience similar changing pattern. In other words, for all these channels there exist transition time at the set $\lbrace 3,4,5,9,10\rbrace$ e.g. $h^{[11]}_1=h^{[11]}_2 \neq h^{[11]}_3 \neq  h^{[11]}_4 \neq h^{[11]}_5=\dots=h^{[11]}_8 \neq h^{[11]}_9 \neq h^{[11]}_{10}$.}
\end{figure}
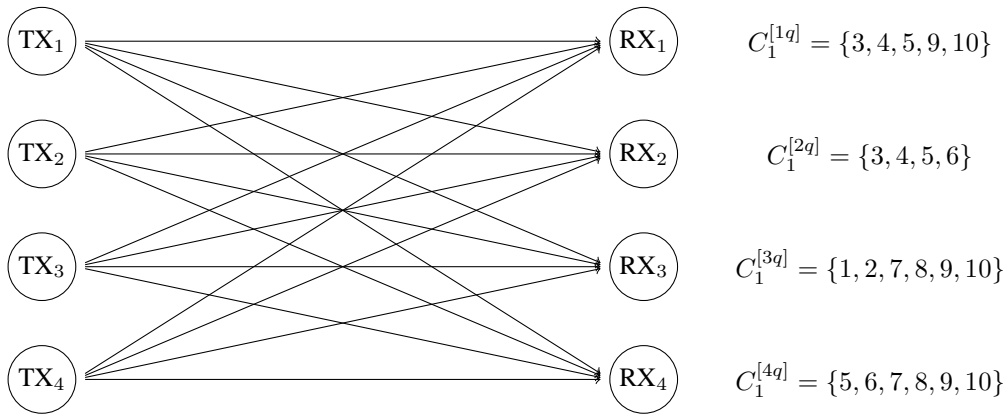
{\bf{Example 2:}} Partial IA with ${C}_{l}^{[pq]}={C}_{l}^{[pq']}, L^{[pq]}=1$

Consider an interference channel in which all the transmitters and receivers are connected to each other using a signaling path (see Figure 3). Let, all the channels ended to the same destination experience similar changing pattern e.g, ${C}_{l=1}^{[11]}={C}_{l=1}^{[12]}={C}_{l=1}^{[13]}={C}_{l=1}^{[14]}$. In this example, our goal is to achieve more than one DoF even by the assumption of ${C}_{l}^{[pq]}={C}_{l}^{[pq']}$. The following steps will provide directions to design proper precoders at different transmitters in details.\\ 
\textit{Step 1:} Consider the following precoder designed vectors at transmitters consisting of $8$ time snapshots:
\begin{equation}
\label{4tvectors2}
\begin{aligned}
&{\bar{\bf{V}}^{[1]}}=\left[ \bf{a} ~\bf{d} ~\bf{f} \right]&\\
&{\bar{\bf{V}}^{[2]}}=\left[ \bf{a} ~\bf{b} ~\bf{e} \right]&\\
&{\bar{\bf{V}}^{[3]}}=\left[ \bf{b} ~\bf{c} ~\bf{d} \right]&\\
&{\bar{\bf{V}}^{[4]}}=\left[ \bf{c} ~\bf{e} ~\bf{f} \right],& 
\end{aligned}
\end{equation}
where $\{\bf{a},\bf{b},\dots,\bf{f}\}$ are the $8 \times 1$ column vectors which are linearly independent. In this case, every basic vector is shared between two different users.\\
\textit{Step 2:} Now, we find the transmission slots in which cross channels ended to the same destination experience constant values. For all the cross channels connected to the first receiver, during time snapshots of $\{1,2\}\cup \{5,6,7,8\}$, all the transmitters experience constant channel values (${\bar{\bf{H}}}^{[1q]}, q=1,2,3,4$).
Similarly, for the second receiver all the channels (${\bar{\bf{H}}}^{[2q]}, q=1,2,3,4$) during time snapshots of $\{1,2\} \cup \{7,8,9,10\}$ have constant values. Finally, for the 3rd and 4th receivers we have constant channel values during $\{1,2,3,4,5,6\}$ and $\{1,2,3,4\}$, respectively.\\
\textit{Step 3:} Using IA, we design vectors $\{\bf{a},\bf{b},\dots,\bf{f}\}$ such that outperform the rate achieved by time sharing method. Since vector $\bf{a}$ was transmitted by the first and second transmitters, it creates interference on 3rd and 4th receivers as ${\bar{\bf{H}}}^{[3q]}{\bf{a}},~q\in \{1,2\}$ and ${\bar{\bf{H}}}^{[4q]}{\bf{a}},~q\in \{1,2\}$. The vectors ${\bar{\bf{H}}}^{[31]}{\bf{a}}$ and ${\bar{\bf{H}}}^{[41]}{\bf{a}}$ should be aligned with the transmitted vectors ${\bar{\bf{H}}}^{[32]}{\bf{a}}$ and ${\bar{\bf{H}}}^{[42]}{\bf{a}}$ from the second transmitter.\\  
Since ${\bar{\bf{H}}}^{[3q]}$ and ${\bar{\bf{H}}}^{[4q]}$ have constant values during $\{ 1,2,3,4,5,6 \} \cap \{1,2,3,4\}=\{1,2,3,4\}$, the non-zero elements of the vector ${\bf{a}}= {\left[ a_1 ~a_2 ~a_3 ~a_4 ~a_5 ~a_6 ~a_7 ~a_8 \right]}^{\mathrm{T}}$, $a_i$ can be selected from the set of $i \in \{1,2,3,4\}$. Similarly, for other vectors $\{\bf{b},\bf{c},\bf{d},\bf{e},\bf{f}\}$ we should have:
\begin{itemize}
  \item If ${\bf{b}}= {\left[ b_1 ~b_2 ~b_3 ~b_4 ~b_5 ~b_6 ~b_7 ~b_8 \right]}^{\mathrm{T}}$, $b_i$ could be non-zero for \\$i\in \left({\{1,2\} \cup \{5,6,7,8\}}\right) \cap \{1,2,3,4\}=\{1,2\}.$
  \item If ${\bf{c}}= {\left[ c_1 ~c_2 ~c_3 ~c_4 ~c_5 ~c_6 ~c_7 ~c_8 \right] }^{\mathrm{T}}$, $c_i$ could be non-zero for \\$i\in \left( \{1,2\} \cup \{5,6,7,8\} \right) \cap \left( \{1,2\} \cup \{7,8,9,10\} \right)=\{1,2\} \cup \{7,8\}.$
  \item If ${\bf{d}}= {\left[d_1 ~d_2 ~d_3 ~d_4 ~d_5 ~d_6 ~d_7 ~d_8\right]}^{\mathrm{T}}$, $d_i$ could be non-zero for \\$i\in \{1,2,3,4\} \cap \left(\{1,2\} \cup \{7,8,9,10\}\right)=\{1,2\}.$ 
  \item If ${\bf{e}}= {\left[e_1 ~e_2 ~e_3 ~e_4 ~e_5 ~e_6 ~e_7 ~e_8\right]}^{\mathrm{T}}$, $e_i$ could be non-zero for \\$i\in \left(\{1,2\}\cup\{5,6,7,8\}\right) \cap \left(\{1,2,3,4,5,6\}\right)=\{1,2\}\cup\{5,6\}.$
 \item If ${\bf{f}}= {\left[f_1 ~f_2 ~f_3 ~f_4 ~f_5 ~f_6 ~f_7 ~f_8\right]}^{\mathrm{T}}$, $f_i$ could be non-zero for \\$i\in \left(\{1,2\}\cup\{7,8,9,10\}\right) \cap \left(\{1,2,3,4,5,6\}\right)=\{1,2\}.$    
\end{itemize}
The vectors $\bf{b}$, $\bf{d}$ and $\bf{f}$ should be linearly independent and have similar constant channel values at the time snapshots of $\{1,2\}$. Since $\lvert \{1,2\} \rvert = 2$, we can select two these vectors among $\bf{b}$, $\bf{d}$ and $\bf{f}$. Therefore, we should omit one of the vectors $\bf{b}$, $\bf{d}$ and $\bf{f}$. We omit the vector $\bf{f}$ and design the remaining vectors as follows:
\begin{equation}
\begin{aligned}
&{\bf{a}}=\left[ 0~0~1~1~0~0~0~0 \right]^{\mathrm{T}}&\\
&{\bf{b}}=\left[ 1~1~0~0~0~0~0~0 \right]^{\mathrm{T}}&\\
&{\bf{c}}=\left[ 0~0~0~0~0~0~1~1 \right]^{\mathrm{T}}&\\
&{\bf{d}}=\left[-1~1~0~0~0~0~0~0 \right]^{\mathrm{T}}&\\
&{\bf{e}}=\left[ 0~0~0~0~1~1~0~0 \right]^{\mathrm{T}}.&\\
\end{aligned}
\end{equation}
\textit{Step 4:} After omitting vector $\bf{f}$, we use the following precoder design for transmitters:
\begin{equation}
\label{4tvectors3}
\begin{aligned}
&{\bar{\bf{V}}^{[1]}}=\left[ \bf{a} ~\bf{d} \right]&\\
&{\bar{\bf{V}}^{[2]}}=\left[ \bf{a} ~\bf{b} ~\bf{e} \right]&\\
&{\bar{\bf{V}}^{[3]}}=\left[ \bf{b} ~\bf{c} ~\bf{d} \right]&\\
&{\bar{\bf{V}}^{[4]}}=\left[ \bf{c} ~\bf{e} \right].& 
\end{aligned}
\end{equation}  
\textit{Step 5:}
Now, we analyze whether all shared basic vectors at their corresponding receivers are linearly independent or not. For example at the first receiver we should show that $\bar{\bf{H}}^{[12]}{\bf{a}}$ and $\bar{\bf{H}}^{[11]}{\bf{a}}$ are linearly independent. It is necessary to show that at the receiver $j$, the shared vectors does not collapse, i.e., all the desired vectors are linearly independent. We analyze all the vectors as follows:
\begin{itemize}
\item The vector $\bf{a}$ is shared between the first and second receivers. The vector $\bf{a}$ has nonzero elements at $\{3,4\}$ time snapshots. Also $\bar{\bf{H}}^{[11]}$ and $\bar{\bf{H}}^{[12]}$ have an altering point at this time set. Therefore, $\bar{\bf{H}}^{[12]}{\bf{a}}$ and $\bar{\bf{H}}^{[11]}{\bf{a}}$ are linearly independent almost surely. Similarly, at the second receiver we can conclude that $\bar{\bf{H}}^{[22]}{\bf{a}}$ and $\bar{\bf{H}}^{[21]}{\bf{a}}$ are linearly independent.
\item The vector $\bf{b}$ is shared among second and third receivers. Since $\bf{b}$ has nonzero element at $\{1,2\}$ time snapshots and both $\bar{\bf{H}}^{[22]}$ and $\bar{\bf{H}}^{[23]}$ don't have any changing element at this times. The vectors $\bar{\bf{H}}^{[22]}{\bf{b}}$ and $\bar{\bf{H}}^{[23]}{\bf{b}}$ are not linearly independent. Therefore, we omit vector $\bf{b}$ from one of these transmitters e.g. second transmitter.
\item Similarly, vectors $\{\bf{c},\bf{e}\}$ satisfy linearly independence conditions at their corresponding receivers.  
\end{itemize}
Therefore, we redesign the precoders as follows:
\begin{equation}
\label{4tvectors4}
\begin{aligned}
&{\bar{\bf{V}}^{[1]}}=\left[ \bf{a} ~\bf{d} ~\bf{g}\right]&\\
&{\bar{\bf{V}}^{[2]}}=\left[ \bf{a} ~\bf{e} \right]&\\
&{\bar{\bf{V}}^{[3]}}=\left[ \bf{b} ~\bf{c} \right]&\\
&{\bar{\bf{V}}^{[4]}}=\left[ \bf{c} ~\bf{e} \right],& 
\end{aligned}
\end{equation}  
where $\bf{g}$ is a vector with random elements.\\
\textit{Step 6:} Now we calculate the achievable DoF for each user. The received signal space at each receiver can be categorize into two separate spaces: desired space and undesired space (interference space). The number of desired signal space at each receiver can be calculated by following strategy:\\
1) $\text{RX}_{1}$: Since ${{C}}^{[1q]}_{1}$ has constant value during intervals of $\{1,2\} \cup \{5,6,7,8\}$, it can not change the vector space of ${\bf{b}}$, $\bf{c}$ and $\bf{e}$. Therefore, the span set of vector ${\bf{b}}$, $\bf{c}$ and $\bf{e}$ equals to that of ${{\bar{\bf{H}}}}^{[1q]}{\bf{b}}$, ${{\bar{\bf{H}}}}^{[1q]}{\bf{c}}$ and ${{\bar{\bf{H}}}}^{[1q]}{\bf{e}}$, respectively. But, due to channel changing pattern the vectors ${\bar{\bf{H}}}^{[11]}{\bf{a}}$ and ${\bar{\bf{H}}}^{[12]}{\bf{a}}$ are linearly independent from each others. In other words, in 8 time slots, the space span at the first receiver is:
\begin{equation}
\bar{S}^{[1]}=\Big{\lbrace} \underbrace{{\bar{\bf{H}}}^{[11]}{\bf{a}},{\bar{\bf{H}}}^{[12]}\bf{a}}_{\text{linearly independent}},{\bar{\bf{H}}}^{[11]}{\bf{d}},{\bar{\bf{H}}}^{[11]}{\bf{g}},\underbrace{{\bar{\bf{H}}}^{[12]}{\bf{b}},{\bar{\bf{H}}}^{[13]}{\bf{b}}}_{\text{align}},\underbrace{{\bar{\bf{H}}}^{[12]}{\bf{e}},{\bar{\bf{H}}}^{[14]}{\bf{e}}}_{\text{align}},\underbrace{{\bar{\bf{H}}}^{[13]}{\bf{c}},{\bar{\bf{H}}}^{[14]}{\bf{c}}}_{\text{align}}\Big{\rbrace}.
\end{equation}
Therefore, at the first receiver we have 3 desired signal (${\bar{\bf{H}}}^{[11]}{\bf{a}}$, ${\bar{\bf{H}}}^{[11]}{\bf{d}}$ and ${\bar{\bf{H}}}^{[11]}{\bf{g}}$) and from 8 dimensions we have assigned three dimensions for desired signal. Finally, we achieve $\frac{3}{8}$ DoF for this user.\\ 
2) $\text{RX}_{2}$: Since ${C}^{[2q]}_{1}$ has constant value during interval of $\{1,2\}\cup\{7,8\}$ it can not change the vector space of ${\bf{c}}$. Therefore, the span set of vector ${\bf{c}}$ is equal to the span set of vector ${{\bar{\bf{H}}}}^{[2q]}{\bf{c}}$. But, due to channel changing pattern the vectors ${\bf{a}}$ and ${\bf{e}}$ at the second receiver are linearly independent. In other words, the space span at the second receiver is:
\begin{equation}
\bar{S}^{[2]}=\Big{\lbrace}{\underbrace{{\bar{\bf{H}}}^{[21]}{\bf{a}},{\bar{\bf{H}}}^{[22]}\bf{a}}_{\text{linearly independent}},{\bar{\bf{H}}}^{[21]}\bf{d},{\bar{\bf{H}}}^{[21]}{\bf{g}},{\bar{\bf{H}}}^{[23]}{\bf{b}},\underbrace{{\bar{\bf{H}}}^{[22]}{\bf{e}},{\bar{\bf{H}}}^{[24]}{\bf{e}}}_{\text{linearly independent}},\underbrace{{\bar{\bf{H}}}^{[23]}{\bf{c}},{\bar{\bf{H}}}^{[24]}{\bf{c}}}_{\text{align}}}\Big{\rbrace},
\end{equation}
therefore, in the second receiver we have two desired vectors (${\bar{\bf{H}}}^{[22]}{\bf{a}}$ and ${\bar{\bf{H}}}^{[22]}{\bf{e}}$). Therefore, from eight dimensions we assigned two dimensions for desired signal. Finally, we achieve $\frac{2}{8}$ DoF for this user.\\
3) $\text{RX}_{3}$: Since ${C}^{[3q]}_{1}$ has constant value during interval of $\{2,3,4,5,6\}$ it can not change the vector space of ${\bf{a}}$. Similar expression can be used for the vector $\bf{e}$. Therefore, the span set of vectors ${\bf{a}}$ and $\bf{e}$ is equal to the span set of vectors (${{\bar{\bf{H}}}}^{[31]}{\bf{a}}$, ${{\bar{\bf{H}}}}^{[32]}{\bf{a}}$) and (${{\bar{\bf{H}}}}^{[32]}{\bf{e}}$, ${{\bar{\bf{H}}}}^{[34]}{\bf{e}}$), respectively. But, due to channel changing pattern the vectors $\bar{\bf{H}}^{[33]}{\bf{c}}$ and $\bar{\bf{H}}^{[34]}{\bf{c}}$ are linearly independent. In other words, the space span at third receiver is:
\begin{equation}
\bar{S}^{[3]}=\Big{\lbrace}{\underbrace{{\bar{\bf{H}}}^{[33]}{\bf{c}},{\bar{\bf{H}}}^{[34]}{\bf{c}}}_{\text{linearly independent}},{\bar{\bf{H}}}^{[33]}{\bf{b}},{\bar{\bf{H}}}^{[31]}\bf{d},{\bar{\bf{H}}}^{[31]}{\bf{g}},\underbrace{{\bar{\bf{H}}}^{[31]}{\bf{a}},{\bar{\bf{H}}}^{[32]}{\bf{a}}}_{\text{align}},\underbrace{{\bar{\bf{H}}}^{[32]}{\bf{e}},{\bar{\bf{H}}}^{[34]}{\bf{e}}}_{\text{align}}} \Big{\rbrace},
\end{equation}
therefore, in the third receiver we have assigned two dimensions for desired signal (${\bar{\bf{H}}}^{[33]}{\bf{b}}$, ${\bar{\bf{H}}}^{[33]}{\bf{c}}$ from 8 dimensions). Finally, we achieve $\frac{2}{8}$ DoF for the third receiver.\\
4) By the similar process, with the designed precoders we can achieve $\frac{2}{8}$ DoF for $4^{th}$ receiver.\\
Therefore, totally we can achieve $\frac{9}{8}$ DoF, which is more than one.\\  
{\bf{Example 3:}} Improving Example 2 solution for ${C}_{l}^{[pq]}={C}_{l}^{[pq']}, L^{[pq]}=1$\\
In Example 2 as it is clear in \eqref{4tvectors2}, we start our solution by sharing every basic vectors e.g., $\bf{a}$, $\bf{b}$, $\bf{c}$, $\bf{d}$, $\bf{e}$ and $\bf{f}$ between two of transmitters. As an example $\bf{a}$ is shared between first and second transmitters. But, in this example we start our solution by sharing every basic vectors to the three of transmitters. Therefore, the number of separated vectors in this example is $\binom{4}{3}=4$.

Similar to Example 2, every channel which is connected to the same destination experience the same changing pattern e.g, ${C}_{l}^{[11]}={C}_{l}^{[12]}={C}_{l}^{[13]}={C}_{l}^{[14]}$. In this example we do not solve this problem to align interferences perfectly, but we align some part of interferences to nearly achieve more than one DoF. Firstly, let us consider the following precoder designed vectors at transmitters which is consisted of 10 time snapshots:
\begin{equation}
\begin{aligned}
&{\bar{\bf{V}}^{[1]}}=\left[ \bf{a}~\bf{c}~\bf{d} \right]^{\mathrm{T}}&\\
&{\bar{\bf{V}}^{[2]}}=\left[ \bf{a}~\bf{b}~\bf{c} \right]^{\mathrm{T}}&\\
&{\bar{\bf{V}}^{[3]}}=\left[ \bf{a}~\bf{b}~\bf{d} \right]^{\mathrm{T}}&\\
&{\bar{\bf{V}}^{[4]}}=\left[ \bf{b}~\bf{c}~\bf{d} \right]^{\mathrm{T}}.& 
\end{aligned}
\end{equation}
In the above relations, $\bf{a}$, $\bf{b}$, $\bf{c}$ and $\bf{d}$ are $10\times 1$ column matrices which are defined as follows:
\begin{equation}
\begin{aligned}
&{\bf{a}}=\left[ 1~1~1~1~0~0~0~0~0~0 \right]^{\mathrm{T}}&\\
&{\bf{b}}=\left[ 0~0~0~0~1~1~1~1~0~0 \right]^{\mathrm{T}}&\\
&{\bf{c}}=\left[ 0~0~1~1~1~1~0~0~0~0 \right]^{\mathrm{T}}&\\
&{\bf{d}}=\left[ 0~0~0~0~0~0~1~1~1~1 \right]^{\mathrm{T}}.& 
\end{aligned}
\end{equation}
Now, let us analyze the signal space at every receiver. The signal space at each receiver can be categorized into two separate spaces: desired space and undesired space or interference space. The received signal at each receiver can be analyzed as follows:\\
1) $\text{RX}_{1}$: Since ${C}^{[1q]}_{1}$ has constant value during interval of $\{5,6,7,8\}$ it can not change the vector space of ${\bf{b}}$. Therefore, the span set of vector ${\bf{b}}$ equals to the span set of vector ${{\bar{\bf{H}}}}^{[1q]}{\bf{b}},~q \neq 1$. But, due to channel changing pattern all the received signals from the basic vectors of ${\bf{a}}$, ${\bf{c}}$ and ${\bf{d}}$ are linearly independent from each other. In other words, the signal received at the first is the span of the following vectors:
\begin{equation}
\bar{S}^{[1]}=\lbrace{\underbrace{\bar{\bf{H}}^{[11]}{\bf{a}},\bar{\bf{H}}^{[12]}{\bf{a}},\bar{\bf{H}}^{[13]}{\bf{a}}}_{\text{linearly independent}},\underbrace{\bar{\bf{H}}^{[11]}{\bf{c}},\bar{\bf{H}}^{[12]}{\bf{c}},\bar{\bf{H}}^{[14]}{\bf{c}}}_{\text{linearly independent}},\underbrace{\bar{\bf{H}}^{[11]}{\bf{d}},\bar{\bf{H}}^{[13]}{\bf{d}},\bar{\bf{H}}^{[14]}{\bf{d}}}_{\text{linearly independent}},\underbrace{\bar{\bf{H}}^{[12]}{\bf{b}},\bar{\bf{H}}^{[13]}{\bf{b},\bar{\bf{H}}}^{[14]}{\bf{b}}}_{\text{align}}}\rbrace.
\end{equation}
Therefore, at first receiver we have 3 from 10 dimensions which are free from interference. Finally the achievable DoF for the first user is equal to $\frac{3}{10}$.

2) $\text{RX}_{2}$: Since ${C}^{[2q]}_{1}$ has constant value during interval of $\{7,8,9,10\}$ it can not change the vector space of ${\bf{d}}$. Therefore, the span set of vector ${\bf{d}}$ is equal to the span set of vector ${{\bar{\bf{H}}}}^{[2q]}{\bf{d}},~q \neq 2$. But, due to channel changing pattern all the received signals from the basic vectors of ${\bf{a}}$, ${\bf{b}}$ and ${\bf{c}}$ are linearly independent from each other. In other words, from ten linearly independent vectors at second receiver we have three dimensions which are free from interference. Finally the achievable DoF for the second user is equal to $\frac{3}{10}$.\\
3) $\text{RX}_{3}$: since ${C}^{[3q]}_{1}$ has constant value during interval of $\{1,2,3,4,5,6\}$ it can not change the vector space of ${\bf{c}}$. Therefore, the span set of vector ${\bf{c}}$ is equal to the span set of vector ${{\bar{\bf{H}}}}^{[3q]}{\bf{c}}$. But, due to channel changing pattern all the received signals from the basic vectors of ${\bf{a}}$, ${\bf{b}}$ and ${\bf{d}}$ are linearly independent from each other. In other words, from ten linearly independent vectors at third receiver we have three dimensions which are free from interference. Finally the achievable DoF at third receiver is equal to $\frac{3}{10}$.\\ 
4) $\text{RX}_{4}$: since ${C}^{[4q]}_{1}$ has constant value during interval of $\{1,2,3,4\}$ it can not change the vector space of ${\bf{a}}$. Therefore, the span set of vector ${\bf{a}}$ is equal to the span set of vector ${{\bar{\bf{H}}}}^{[4q]}{\bf{a}}$. But, due to channel changing pattern all the received signals from the basic vectors of ${\bf{b}}$, ${\bf{c}}$ and ${\bf{d}}$ are linearly independent from each other. In other words, from ten linearly independent vectors at 4th receiver we have three dimensions which are free from interference. Finally the achievable DoF for the 4th user is equal to $\frac{3}{10}$.\\       
Therefore, in this case we can totally achieve $\left( {\frac{12}{10}>1} \right)$ DoF which is greater than what is achieve in Example 2.
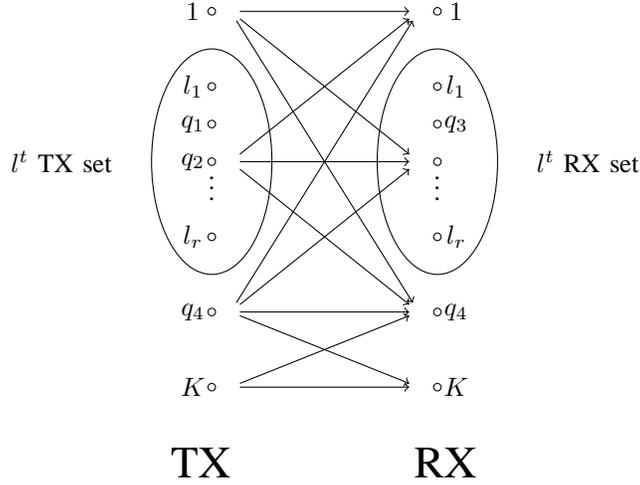
\begin{figure}
  \centering
  \begin{tikzpicture}
  \draw (-12,4) ellipse (0.8cm and 1.5cm);
  \draw (-9,4) ellipse (0.8cm and 1.5cm);
  \draw (-12,6) circle (0.05cm);
  \node            (a) at (-12.25,6){$1$};
  \draw (-12,5) circle (0.05cm);
  \node            (b) at (-12.25,5){$l_1$};
  \draw (-12,4.5) circle (0.05cm);
  \node            (c) at (-12.25,4.5){$q_1$};
  \node            (c) at (-12.25,4){$q_2$};
  \draw (-12,4) circle (0.05cm);
  \node            (d) at (-12,3.75){$\vdots$};
  \draw (-12,3) circle (0.05cm);
  \node            (e) at (-12.25,3){$l_r$};
  \draw (-12,2) circle (0.05cm);
  \node            (f) at (-12.25,2){$q_4$};
  \draw (-12,1) circle (0.05cm);
  \node            (f) at (-12.25,1){$K$};
  \draw (-9,6) circle (0.05cm);
  \node            (g) at (-8.75,6){$1$};
  \draw (-9,5) circle (0.05cm);
  \node            (h) at (-8.75,5){$l_1$};
  \draw (-9,4.5) circle (0.05cm);
  \node              at (-8.75,4.5){$q_3$};
  \draw (-9,4) circle (0.05cm);
  \node            (j) at (-9,3.75){$\vdots$};
  \draw (-9,3) circle (0.05cm);
  \node            (k) at (-8.75,3){$l_r$};
  \draw (-9,2) circle (0.05cm);
  \node            (l) at (-8.75,2){$q_4$};
  \draw (-9,1) circle (0.05cm);
  \node            (l) at (-8.75,1){$K$};
  \node            (m) at (-7,4){$l^{t}$ RX set};
  \node            (n) at (-14,4){$l^{t}$ TX set};
  \node            (o) at (-12.25,0){\begin{LARGE}
  TX
  \end{LARGE}};
  \node            (p) at (-9,0){\begin{LARGE}
  RX
  \end{LARGE}};
  \node            (tk) at (-11.75,1){};
  \node            (rk) at (-9.25,1){};
  \node            (t1) at (-11.75,2){};
  \node            (tlq) at (-11.75,4){};
  \node            (tq4) at (-11.75,6){};
  \node            (r1) at (-9.25,2){};
  \node            (rlq) at (-9.25,4){};
  \node            (rq4) at (-9.25,6){};
  \draw[->] (t1) edge (r1);
  \draw[->] (t1) edge (rlq);
  \draw[->] (t1) edge (rq4);
  \draw[->] (tlq) edge (r1);
  \draw[->] (tlq) edge (rlq);
  \draw[->] (tlq) edge (rq4);
  \draw[->] (tq4) edge (r1);
  \draw[->] (tq4) edge (rlq);
  \draw[->] (tq4) edge (rq4);
  \draw[->] (t1) edge (rk);
  \draw[->] (tk) edge (r1);
  \draw[->] (tk) edge (rk);    
  \end{tikzpicture}
  \caption{In this figure we show number of transmitters and receivers of the set of $l^{t}=\{l_1,l_2,\dots,l_r\}$ with the closed circular shape. The complimentary transceivers are out of this circular shape which are modeled by the set of $\{1,\dots,K\}-l^{t}$. Also there is a signaling path between all transmitters and receivers but to avoid having so crowded figure we show a few of them.}
\end{figure}
\subsection{Outer-bound on Achievable DoF when $C_l^{[pq]}=C_l^{[pq']}, L^{[pq]}=1$}
In Example 2 when the changing pattern of the channels with the same destinations are similar, we proposed an algorithm to achieve $\frac{9}{8}$ DoF. In the third example with changing the number of shared basic vectors between transmitters we achieve $\frac{12}{10}$ DoF which has better performance than what is achieve in second example. In this subsection we want to find an outer-bound on achievable DoF when $C_l^{[pq]}=C_l^{[pq']}, L^{[pq]}=1$. 

Let consider the set $l^{t}=\{l_1,\dots,l_r\}\subseteq \{1,\dots,K\}$ where $\lvert {l^{t}} \rvert = r$ and $1 \leq t \leq \binom{K}{r}$. It means that we can choose $r$ different transmitters from the set $\{1,\dots,K\}$ to generate the set of $l^{t}$. We assume every basic vector from each transmitter aligns with interference generated from $r-1$ transmitters at $K-r$ receivers. In other words, if ${\bf{v}}^{[q]}$ is one of the basic vectors of transmitter $q$, we have:
\begin{equation}
{\bar{\bf{H}}}^{[pq]}{\bf{v}}^{[q]} \prec {\bar{\bf{H}}}^{[pq']}\bar{\bf{V}}^{[q']}.
\end{equation}
Where, $(q, q' \in l^{t},~q \neq q'$ and $p \in \{1,\dots,K\}-l^{t}$.

\textit{\textbf{Remark:}} If we have ${\bar{\bf{H}}}^{[qq]}{\bf{v}}^{[q]}\notin \mathrm{span} \left( {\bar{\bf{H}}}^{[qq']}\bar{\bf{V}}^{[q']} \right),~q \neq q'$, it could be possible to separate desired and interference signals,  otherwise the desired signal space is polluted by interference.

\begin{lem}
\it{If ${\bf{v}}^{[q]}$ was aligned with the transmitted signal of $\mathrm{TX}_i$, $i \in l^{t}$ at the $\mathrm{RX}_j$, $j \in \{1,\dots,K\}-l^{t}$, it could not be aligned with the interference generated from $\mathrm{TX}_i$, $i \in l^{t}$ at the $\mathrm{RX}_{j^{'}}$, ${j^{'}} \in l^{t}-q$.}
\end{lem}
\begin{IEEEproof}
(Proof by Contradiction.) Suppose $\mathrm{TX}_{q_{1}}, q_{1}\in l_t$ and $\mathrm{TX}_{q_{2}}, q_{2}\in l_t $ are two transmitters. Also, $\mathrm{RX}_{q_3},{q_3}\in l_t$ and $\mathrm{RX}_{q_4},{q_4}\in \{1,\dots,K\}-l^{t}$ are two receivers (see Figure 4). From the assumption of this lemma, we have:
\begin{equation}
{\bar{\bf{H}}}^{[{q_{4} q_{1}}]}{\bf{v}}^{[q_1]} \in \mathrm{span}\left({{\bar{\bf{H}}}^{[q_4 q_3]}\bar{\bf{V}}^{[q_3]}}\right).
\end{equation}
From Lemma 2 of \cite{7}, since ${\bar{\bf{H}}}^{[q_4 q_1]}$ and ${\bar{\bf{H}}}^{[q_4 q_3]}$ are diagonal and have the same changing pattern, ${\bf{v}}^{[q_1]} \in \mathrm{span}\left({\bar{\bf{V}}^{[q_3]}}\right)$.\\
Assume to the contrary, that:
\begin{equation}
\left\{ {\exists q_3 \in l_t: \mathrm{span}\left( {\bar{\bf{H}}}^{[q_{3}q_{1}]}{\bf{v}}^{[q_1]}\right)\in \mathrm{span}\left( {\bar{\bf{H}}}^{[q_{3}q_{2}]}{\bar{\bf{V}}}^{[q_2]}\right)} \right\}.
\end{equation}   
From this assumption, we have:
\begin{equation}
\mathrm{span}\left({{\bar{\bf{H}}}}^{[q_3 q_1]}{\bf{v}}^{[q_1]}\right) \in \mathrm{span} \left({{\bar{\bf{H}}}^{[q_3 q_3]}} \left({\bar{\bf{H}}}^{[q_3 q_3]}\right)^{-1}{\bar{\bf{H}}}^{[q_3 q_2]}{\bar{\bf{V}}^{[q_2]}}\right).
\end{equation}
Since ${{\bar{\bf{H}}}^{[q_3 q_1]}}$ and ${{\bar{\bf{H}}}^{[q_3 q_3]}}$ have similar changing pattern, we get:
\begin{equation}
\mathrm{span}\left({\bf{v}}^{[q_1]}\right) \in \mathrm{span} \left(\left({\bar{\bf{H}}}^{[q_3 q_3]}\right)^{-1}{\bar{\bf{H}}}^{[q_3 q_2]}{\bar{\bf{V}}^{[q_2]}}\right).
\end{equation}
Hence, as ${\bf{v}}^{[q_1]} \in \mathrm{span}\left({\bar{\bf{V}}^{[q_3]}}\right)$, we have:
\begin{equation}
\text{dim} \left( {\bar{\bf{V}}^{[q_{3}]}} \cap \left({\bar{\bf{H}}}^{[q_{3} q_{3}]}\right)^{-1}{\bar{\bf{H}}}^{[q_{3} q_{2}]}{\bar{\bf{V}}^{[q_{2}]}} \right) >0,
\end{equation}
or equivalently:
\begin{equation}
\text{dim} \left( {\bar{\bf{H}}}^{[q_{3} q_{3}]}{\bar{\bf{V}}^{[q_{3}]}} \cap {\bar{\bf{H}}}^{[q_{3} q_{2}]}{\bar{\bf{V}}^{[q_{2}]}} \right) >0.
\end{equation}
The above relation shows that the desired signal ${\bar{\bf{H}}}^{[q_{3} q_{3}]}{\bar{\bf{V}}^{[q_{3}]}}$ at $\mathrm{RX}_{q_3}$ has been polluted by the interference of $\mathrm{TX}_{q_2}$. Hence, the assumption of $\left\{ \exists q_3 \in l_t: \mathrm{span}\left( {\bar{\bf{H}}}^{[q_{3}q_{1}]}{\bf{v}}^{[q_1]}\right)\in \mathrm{span}\left( {\bar{\bf{H}}}^{[q_{3}q_{2}]}{\bar{\bf{V}}}^{[q_2]}\right) \right\}$ leaves us with a contradiction. This completes the proof.
\end{IEEEproof}
{\bf{Definition:}} $d_{{i_1}{i_2}\dots{i_r}},~i_1 \neq i_2 \neq \dots \neq i_r$ shows the number of dimensions which is jointly occupied by $\mathrm{TX}_{i_1}$, $\mathrm{TX}_{i_2}$,... and $\mathrm{TX}_{i_r}$ at $\mathrm{RX}_j$, where $j\notin \{i_1,i_2,\dots,i_r\}$. Also for every permutation of $i'_1,\dots,i'_r \in \{i_1,i_2,\dots,i_r\}$ we have:
\begin{equation}
d_{{i_1}{i_2}\dots{i_r}}=d_{{i'_1}{i'_2}\dots{i'_r}}.
\end{equation}
\begin{thm}
{\it For the $K-$user interference channel with $C_l^{[pq]}=C_l^{[pq']}, L^{[pq]}=1$, we cannot achieve more than $\max_{r}{\frac{Kr}{r^2-r+K}},~r\in \mathbb{N}$ DoF.}
\end{thm}
\begin{IEEEproof}
The proof follows from the following basic relation on the DoF of the BIA in $K-$user interference channel problem. The jointly interference signal from $\mathrm{TX}_{i_1}$, $\mathrm{TX}_{i_2}$,... and $\mathrm{TX}_{i_r}$ occupy $d_{{i_1}{i_2}\dots{i_r}}$ dimensions at $\mathrm{RX}_{j},~j\notin \{{i_1},{i_2},\dots,{i_r}\}$ . In other words, every shared vector among $r$ different users e.g. $\mathrm{TX}_{i_1}$, $\mathrm{TX}_{i_2}$,... and $\mathrm{TX}_{i_r}$ occupy just only one dimension at $j^{th}$ receiver. Also we know the total number of dimensions is $n$. Therefore, at $\mathrm{RX}_{j},~j\notin \{{i_1},{i_2},\dots,{i_r}\}$, we have:
\begin{equation}
\label{14}
d_1+d_2+\dots+d_K-\left(r-1\right) \sum_{i_1,\dots,i_r}{d_{i_1,\dots,i_r}} \leq n,~i_1,\dots,i_r \in \{1,\dots,K\}-\{j\},
\end{equation}
where, the coefficient $(r-1)$ comes from this fact that $d_{i_1,\dots,i_r},~i_1,\dots,i_r \in \{1,\dots,K\}-\{j\}$ just only occupy one dimension at $j^{th}$ receiver while it counts $r$ times when we calculate $d_1+d_2+\dots+d_K$. Similarly, at all the receivers, we have:
\begin{equation}
\label{15}
\begin{aligned}
&\mathrm{RX}_{1}:~d_1+d_2+\dots+d_K-\left(r-1\right) \sum_{i_1,\dots,i_r}{d_{i_1,\dots,i_r}} \leq n,~i_1,\dots,i_r \in \{1,\dots,K\}-\{1\}&\\
&\mathrm{RX}_{2}:~d_2+d_1+\dots+d_K-\left(r-1\right) \sum_{i_1,\dots,i_r}{d_{i_1,\dots,i_r}} \leq n,~i_1,\dots,i_r \in \{1,\dots,K\}-\{2\}&\\
&\vdots&\\
&\mathrm{RX}_{K}:~d_K+d_1+\dots+d_{K-1}-\left(r-1\right) \sum_{i_1,\dots,i_r}{d_{i_1,\dots,i_r}} \leq n,~i_1,\dots,i_r \in \{1,\dots,K\}-\{K\}.&\\
\end{aligned}
\end{equation}
by summing all the above relations we have:
\begin{equation}
\label{DOFlower}
K \sum_{i=1}^{K}{d_i}+\left(K-1\right)\left(r-1\right)\sum_{i_1,\dots,i_r}{d_{i_1,\dots,i_r}} \leq Kn,
\end{equation} 
in addition, it is clear that:
\begin{equation} 
r \sum_{i_1,\dots,i_r}{d_{i_1,\dots,i_r}} \geq \sum_{i=1}^{K}{d_i}.
\end{equation}
Since for $r \geq 1$, the value of $\left( K-1 \right) \geq \left(K-r\right)$, so we have:
\begin{equation} 
\left( K-1 \right) r \sum_{i_1,\dots,i_r}{d_{i_1,\dots,i_r}} \geq \left(K-r\right) \sum_{i=1}^{K}{d_i},
\end{equation}
which shows that:
\begin{equation} 
\sum_{i_1,\dots,i_r}{d_{i_1,\dots,i_r}} \geq \frac{\left(K-r\right)}{\left( K-1 \right) r } \sum_{i=1}^{K}{d_i}.
\end{equation}
Therefore, from \eqref{DOFlower} we have:
\begin{equation}
\label{DOFlower2}
K \sum_{i=1}^{K}{d_i}+\left(K-1\right)\left(r-1\right)\frac{\left(K-r\right)}{\left( K-1 \right) r } \sum_{i=1}^{K}{d_i} \leq Kn.
\end{equation}
After some manipulation on \eqref{DOFlower2} we get:
\begin{equation}
\frac{\sum_{i=1}^{K}{d_i}}{n} \leq \frac{Kr}{r^2-r+K} \leq \max_{r}{\frac{Kr}{r^2-r+K}},
\end{equation}
which completes the proof.
\end{IEEEproof} 
Theorem 3 shows that in the case of $C_{l}^{[pq]}=C_{l}^{[pq']}$, at $\mathrm{RX}_{p}$ we have three types of received vectors:
\begin{itemize}
\item Type I: Aligned interference vectors where generated by $\mathrm{TX}_j, j\in \mathcal{K}_1$ and $\mathcal{K}_1 \subset \{1,\dots,K\}-\{p\}, \lvert \mathcal{K}_1 \rvert=r$. The number of joint occupied dimensions by these transmitters at $\mathrm{RX}_p$ can be calculated as follows:  
\begin{equation}
\label{d1}
\text{dim}\left({\bigcup_{q \in \mathcal{K}_1}{{{\bar{\bf{H}}}^{[pq]}\bar{\bf{V}}^{[q]}}}}\right)=\binom{K-1}{r}.
\end{equation}   
\item Type II: Linearly independent interference vectors which are shared among $\mathrm{TX}_p$ and $r-1$ different transmitters of $\mathrm{TX}_j, j\in \mathcal{K}_2, \mathcal{K}_2 \subset \{1,\dots,K\}-\{p\}$ and $\lvert \mathcal{K}_2 \rvert=r-1$. These vectors should be linearly independent at $\mathrm{RX}_{p}$. Since all these vectors are jointly linearly independent, they occupied $(r-1) \binom{K-1}{r-1}$ dimensions from $\mathrm{RX}_{p}$ receiver. In other words:
\begin{equation}
\label{d2}
\text{dim}\left({\bigcup_{q \in \mathcal{K}_2}{{{\bar{\bf{H}}}^{[p q]}\bar{\bf{V}}^{[q]}}}}\right)=(r-1)\binom{K-1}{r-1}.
\end{equation}
\item Type III: Desired signal vectors shared among $\mathrm{TX}_p$ and $r-1$ different transmitters of $\mathrm{TX}_j, j\in \mathcal{K}_2$ and $\lvert \mathcal{K}_2 \rvert=r-1$. The number of this type of vectors can be calculated as follows:
\begin{equation}
\label{d3}
\mathrm{dim}\left({{{\bar{\bf{H}}}^{[pp]}\bar{\bf{V}}^{[p]}}}\right)=\binom{K-1}{r-1}.
\end{equation}
\end{itemize}
Therefore, the total number of used dimensions can be calculated by summing \eqref{d1}, \eqref{d2} and \eqref{d3}:
\begin{equation}
n=\binom{K-1}{r}+\left(r-1\right)\binom{K-1}{r-1}+\binom{K-1}{r-1}=\binom{K-1}{r}+r\binom{K-1}{r-1}.
\end{equation} 
The number of dimensions of desired signal space at each receiver is $\binom{K-1}{r-1}$, which equals to the number of type III vectors. Hence, the total achievable DoF can be calculated as follows:
\begin{equation}
\frac{d_1+d_2+\dots+d_k}{n}=\frac{K\binom{K-1}{r-1}}{\binom{K-1}{r}+r\binom{K-1}{r-1}}=\frac{Kr}{r^2-r+K},
\end{equation} 
Since $r$ is a designing parameter, we want to find $r$ in such away that maximizes total achievable DOF of $d(r)=\frac{Kr}{r^2-r+K}$. 
To find the maximum value of $d(r)$, we analyze the continuous function of $f(x)=\frac{Kx}{x^2-x+K}$. The first derivation of this function has just one positive root of $x=\sqrt{K}$ which shows that it has just only one extremum point. Also, it can easily show that for $x\geq 0$, the function $f(x)$ is greater than or equals to zero. Since $f(x=0)=0$ and $f(x\rightarrow \infty ) \rightarrow 0^{+}$, the function $f(x)$ for $x\geq 0$  is something like Figure 5.
\begin{figure}
  \centering
  \includegraphics[width=0.75\textwidth]%
    {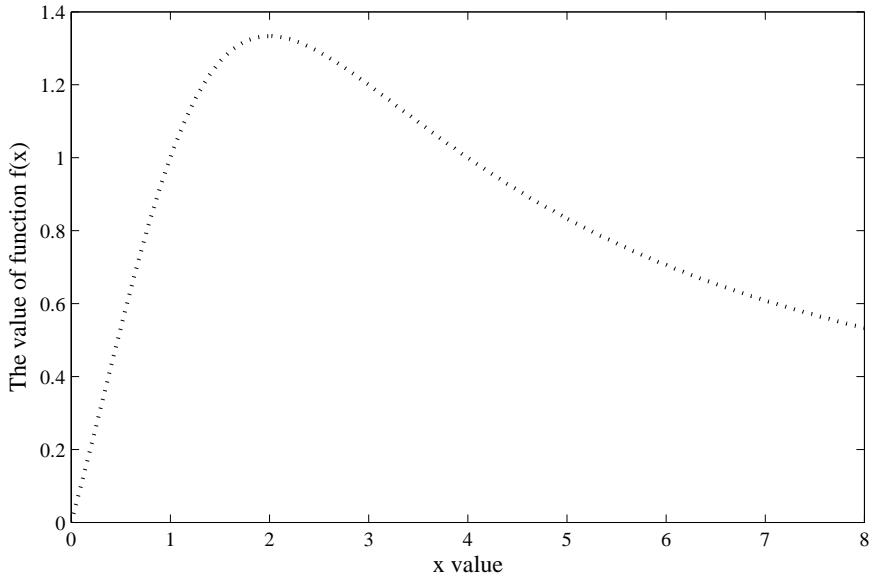}
  \caption{The function $f(x)=\frac{Kx}{x^2-x+K}$ versus continuous variable of $x$ and $K=4$.}
\end{figure}
Therefore, the maximum value of the $d(r)$ can be achieved by finding out the minimum value of $r\in \mathbb{N}$ such that:
\begin{equation}
d(r+1)-d(r)\leq 0.
\end{equation} 
In order to find $r$ which satisfied $d(r+1)-d(r)\leq 0$, we have:
\begin{equation}
\begin{aligned}
d(r+1)-d(r)&=\frac{K(r+1)}{\underbrace{(r+1)^2-(r+1)+K}_{>0}}-\frac{Kr}{\underbrace{r^2-r+K}_{>0}}&\\
&=\frac{K(r+1)(r^2-r+K)-Kr\left((r+1)^2-(r+1)+K \right)}{\underbrace{\left((r+1)^2-(r+1)+K\right)\left(r^2-r+K\right)}_{>0}}\\
&=\frac{-K\left( r^2+r-K \right)}{\underbrace{\left((r+1)^2-(r+1)+K\right)\left(r^2-r+K\right)}_{>0}} \leq 0&\\
&\Rightarrow r \geq \frac{\sqrt{1+4K}-1}{2},&
\end{aligned}
\end{equation} 
Therefore, the minimum value of $r \in \mathbb{N}$ which satisfies the above equation is $ r^{*}=\left \lceil{\frac{\sqrt{1+4K}-1}{2}} \right \rceil$. 
The exact value of $d(r^{*})$ is shown in the Figure 6. This result shows that when the number of user is $K=1,2$, the maximum achievable DoF is one which satisfy our previous knowledge about two user interference channel (we know that without any knowledge of CSI for the 2-user interference channel we can achieve maximum DoF of one). Also in the case of blind IA we can conclude that when the number of users tend to infinite we can achieve maximum DoF of $\frac{\sqrt{K}}{2}$, which has an interesting result compare to DoF of $\frac{K}{2}$ in the case of perfect CSI.
\begin{figure}
  \centering
  \includegraphics[width=0.75\textwidth]%
    {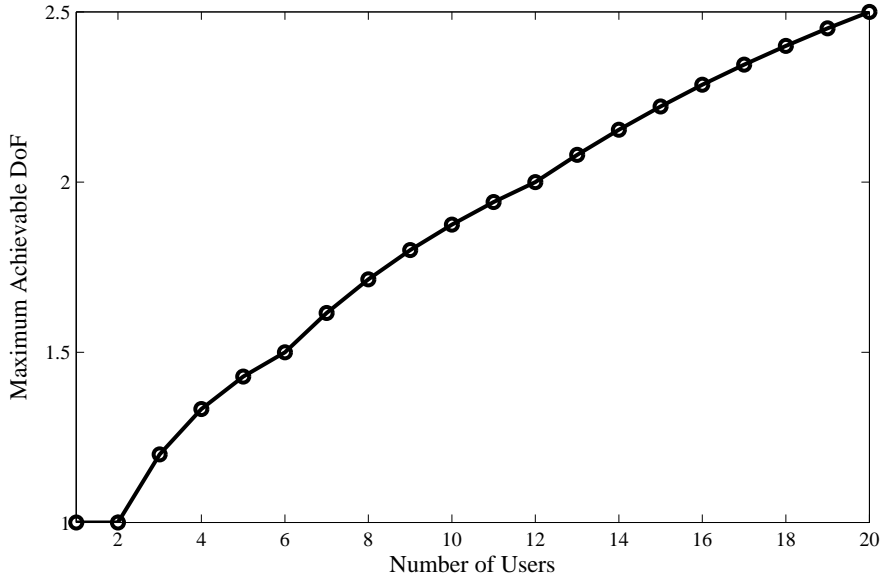}
  \caption{The maximum value of $d(r^{*})$ versus number of users $K$.}
\end{figure}
\section{3-User interference alignment with partial unknown interference channel coefficients}
Consider an interference channel with 3 transmitters ${\left \{\mathrm{TX}_k \right \}_{k=1}^{3}}$ and 3 receivers ${\left \{\mathrm{RX}_k  \right \}_{k=1}^{3}}$. Let the interference channel be 15-tuples $\left({\bar{\bf{H}}}^{[11]} ,...,{\bar{\bf{H}}}^{[33]} ,\bar{\bf{X}}^{[1]} ,...,\bar{\bf{X} }^{[3]} ,\bar{\bf{Y}}^{[1]} ...,\bar{\bf{Y}}^{[3]} \right)$, where, $\left(\bar{\bf{X}}^{[1]} ,...,\bar{\bf{X}}^{[3]} \right)$ and
$\left(\bar{\bf{Y}}^{[1]} ,...,\bar{\bf{Y}}^{[3]} \right)$ are 3 finite inputs and outputs of the channel, respectively. In the deterministic interference channel, the input of $\mathrm{TX}_k$ at a specific time duration is represented by 
$\bar{\bf{X}}^{[k]} =[{X}_{1} ,....,{X}_{n} ]^{\mathrm{T}}$.
For a specific case where the thermal noise power is zero, ${\bar{\bf{H}}}^{[pq]} ,p\neq q$
is a collection of such a $\mathrm{diag}\left(\left[{{h}_{1}^{[pq]} ,{h}_{2}^{[pq]} ,...,{h}_{n}^{[pq]}}\right]\right)$
matrices where map $\bar{\bf{X}}^{[q]}$ to received signal at $p^{th}$ receiver and represent channel model. The direct channels can be modeled with a matrix but not only diagonal (because of its permutation and memorial characteristics).  Therefore, the received signal at $\mathrm{RX}_p$ can be modeled as follows:
\begin{equation}
\bar{\bf{Y}}^{[p]} =\sum_{{q=1},q\neq p}^3 \bar{{\bf H} }^{[pq]} \bar{\bf{X}}^{[q]} +{\bar{\bf{H}}}^{[pp]} \bar{\bf{X}}^{[p]} +\bar{\bf{Z}}^{[p]},
\end{equation}
where, $\lim_{n \to \infty}\frac{1}{n}\mathrm{tr}\{\bar{\bf{X}}^{[q]} \left(\bar{\bf{X}}^{[q]}\right)^{\mathrm{H}}\}=\mathrm{SNR}$, $\lim_{n \to \infty}\frac{1}{n}\mathrm{tr}\{{\bar{\bf{H}}}^{[pq]}\left({\bar{\bf{H}}}^{[pq]}\right)^\mathrm{H}\}=1$. The column matrix $\bar{\bf{Z}}^{[p]}$ is an $n\times1$ column matrix, shows additive white Gaussian noise where, $\lim_{n \to \infty}\frac{1}{n}\mathrm{tr}\{\bar{\bf{Z}}^{[p]} \left(\bar{\bf{Z}}^{[p]}\right)^\mathrm{H}\}=1$.
\subsubsection{Interference Channel Model}
Through this section, we consider fast fading interference channel, where the channel states change in time duration of $T_c$. Also, the symbol duration is $T_s$ which is equal to the channel coherence time of $T_c$. Therefore, during transmission time of $nT_s$, the interference channel can be modeled as follows:
\begin{equation}
\label{channel model}
 \bar{\bf{H}}^{[pq]}=\mathrm{diag} \left(\left[{{h}_{{1}}^{[pq]}, {h}_{2}^{[pq]},\dots ,{h}_{n}^{[pq]}}\right]\right), p\neq q
\end{equation}
where ${h}_{j}^{[pq]} $, shows the $j^{th}$ point of altering channel state between $\mathrm{TX}_p$ and $\mathrm{RX}_q$. All ${h }_{{j}}^{[pq]}$ are $i.i.d$ random variables with a specific distribution and bounded between a nonzero and a finite maximum values.
\subsubsection{Direct Channels Model}
\subsubsection*{ Channel with memory characteristics}
Considering peer to peer signaling scheme where the transmitted signal reach to the receiver by more than one signaling path. Assume $x(t)$, is the input signal to this channel at $t^{th}$ time, the received signal $y(t)$ without considering additive noise effects can be modeled as follows:
\begin{equation}
y(t)=\sum_{m=1}^{M}{h_m (t) x(t-mT_s)}
\end{equation}
where $h_m$ and $M$ are the $m^{th}$ received signal path gain and memory length of the channel, respectively. Also, we can assume transmitted signal has constant value during independent transmissions $T_s$. Therefore, this channel can be modeled with the lower triangular matrix. The elements of this matrix are random variables with a specific distribution and are bounded between a nonzero and a finite maximum values. In other words, we can assume the matrix $\bar{\bf{H}}^{[pp]}$ is a lower triangular matrix.\\ 
\subsubsection*{ Channel with permutation characteristics}
In this case every transmitted signal in each time snapshot is received in another time snapshot. We can model the matrix $\bar{\bf{H}}^{[pp]}$ with a square matrix. This matrix can be obtained from permutation of the rows of a diagonal matrix. By a permutation channel over a specific set, we understand the channel whose inputs are sequences of this set, but the outputs are the random permutation of the elements of this set. In this specific case $M$ shows the maximum permutation distance. In other words, for the transmitted signal of $x_1,x_2,\dots,x_{n-M},\dots,x_n$ and permutation distance of $M$ the received signal may be $x_1,x_2,\dots,x_n,\dots,x_{n-M}$. \\
\begin{lem}
\label{lemma_I_one_one} \it{If $\bar{{\bf H} }^{[pq]} $ is a diagonal matrix with unknown elements of ${h^{[pq]}_{j}} , j\in U^{[pq]}$ the matrix $\bar{{\bf H}}^{[pq]} $ can be represented by $\sum_{i=1}^{\lvert{U^{[pq]}}\rvert+1}\beta^{[pq]}_{i}{\bf{\bar{I}}}{\bar{\bf{Q}}^{[pq]_i} }$ where, $\beta^{[pq]}_i {\bf{\bar{I}}}$ is a diagonal matrix with the same elements of $\beta^{[pq]}_i$ and ${\bar{\bf{Q}}^{[pq]_i} }=\mathrm{diag}\left(\left[{{q}_{1}^{[pq]_i},\dots,{q}_{n}^{[pq]_i}}\right]\right)$ where, ${q}^{[pq]_i}_{j}=\gamma^{[pq]_i}_{j}, j\in U^{[pq]}$ and ${q}^{[pq]_i}_{j}={h^{[pq]}_{j}} , j \notin U^{[pq]}$ also $\gamma^{[pq]_i}_{j}$ is a random number generated from arbitrary distribution.}
\end{lem}
\begin{IEEEproof}
 The proof is similar to the proof of the Lemma 1.
\end{IEEEproof}
This lemma shows that every channel matrix with $\lvert {U^{[pq]}} \rvert$ unknown elements can be represented by summation of $\lvert {U^{[pq]}} \rvert+1$ basic matrices e.g. ${\bar{\bf{Q}}^{[pq]_i} }$.\\
\begin{lem}
\label{lemma_two} \it{If ${\bar{\bf{H}}}^{[pq]} =\sum_{j=1}^{\lvert U^{[pq]} \rvert +1}{\beta^{[pq]}_j{\bf{\bar{I}}}{\bar{\bf{Q}}^{[pq]_j}}}$, in order to satisfies IA conditions, it is sufficient we find such $\bar{\bf{V}}^{[p]}$ precoders at $\mathrm{TX}_p$ that satisfies following conditions:
\begin{align}
&{\bar{\bf{Q}}^{[pq]_{i_1}}}\bar{\bf{V}}^{[q]} \prec{\bar{\bf{Q}}^{[p1]_{i_2}} }\bar{\bf{V}}^{[1]} ,~p,q \in \{1,2,3\},~p\neq 1& \\
&\mathrm{span}\left({\bar{\bf{Q}}^{[12]_{i_3}} }\bar{\bf{V}}^{[2]} \right)=\mathrm{span}\left({\bar{\bf{Q}}^{[13]_{i_4}} }\bar{\bf{V}}^{[3]} \right)&.
\end{align}
where, $i_1 \in \{1,\dots,\lvert {U^{[pq]}} \rvert+1\},~p \neq 1$, $i_2 \in \{1,\dots,\lvert {U^{[p1]}} \rvert+1\},~p \neq 1$, $i_3 \in \{1,\dots,\lvert {U^{[12]}} \rvert+1\}$ and $i_4 \in \{1,\dots,\lvert {U^{[13]}} \rvert+1\},~p \neq 1$.}
\end{lem}
\begin{IEEEproof}
Starting from the definition of the span of a matrix if $\lbrace {{\bf{v}}_1^{[q]},{\bf{v}}_2^{[q]},\dots,{\bf{v}}_{d_q}^{[q]} }\rbrace$ are the basic vectors of $\bar{\bf{V}}^{[q]},~q=2,3$ we have: 
\begin{equation}
\mathrm{span}\left(\bar{\bf{V}}^{[q]} \right)=\left\lbrace \sum_{l=1}^{f}\lambda_{l}{\bf{v}}_{l}^{[q]}\vert \lambda_{l}\in \mathbb{R}, f\in \mathbb{N}\right\rbrace.
\end{equation}
Therefore, from Lemma 3 we have:
\begin{equation}
\begin{aligned}
\mathrm{span}& \left({\bar{\bf{H}}}^{[pq]}\bar{\bf{V}}^{[q]} \right)=\left\lbrace \sum_{l=1}^{f}\lambda_{l}{\bar{\bf{H}}}^{[pq]}{\bf{v}}_{l}^{[q]}\vert \lambda_{l}\in \mathbb{R}, f\in \mathbb{N} \right\rbrace \\
&=\left\lbrace \sum_{l=1}^{f}\sum_{j=1}^{\lvert{U^{[pq]}}\rvert+1}\beta^{[pq]}_j{\bf{\bar{I}}}{\bar{\bf{Q}}^{[pq]_j} }\lambda_{l}{\bf{v}}_{l}^{[q]}\vert \lambda_{l}\in \mathbb{R}, f\in \mathbb{N} \right\rbrace,\\
\end{aligned}
\end{equation}
finally:
\begin{equation}
\label{eqn_11}
\begin{aligned}
\mathrm{span}&\left({\bar{\bf{H}}}^{[pq]}\bar{\bf{V}}^{[q]} \right)\\
&=\left\lbrace \sum_{l=1}^{f'}\sum_{j=1}^{\lvert{U^{[pq]}}\rvert+1}{\bar{\bf{Q}}^{[pq]_j} } \lambda_{l}^{'}{\bf{v}}_{l}^{[q]}\vert \lambda_{l}^{'}\in \mathbb{R}, f'\in \mathbb{N} \right\rbrace.
\end{aligned}
\end{equation}
The above relation shows the space span by the vectors of the set $\left\lbrace {\bar{\bf{Q}}^{[pq]_j} {\bf{v}}_{l}^{[q]}\vert 1\leq l\leq d_q, 1\leq j\leq \lvert{U^{[pq]}}\rvert+1}\right\rbrace $ is equal to the span of the vectors in the set of $\lbrace{{\bar{\bf{H}}^{[pq]} {\bf{v}}_{l}^{[q]}\vert 1\leq l\leq d_q}\rbrace}$. Therefore, all the IA conditions are satisfied.
\end{IEEEproof}
\begin{thm}
\label{theorem_two} \it{In the 3-user fast fading channel, if half of the entire CSI is not available at both transmitters and receivers we can achieve $\frac{3}{2}$ DoF.}
\end{thm}
\begin{IEEEproof}
To prove achievability we first consider a transmission scenario which is consisted of $n$ time snapshots. Referring Section II, in this case we have a signaling path between all the transmitters and receivers. All the cross signaling paths are modeled by the matrix ${\bar{\bf{H}}}^{[pq]}=\mathrm{diag}\left({\left[ h^{[pq]}_{1},\dots,h^{[pq]}_{n}\right]}\right),~p\neq q\in\{1,2,3\}$ and $U^{[pq]}=\left\lbrace u_{1}^{[pq]},\dots,u^{[pq]}_{\lvert U^{[pq]} \rvert} \right\rbrace $ shows the time snapshots in which the exact channel value ${h^{[pq]}_{j}} , j\in U^{[pq]}$ between $\mathrm{TX}_p$ and $\mathrm{RX}_q$ is unknown. We assume all the direct channel matrices ${\bar{\bf{H}}}^{[pp]}$ is available at $\mathrm{RX}_p$. Considering all the channel matrices ${\bar{\bf{H}}}^{[pq]} $ are full rank almost surely. Our objective is to find proper encoding $e_{q}\left(M^{[q]},\bar{\bf{X}}^{[q]} |\bf{\Theta}' \right)$ and decoding $d_{q}\left(\bar{\bf{Y}}^{[q]} |{\bf{\Theta}}',\bar{\bf{H}}^{[pp]} \right), p,q\in{1,\dots,K}$
functions at both transmitters and receivers, respectively that satisfied IA conditions in \eqref{align1}, \eqref{align2}. Also ${\bf{\Theta}}'$ shows our partial knowledge from CSI. In this case the received signal at $\mathrm{RX}_p$ like \eqref{received signal model} can be modeled as follows:
\begin{equation}
\label{RX}
\bar{\bf{Y}}^{[p]} =\sum_{q=1}^{K} {\bar{\bf{H}}^{[pq]} \bar{\bf{X}}^{[q]}}+\bar{\bf{Z}}^{[p]}.
\end{equation}
\newtheorem{theorem_I_one}{Lemma}
Now, we want to design such precoder vectors $\bar{\bf{V}}^{[1]} $, $\bar{\bf{V}}^{[2]} $ and $\bar{\bf{V}}^{[3]} $ so that the IA conditions are satisfied. Referring Lemma 4 without losing generality the IA conditions can be expressed as follows:
\begin{equation}
\label{eqn_13}
\begin{aligned}
&{{\mathrm{span}}}\left(\bar{\bf{Q}}^{[23]_{i_2}} \bar{\bf{V}}^{[3]} \right)\subseteq {{\mathrm{span}}}\left(\bar{\bf{Q}}^{[21]_{i_1}} \bar{\bf{V}}^{[1]} \right)&\\
&{{\mathrm{span}}}\left(\bar{\bf{Q}}^{[32]_{i_4}} \bar{\bf{V}}^{[2]} \right)\subseteq {{\mathrm{span}}}\left(\bar{\bf{Q}}^{[31]_{i_3}} \bar{\bf{V}}^{[1]} \right)&\\
\end{aligned}
\end{equation}
and
\begin{equation}
\label{eqn_14}
{{\mathrm{span}}}\left(\bar{\bf{Q}}^{[12]_{i_5}} \bar{\bf{V}}^{[2]} \right)={{\mathrm{span}}}\left(\bar{\bf{Q}}^{[13]_{i_6}} \bar{\bf{V}}^{[3]} \right).
\end{equation}     
where, $ i_l \in \lbrace {1,\dots,\lvert U^{[pq]}\rvert+1} \rbrace $ e.g. $ i_1 \in \lbrace {1,\dots,\lvert U^{[21]} \rvert+1} \rbrace $. The relations \ref{eqn_13} and \ref{eqn_14} can equivalently be presented as: 
\begin{equation}
\label{eqn_15}
{{\mathrm{span}}}\left({{\bf{B}}_{\left({i_1,i_2}\right)}}\right)={\mathrm{span}}\left({{\bf{T}}_{\left({i_1,i_2,i_3,i_4,i_5,i_6}\right)}{\bf{C}}_{\left({i_3,i_4}\right)}}\right)
\end{equation}
\begin{equation}
\label{eqn_16}
{{\mathrm{span}}}\left({\bf{B}}_{\left({i_1,i_2}\right)}\right)\subseteq{{\mathrm{span}}}\left({\bf{A}}\right)
\end{equation}
\begin{equation}
\label{eqn_17}
{{\mathrm{span}}}\left({\bf{C}}_{\left({i_3,i_4}\right)}\right)\subseteq{{\mathrm{span}}}\left({\bf{A}}\right)
\end{equation}
where,
\begin{equation}
{\bf{A}}={\bar{\bf{V}}}^{[1]} ~~~~~~~~~~~~~~~~~~~~~~~~~~~~~~~~~~~~~~~~~
\end{equation}
\begin{equation}
{\bf{B}}_{\left({i_1,i_2}\right)}=\left(\bar{\bf{Q}}^{[21]_{i_1}} \right)^{-1} \bar{\bf{Q}}^{[23]_{i_2}} {\bar{\bf{V}}}^{[3]} ~~~~~~~~~~~~~~~~~~
\end{equation}
\begin{equation}
{\bf{C}}_{\left({i_3,i_4}\right)}=\left(\bar{\bf{Q}}^{[31]_{i_3}} \right)^{-1} \bar{\bf{Q}}^{[32]_{i_4}} {\bar{\bf{V}}}^{[2]} ~~~~~~~~~~~~~~~~~~
\end{equation}
\begin{equation}
\begin{aligned}
{\bf{T}}_{\left({i_1,\dots,i_6}\right)}=\bar{\bf{Q}}^{[12]_{i_5}} &\left(\bar{\bf{Q}}^{[21]_{i_1}} \right)^{-1}\bar{\bf{Q}}^{[23]_{i_2}} &\\
&\left(\bar{\bf{Q}}^{[32]_{i_4}} \right)^{-1}\bar{\bf{Q}}^{[31]_{i_3}} \left(\bar{\bf{Q}}^{[13]_{i_6}} \right)^{-1},&
\end{aligned}
\end{equation}
Let ${\bf{W}}=\left[1~1 \dots1\right]^\mathrm{T}$ be an $n\times1$ column matrix and the matrix ${\bar{\bf{\Gamma}}}=\mathrm{diag}\left(\left[{\Gamma_{1},\dots,\Gamma_{n}}\right]\right)$ is defined as follows:
\begin{equation}
\left\{
\begin{matrix}
\Gamma_{rr}=1,~&\mathrm{if}&~r\notin \bigcup_{p,q} U^{[pq]}, p\neq q\\ \\
\Gamma_{rr}=\gamma_{r},~&\mathrm{if}&~r\in \bigcup_{p,q} U^{[pq]}, p\neq q
\end{matrix}\right.
,
\end{equation}
where $\gamma_{r}$ is a random variable with an arbitrarily distribution, $n=2L+2\epsilon+1, \epsilon \in \mathbb{N}$ and $L=\lvert{\bigcup_{p,q} U^{[pq]}}\rvert+1, p\neq q$.
Now, we select ${\bf{A}}$, ${\bf{B}}$ and ${\bf{C}}$ such that:
\begin{align}
{{\mathrm{span}}}\left({\bf{A}}\right)=&{{\mathrm{span}}}\left\lbrace \prod_{i,j} {\bf{T}}^{i}{\bf{\Gamma}}^{j}{\bf{W}}:0\leq i \leq\epsilon,1\leq j \leq L \right\rbrace,&\\
{{\mathrm{span}}}\left({\bf{B}}_{\left({i_1,i_2}\right)}\right)=&{{\mathrm{span}}}\left\lbrace \prod_{i,j} {\bf{T}}^{i}{\bf{\Gamma}}^{j}{\bf{W}}:1\leq i \leq\epsilon,1\leq j \leq L \right\rbrace,&\\
{{\mathrm{span}}}\left({{\bf{C}}_{\left({i_3,i_4}\right)}}\right)=&{{\mathrm{span}}}\left\lbrace \prod_{i,j} {\bf{T}}^{i}{\bf{\Gamma}}^{j}{\bf{W}}:0\leq i \leq\epsilon-1,1\leq j \leq L \right\rbrace.&
\end{align}
where, ${\bf{T}}={\bf{T}}_{\left(1,\dots,1\right)}$. In the next lemma we show that the designed vector sets satisfied IA constrains.
\begin{lem}
\label{lem5} \it{The designed ${\bf{A}}$, ${\bf{B}}_{\left({i_1,i_2}\right)}$ and ${\bf{C}}_{\left({i_3,i_4}\right)}$ satisfy our modified IA conditions \eqref{eqn_15}, \eqref{eqn_16} and \eqref{eqn_17}. }
\end{lem}
\begin{IEEEproof}
First we show that:
\begin{equation}
 {\bf{T}}_{\left({i_1,\dots,i_6}\right)}{\bf{\Gamma}}^{j'}{\bf{W}}=\sum_{j=1}^{L}{{{\alpha_{j}}\bf{T}}{\bf{\Gamma}}^{j}{\bf{W}}}.
\end{equation} 
The matrices ${\bf{T}}_{\left({i_1,\dots,i_6}\right)}{\bf{\Gamma}}^{j'}{\bf{W}}$ and $ {\bf{T}}{\bf{\Gamma}}^{j}{\bf{W}}$ have equal values in the rows of the set $\{1,\dots,n\}-\bigcup_{p,q} U^{[pq]}$ (the rows in the set of $\bigcup_{p,q} U^{[pq]}$ are not equivalent). Therefore, we can find a linear combination of the vectors in the set of $\lbrace{{\bf{T}}{\bf{\Gamma}}^{j}{\bf{W}}}\rbrace$ to be equivalent with ${\bf{T}}_{\left({i_1,\dots,i_6}\right)}{\bf{\Gamma}}^{j'}{\bf{W}}$. Let us check the relation \eqref{eqn_15}:
 \begin{align}
 {\mathrm{span}}\left({{\bf{T}}_{\left({i_1,i_2,i_3,i_4,i_5,i_6}\right)}{\bf{C}}_{\left({i_3,i_4}\right)}}\right)&={{\mathrm{span}}}\left\lbrace {\bf{T}}_{\left({i_1,i_2,i_3,i_4,i_5,i_6}\right)}\prod_{i,j} {\bf{T}}^{i}{\bf{\Gamma}}^{j}{\bf{W}}:0\leq i \leq\epsilon-1,1\leq j \leq L \right\rbrace&\\
 &={{\mathrm{span}}}\left\lbrace \prod_{i,j} {\bf{T}}^{i}{\bf{T}}_{\left({i_1,i_2,i_3,i_4,i_5,i_6}\right)}{\bf{\Gamma}}^{j}{\bf{W}}:0\leq i \leq\epsilon-1,1\leq j \leq L \right\rbrace&\\
 &={{\mathrm{span}}}\left\lbrace \prod_{i,j} {{\bf{T}}^{i}}{\sum_{j=1}^{L}{{{\alpha_{j}}\bf{T}}{\bf{\Gamma}}^{j}{\bf{W}}}}:0\leq i \leq\epsilon-1,1\leq j \leq L \right\rbrace&\\
 &={{\mathrm{span}}}\left\lbrace \sum_{j=1}^{L}{{\alpha_{j}} \prod_{i,j} {{\bf{T}}^{i}}{{\bf{\Gamma}}^{j}{\bf{W}}}}:1\leq i \leq\epsilon,1\leq j \leq L \right\rbrace&\\
 &={{\mathrm{span}}}\left\lbrace { \prod_{i,j} {{\bf{T}}^{i}}{{\bf{\Gamma}}^{j}{\bf{W}}}}:1\leq i \leq\epsilon,1\leq j \leq L \right\rbrace&\\
 &={{\mathrm{span}}}\left({\bf{B}}_{\left({i_1,i_2}\right)}\right).&
  \end{align}
In the similar way, we can show that relations \eqref{eqn_16} and \eqref{eqn_17} are satisfied.
\end{IEEEproof}
Since Lemma 5 for all the values of $\{{i_1,\dots,i_6}\}$ are satisfied, in the rest of the paper for simplifying the notation we use $\bf{B}$ and $\bf{C}$ instead of ${\bf{B}}_{\left({i_1,i_2}\right)}$ and ${\bf{C}}_{\left({i_3,i_4}\right)}$, respectively. 

Now, we should find the number of dimension which is occupied by each user. In the next lemma the number of active dimension for each user is calculated.
\begin{lem}
\label{theorem_three} \it{The dimension of spaces spanned by the matrices ${\bf{A}}$, ${\bf{B}}$ and ${\bf{C}}$ are as follows:
 \begin{align}
  \mathrm{rank}\left({\bf{A}}\right)&=L+\epsilon+1&\\
  \mathrm{rank}\left({\bf{B}}\right)&=L+\epsilon~{\bf{1}}(\epsilon)&\\
  \mathrm{rank}\left({\bf{C}}\right)&=L+\epsilon~{\bf{1}}(\epsilon)&  
 \end{align}
where, ${\bf{1}}(\epsilon=1)=0$ and ${\bf{1}}(\epsilon>1)=1$.} 
\end{lem}
\begin{IEEEproof}
Since all the relations of $\mathrm{rank}\left({\bf{A}}\right)=L+\epsilon+1$, $\mathrm{rank}\left({\bf{B}}\right)=L+\epsilon~{\bf{1}}(\epsilon)$ and $\mathrm{rank}\left({\bf{C}}\right)=L+\epsilon~{\bf{1}}(\epsilon)$ have similar way of proof, to avoid repetition we focus on the proof of $\mathrm{rank}\left({\bf{A}}\right)=L+\epsilon+1$ and all the other equality have the similar way of proof. Let,
\begin{equation}
\label{EE}
{\bf{E}}=\left[
\begin{aligned}
&{\bf{t_1}}&~&{\bf{t_2}}&~&\dots&~&{\bf{t_{\epsilon+1}}}&~&{\bf{0}}&~&{\bf{0}}&~&\dots&~&{\bf{0}}&\\
&{\bf{0'}}&~&{\bf{0'}}&~&\dots&~&{\bf{0'}}&~&{\bf{e_1}}&~&{\bf{e_2}}&~&\dots&~&{\bf{e_{L}}}&\\
\end{aligned}
\right]
\end{equation}
where, ${\bf{e_i}}={\bf{e_1}}^{i}, 1\leq i \leq \epsilon+1$ and ${\bf{t_i}}={\bf{t_1}}^{i}, 1\leq i \leq L$ are two column matrices with the size of $(L+2\epsilon+1)\times 1$ and $L\times 1$, respectively. Also, the matrices ${\bf{0}}$ and ${\bf{0'}}$ are all zero column matrices with the size of $(L+2\epsilon+1)\times 1$ and $L\times 1$, respectively. The terms ${\bf{e_1}}^{i}$ and ${\bf{t_1}}^{i}$ are defined as follows:
\begin{equation}
\begin{aligned}
{\bf{e_1}}^{i}&=&\underbrace{{\bf{e_1}}\odot\dots\odot{\bf{e_1}}}_{\textrm{i times}}\\
{\bf{t_1}}^{i}&=&\underbrace{{\bf{t_1}}\odot\dots\odot{\bf{t_1}}}_{\textrm{i times}}.
\end{aligned}
\end{equation}
where, ${\bf{P}}\odot {\bf{Q}}$ shows Hadamard product between two matrices ${\bf{P}}$ and ${\bf{Q}}$. Assume two time snapshot sets of ${\bf{\Omega}}=\bigcup_{p,q}{U^{[pq]}}, (p\neq q)=\lbrace{\Omega_{1},\Omega_{2},\dots,\Omega_{L}}\rbrace,~\Omega_{1}<\Omega_{2}<\dots<\Omega_{L}$ and ${\bf{\Omega}'}=\lbrace{1,2,\dots,n}\rbrace-{\bf{\Omega}}=\lbrace{\Omega_{1}',\Omega_{2}',\dots,\Omega_{n-L}'}\rbrace$ that $\Omega_{1}'<\Omega_{2}'<\dots<\Omega_{n-L}'$. If we define ${\bf{T}}=\mathrm{diag} \left( \left[{T_1,\dots,T_n}\right] \right)$ the column matrices ${\bf{t}_1}$ and ${\bf{e}_1}$ are represented as follows:
\begin{equation}
\begin{aligned}
{\bf{e}_1}&={\left[T_{\Omega_{1}},T_{\Omega_{2}},\dots,T_{\Omega_{L}}\right]}^\mathrm{T}&\\
{\bf{t}_1}&={\left[\Gamma_{\Omega_{1}'},\Gamma_{\Omega_{2}'},\dots,\Gamma_{\Omega_{n-L}'}\right]}^\mathrm{T}&
\end{aligned}
\end{equation}
Now, we show that ${\mathrm{span}}\left({\bf{A}}\right) \subseteq {{\mathrm{span}}} \left({\bf{E}}\right)$. In order to prove ${{\mathrm{span}}}\left({\bf{A}}\right) \subseteq {{\mathrm{span}}}({\bf{E}})$, we should show that all the members of the set $\lbrace{ \prod_{i,j} {\bf{T}}^{i} {\bf{\Gamma}}^{j} {\bf{W}}: 0 \leq i \leq \epsilon, 1 \leq j \leq L }\rbrace$ can be generated through the basic operations on the columns and the rows of the matrix ${\bf{E}}$, e.g. interchanging, adding, subtracting and multiplying constant numbers. For every values of $i$ and $j$ the column matrix ${\bf{T}}^{i}{\bf{\Gamma}}^{j}{\bf{W}} \triangleq {\bf{P}}^{[ij]}$ can be represented as follows:
\begin{equation}
{\bf{T}}^{i}{\bf{\Gamma}}^{j}{\bf{W}} \triangleq {\bf{P}}^{[ij]}={\left[ T_{1}^{i},\dots,T_{\Omega_{1}}^{i}\gamma_{\Omega_{1}}^{j},\dots,T_{\Omega_{L}}^{i}\gamma_{\Omega_{L}}^{j},\dots,T_{n}^{i} \right]}^\mathrm{T}.
\end{equation}
Let ${\bf{P}}^{[ij]}={\left[ P_1,\dots,P_n\right]}^{\mathrm{T}}$, also assume $G_1$ is a matrix with the following definition:
\begin{equation}
G_1 \triangleq \left[{P_{j_1},\dots,P_{j_{L}}}\right]^\mathrm{T},~j_{l} \in {\bf{\Omega}}
\end{equation}
where, $G_1$ can be represented by the linear combination of vectors ${\bf{e}_i}, i=\lbrace 1,\dots,L\rbrace$ as follows:
\begin{equation}
G_1=\sum_{i=1}^{L}{\alpha_{i}{\bf{e}_i}}.
\end{equation}
Similarly, we define the matrix $G_2$ as follows:
\begin{equation}
G_2 \triangleq \left[{P_{j_1},\dots,P_{j_{L}}}\right]^{\mathrm{T}},~j_{l} \in {\bf{\Omega}}',
\end{equation}
all the rows of the matrix $G_2$ are equal to the matrix ${\bf{P}}^{[ij]}$ with the row number of the set $\Omega'$. This matrix can be represented by ${\bf{t}_i}, i=\lbrace 1,\dots,n-L\rbrace$ from the matrix ${\bf{E}}$ defined at \eqref{EE}. Therefore, all the members of the set $\bf{A}$ can be generated from the linear combination of the columns of matrix ${\bf{E}}$, so we have:
\begin{equation}
\mathrm{span}{\left({\bf{A}}\right)}\subseteq\mathrm{span}{\left({\bf{E}}\right)},
\end{equation}
in the similar way, we can prove that $\mathrm{span}\left({\bf{E}}\right)\leq \mathrm{span}({\bf{A}})$. Since the matrix ${\bf{E}}$ has $L+\epsilon+1$ independent columns, we get:
\begin{equation}
\mathrm{rank}({\bf{E}})=\mathrm{rank}({\bf{A}})=L+\epsilon+1,
\end{equation}
 which proves this lemma. 
\end{IEEEproof}
\begin{lem}
\label{lemma_four} \it{If $\Omega_{i+1}-\Omega_{i}<M$ then:
\begin{equation}
 rank\left(\left[{{\bar{\bf{H}}}}^{[11]} \bar{\bf{V}}^{[1]}~{{\bar{\bf{H}}}}^{[12]} \bar{\bf{V}}^{[2]}\right]\right)=2(L+\epsilon)+1
\end{equation}}
\end{lem}
\begin{IEEEproof}
By multiplying matrix $\left[{{\bar{\bf{H}}}}^{[11]} \bar{\bf{V}}^{[1]}~{\bar{\bf{H}}}^{[12]} \bar{\bf{V}}^{[2]}\right]$ by the matrix $\left({{\bar{\bf{H}}}}^{[12]} \right)^{-1}$ and applying some simplification, we should show that $\left[{\bf{\bar{H}}}{\bf{A}}~{\bf{C}}\right]$ is a full rank matrix where, ${\bf{\bar{H}}}={\bar{\bf{H}}}^{[11]} \left({{\bar{\bf{H}}}}^{[12]} \right)^{-1}\left(\bar{\bf{Q}}^{[32]_1} \right)^{-1}\bar{\bf{Q}}^{[31]_1} $. Since $\left({{\bar{\bf{H}}}}^{[12]} \right)^{-1}\left(\bar{\bf{Q}}^{[32]_1} \right)^{-1}\\ \bar{\bf{Q}}^{[31]_1} $ is a diagonal matrix with random elements, without losing generality of our problem, the matrix ${\bf{\bar{H}}}$ can be represented by a matrix with the same structure of ${\bar{\bf{H}}}^{[11]} $ as follows:
\begin{equation}
{\bf{\bar{H}}}=
\begin{bmatrix}
&H_{11}&&\dots&&0&\\
&\vdots&&\ddots&&\vdots&\\
&H_{n1}&&\dots&&H_{nn}&
\end{bmatrix}
\end{equation}
where $H_{m_1 m_2}$ has non-zero elements for the $0\leq m_{1}-m_{2}\leq M$. Therefore, the space spanned by the matrix ${\bf{\bar{H}}}{\bf{A}}$ can be calculated as follows:
\begin{equation}
{{\mathrm{span}}}\left({\bf{\bar{H}}}{\bf{A}}\right)=\Bigg\{{\sum_{l=1}^{f}\lambda_{l}\prod_{i,j}{\bf{\bar{H}}}{\bf{\Gamma}}^{j}{\bf{T}}^{i}{\bf{W}}\vert f\in\mathbb{N},~\lambda_{l}\in {\bf{R}},0\leq i \leq\epsilon,1\leq j \leq L}\Bigg\},
\end{equation}
finally,
\begin{equation}
{{\mathrm{span}}}\left({\bf{\bar{H}}}{\bf{A}}\right)=\Bigg\{{\sum_{l=1}^{f}\lambda_{l}\prod_{i,j}{\bf{\bar{H}}'}_{j}{\bf{T}}^{i}{\bf{W}}\vert f\in\mathbb{N},~\lambda_{l}\in {\bf{R}},0\leq i \leq\epsilon,1\leq j \leq L}\Bigg\}.
\end{equation}
Since for $0\leq m_{1}-m_{2}\leq M$ the ${\bf{\bar{H}}'}_{j}$ is a full rank matrix, the basic vectors of ${\bf{\bar{H}}}{\bf{A}}$ are linearly independent from the basic vectors of ${\bf{C}}$, which concludes the proof of this lemma.  
\end{IEEEproof}
Lemma 7 shows that the signal received to the first receiver is linearly independent from the interference subspace. Similar method of proof can be used for the second and third receivers. Therefore, we show that $\left(d_1,d_2,d_3\right)$ lies in the DoF region of 3-user interference channel while we do not know about the $\frac{L}{2(L+\epsilon)+1}$ portion of total CSI, where:
\begin{equation}
\begin{aligned}
(d_1,&d_2,d_3)=&\\
&\left(\frac{L+\epsilon+1}{2(L+\epsilon)+1},\frac{L+\epsilon}{2(L+\epsilon)+1},\frac{L+\epsilon}{2(L+\epsilon)+1}\right)&\\
\end{aligned}
\end{equation}
As $L\rightarrow\infty$, the value of $\frac{L}{2(L+\epsilon)+1}$ and triple $\left(d_1,d_2,d_3\right)$ go to $\frac{1}{2}$ and $\left(\frac{1}{2},\frac{1}{2},\frac{1}{2}\right)$, respectively.
\end{IEEEproof}
The result of this theorem can be easily extended to the $K-$user interference channel which is prepared in Appendix C.

\textbf{\textit{definition:}}$\Upsilon$ is a fraction of time in which all the transmitters have access to perfect CSIT. We define $\Upsilon_K$ for the $K-$user interference channel as follows:
\begin{equation}
\Upsilon_K \triangleq 1-\frac{\vert\cup_{pq}{U^{[pq]},p\neq q}\vert}{n}
\end{equation}
\section{Lower Bound on the Minimum $\Upsilon_K$ to Achieve Maximum DoF of $\frac{K}{2},~K\geq3$}
For the case where $K=2$, it is clear that using time sharing among transmitters (half of the total time slots for the first user and what remains for the second one) every user easily achieves $\frac{1}{2}$ DoF. Therefore, in this case, the minimum value of $\Upsilon_2$ to achieve 1 DoF is zero and there is no need of CSI either at transmitters or receivers. The interesting case is when $K>2$. In the Theorem 4, we propose an achievable scheme to achieve maximum DoF of $\frac{3}{2}$, when $\Upsilon_3=\frac{1}{2}$. In Appendix C, we generalized this theorem to more general problem of the $K-$user interference channel. Now, we present a lower bound for the $\Upsilon_K$ and we show that $\Upsilon_K=\frac{1}{2}$ is the minimum value being needed to achieve maximum DoF of $\frac{K}{2}$.
\begin{thm}
\label{theorem_3} \it{If $\Upsilon_K$ and $\Upsilon_{K-1}$ are the minimum fractions of time values for $K-$user and $(K-1)-$user interference channels with the same channel distributions to achieve maximum DoF then $\Upsilon_{K} \geq \Upsilon_{K-1}$.}
\end{thm}
\begin{IEEEproof}
Assume $\Upsilon_{K}<\Upsilon_{K-1}$, by omitting $\mathrm{TX}_K$ and $\mathrm{RX}_K$ from $K-$user interference channel, we find degraded version of $(K-1)-$user network. This degraded network can also achieve its maximum achievable DoF with $\Upsilon_{K}$ portion of CSI. Therefore, with the $\Upsilon_{K}$, which is less than $\Upsilon_{K-1}$, the maximum DoF of $\frac{K-1}{2}$ is also achievable. This contradicts $\Upsilon_{K}<\Upsilon_{K-1}$, thus $\Upsilon_{K} \geq \Upsilon_{K-1}$, where concludes the proof of this theorem. 
\end{IEEEproof}
\begin{thm}
\label{theorem_4} For $K=3$, $\frac{1}{2}$ is the minimum value for $\Upsilon_3$ to achieve maximum DoF of $\frac{3}{2}$. 
\end{thm}
\begin{IEEEproof}
 Consider the channel output at the first receiver is denoted as follows:
\begin{equation}
Y^{n}_1=\left(Y^{n}_{1,P},Y^{n}_{1,NP}\right),
\end{equation}
$Y^{n}_{1,P}$ is the channel output at the first receiver where the perfect channel state is presented. $Y^{n}_{1,NP}$ is the channel outputs in which the perfect values of channel state is not presented.
Next, we add up one artificial receiver that is statistically similar to the first one. The output of this artificial receiver is denoted as follows:
\begin{equation}
\widehat{Y}^{n}_1=\left(Y^{\frac{n}{2}}_{1,P},\widehat{Y}^{\frac{n}{2}}_{1,NP}\right)
\end{equation}
where, during $\frac{n}{2}$ time slots $Y^{\frac{n}{2}}_{1,P}$ is exactly similar to what is received at first receiver also $\widehat{Y}^{\frac{n}{2}}_{1,NP}$ has the same distribution of $Y^{n}_{1,NP}$ but is not equal. Let $\bf{\Theta}$ be the total channel state information of interference channel, we can upper bound $R_1$ as follow:
\begin{equation}
\begin{aligned}
nR_1&=H\left(M^{[1]}\right)&\\
&=H\left(M^{[1]}\vert{\bf{\Theta}}\right)&\\
&\overset{(a)}{\leq} I\left(M^{[1]};Y^{n}_{1}\vert{\bf{\Theta}}\right)+n\varepsilon_{n}&\\
&=I\left(M^{[1]};Y^{n}_{1,P},Y^{n}_{1,NP}\vert \bf{\Theta}\right)+n\varepsilon_{n}&\\
&=h\left(Y^{n}_{1,P},Y^{n}_{1,NP}\vert {\bf{\Theta}}\right)-h\left(Y^{n}_{1,P},Y^{n}_{1,NP}\vert M^{[1]},{\bf{\Theta}}\right)+n\varepsilon_{n}&\\
&\overset{(b)}{\leq} n\log{(P)}-\underbrace{h\left(Y^{n}_{1,P}\vert M^{[1]},{\bf{\Theta}}\right)}_{\leq no\left(\log(P)\right)}-h\left(Y^{n}_{1,NP} \vert M^{[1]},{\bf{\Theta}},Y^{n}_{1,P}\right)+n\varepsilon_{n}&
\end{aligned}
\end{equation}
where, (a) comes from Fano's inequality and (b) comes from this fact that by knowing $M^{[1]}$ and $\bf{\Theta}$ we can estimate $Y^{n}_{1,P}$ within noise power. Let us divide $M^{[1]}$ into two sets of $M'^{[1]}=\lbrace 1,2,\dots,2^{nR'_1}\rbrace$ and $M''^{[1]}=\lbrace 1,2,\dots,2^{nR''_1}\rbrace$ where $R'_1=R''_1=\frac{R_1}{2}$ for artificial receiver we get ($M^{[1]}=M'^{[1]} \times M''^{[1]}$):
\begin{equation}
\begin{aligned}
nR'_1=n\frac{R_1}{2}&=H(M'^{[1]})&\\
&\leq I\left(M'^{[1]};\widehat{Y}^{\frac{n}{2}}_1\vert{\bf{\Theta}}\right)+\frac{n}{2}\varepsilon_{\frac{n}{2}}&\\
&=I\left(M'^{[1]};\widehat{Y}^{\frac{n}{2}}_{1,P},\widehat{Y}^{\frac{n}{2}}_{1,NP}\vert \bf{\Theta}\right)+\frac{n}{2}\varepsilon_{\frac{n}{2}}&\\
&=h\left(\widehat{Y}^{\frac{n}{2}}_{1,P},\widehat{Y}^{\frac{n}{2}}_{1,NP}\vert {\bf{\Theta}}\right)-h\left(\widehat{Y}^{\frac{n}{2}}_{1,P},\widehat{Y}^{\frac{n}{2}}_{1,NP}\vert M'^{[1]},{\bf{\Theta}}\right)+\frac{n}{2} \varepsilon_{\frac{n}{2}}&\\
&\overset{(a)}{\leq} \frac{n}{2}\log{(P)}-\underbrace{h\left(Y^{\frac{n}{2}}_{1,P}\vert M'^{[1]},{\bf{\Theta}}\right)}_{\leq \frac{n}{2}o\left(\log(P)\right)}-h\left(Y^{\frac{n}{2}}_{1,NP} \vert M'^{[1]},{\bf{\Theta}},Y^{\frac{n}{2}}_{1,P}\right)+\frac{n}{2}\varepsilon_{\frac{n}{2}}&
\end{aligned}
\end{equation}
adding up all the above bounds we have:
\begin{equation}
\label{entropy2}
\begin{aligned}
\frac{3}{2} n R_{1}&\leq \frac{3}{2} n \log{(P)}&\\
&-h(Y^{n}_{1,p}\vert M^{[1]},{\bf{\Theta}})-h\left(Y^{n}_{1,NP} \vert M^{[1]},{\bf{\Theta}},Y^{n}_{1,P}\right)-\underbrace{h\left(\widehat{Y}^{\frac{n}{2}}_{1,NP} \vert M'^{[1]},{\bf{\Theta}},Y^{\frac{n}{2}}_{1,P}\right)}_{\geq h\left(\widehat{Y}^{\frac{n}{2}}_{1,NP} \vert M^{[1]},{\bf{\Theta}},Y^{n}_{1,P},Y^{n}_{1,NP}\right)}-no\left(\log{(P)}\right)+n\varepsilon '_{n}&\\
&\leq \frac{3}{2} n \log{(P)}&\\
&-h(Y^{n}_{1,p}\vert M^{[1]},{\bf{\Theta}})-h\left(Y^{n}_{1,NP} \vert M^{[1]},{\bf{\Theta}},Y^{n}_{1,P}\right)- h\left(\widehat{Y}^{\frac{n}{2}}_{1,NP} \vert M^{[1]},{\bf{\Theta}},Y^{n}_{1,P},Y^{n}_{1,NP}\right)-no\left(\log{(P)}\right)+n\varepsilon '_{n}&\\
\end{aligned}
\end{equation}
Then we can write the following bounds for the receivers $2,3\dots,K$ as follows:
\begin{equation}
\label{entropy3}
\begin{aligned}
&n\left(R_2+R_3\right)&\\
&\leq H\left(M^{[2]},M^{[3]}\right)&\\
&\overset{(a)}{=}H\left(M^{[2]},M^{[3]}\vert M^{[1]},{\bf{\Theta}}\right)&\\
&\overset{(b)}{\leq} I\left(M^{[2]},M^{[3]};Y^{n}_{1},Y^{n}_{2},Y^{n}_{3},\widehat{Y}^{\frac{n}{2}}_{1,NP}\vert M^{[1]},{\bf{\Theta}}\right)+n\varepsilon_n&\\
&=h\left(Y^{n}_{1,P},Y^{n}_{2},Y^{n}_{3},{Y}^{n}_{1,NP},\widehat{Y}^{\frac{n}{2}}_{1,NP}\vert M^{[1]},{\bf{\Theta}}\right)&\\
&~~~-h\left(Y^{n}_{1,P},Y^{n}_{2},Y^{n}_{3},{Y}^{n}_{1,NP},\widehat{Y}^{\frac{n}{2}}_{1,NP}\vert M^{[1]},M^{[2]},M^{[3]},{\bf{\Theta}}\right)+n\varepsilon_n&\\
&\leq h\left(Y^{n}_{1,P},Y^{n}_{2},Y^{n}_{3},{Y}^{n}_{1,NP},\widehat{Y}^{\frac{n}{2}}_{1,NP}\vert M^{[1]},{\bf{\Theta}}\right)+n\varepsilon_n&\\
&= h\left(Y^{n}_{1,P},{Y}^{n}_{1,NP},\widehat{Y}^{\frac{n}{2}}_{1,NP}\vert M^{[1]},{\bf{\Theta}}\right)+h\left(Y^{n}_{2},Y^{n}_{3}\vert Y^{n}_{1,P},{Y}^{n}_{1,NP},\widehat{Y}^{\frac{n}{2}}_{1,NP},M^{[1]},{\bf{\Theta}}\right)+n\varepsilon_n&\\
&=h\left(Y^{n}_{1,P}\vert M^{[1]},{\bf{\Theta}}\right)+h\left({Y}^{n}_{1,NP},\widehat{Y}^{\frac{n}{2}}_{1,NP}\vert M^{[1]},{\bf{\Theta}},Y^{n}_{1,P}\right)&\\
&~~~+h\left(Y^{n}_{2},Y^{n}_{3}\vert Y^{n}_{1,P},{Y}^{n}_{1,NP},\widehat{Y}^{\frac{n}{2}}_{1,NP},M^{[1]},{\bf{\Theta}}\right)+n\varepsilon_n&\\
&=h\left( Y^{n}_{1,P}\vert M^{[1]},{\bf{\Theta}}\right)+h\left( {Y}^{n}_{1,NP}\vert M^{[1]},{\bf{\Theta}},Y^{n}_{1,P}\right)&\\
&~~~+h\left(\widehat{Y}^{\frac{n}{2}}_{1,NP}\vert M^{[1]},{\bf{\Theta}},Y^{n}_{1,P},{Y}^{n}_{1,NP}\right)+h\left(Y^{n}_{2},Y^{n}_{3}\vert Y^{n}_{1,P},{Y}^{n}_{1,NP},\widehat{Y}^{\frac{n}{2}}_{1,NP},M^{[1]},{\bf{\Theta}}\right)+n\varepsilon_n&
\end{aligned}
\end{equation}
since $M^{[2]}$, $M^{[3]}$ are independent from $M^{[1]}$ and ${\bf{\Theta}}$, (a) comes from the conditional entropy, (b) comes from Fano's inequality. By adding relations \eqref{entropy2} and \eqref{entropy3} we have:
\begin{equation}
\label{entropy4}
\begin{aligned}
&n\left(\frac{3}{2} R_1+R_2+R_3\right)&\\
&~~~~~~\leq n\frac{3}{2}\log{(P)}+h\left(Y^{n}_{2},Y^{n}_{3}\vert Y^{n}_{1,P},{Y}^{n}_{1,NP},\widehat{Y}^{\frac{n}{2}}_{1,NP},M^{[1]},{\bf{\Theta}}\right)+n{\varepsilon_n}'.&\\
\end{aligned}
\end{equation}
Then $Y^{n}_2,Y^{n}_{3}$ can be divided into two parts as follows:
\begin{equation}
\begin{aligned}
&Y^{n}_2=\left(Y^{n}_{2,(1,P)},Y^{n}_{2,(1,NP)}\right)&\\
&Y^{n}_3=\left(Y^{n}_{3,(1,P)},Y^{n}_{3,(1,NP)}\right)&\\
\end{aligned}
\end{equation}\\
where, $Y^{n}_{k,(1,P)}$ and $Y^{n}_{k,(1,NP)}$ are channel outputs at $\mathrm{RX}_k$ for those instances in which perfect CSIT is presented and not presented, respectively.

The term $h\left(Y^{n}_{2},Y^{n}_{3}\vert Y^{n}_{1,P},{Y}^{n}_{1,NP},\widehat{Y}^{\frac{n}{2}}_{1,NP},M^{[1]},{\bf{\Theta}}\right)$ can be bounded as follows:\\
\begin{equation}
\label{entropy5}
\begin{aligned}
h \Big{(}Y^{n}_{2},Y^{n}_{3} & \vert Y^{n}_{1,P},{Y}^{n}_{1,NP},\widehat{Y}^{\frac{n}{2}}_{1,NP},M^{[1]},{\bf{\Theta}}\Big{)} &\\
&=h\left( Y^{n}_{2,P},Y^{n}_{2,NP},Y^{n}_{3,P},Y^{n}_{3,NP}\vert Y^{n}_{1,P},{Y}^{n}_{1,NP},\widehat{Y}^{\frac{n}{2}}_{1,NP},M^{[1]},{\bf{\Theta}}\right)&\\
&\leq h\left( Y^{n}_{2,(1,P)},Y^{n}_{3,(1,P)}\vert Y^{n}_{1,P},\widehat{Y}^{\frac{n}{2}}_{1,NP},{Y}^{n}_{1,NP},M^{[1]},{\bf{\Theta}}\right)&\\
&~~~+h\left(Y^{n}_{2,(1,NP)},Y^{n}_{3,(1,NP)}\vert Y^{n}_{1,P},\widehat{Y}^{\frac{n}{2}}_{1,NP},{Y}^{n}_{1,NP},M^{[1]},{\bf{\Theta}}\right)&\\
&\overset{(a)}{\leq} h\left(Y^{n}_{2,(1,P)},Y^{n}_{3,(1,P)}\vert {Y}^{\frac{n}{2}}_{1,P}\right)&\\
&~~~+h\left(Y^{n}_{2,(1,NP)},Y^{n}_{3,(1,NP)}\vert Y^{n}_{1,P},\widehat{Y}^{\frac{n}{2}}_{1,NP},{Y}^{n}_{1,NP},M^{[1]},{\bf{\Theta}}\right)&\\
&\overset{(b)}{=}h\left(Y^{n}_{2,(1,P)},Y^{n}_{3,(1,P)}\right)-h\left({Y}^{\frac{n}{2}}_{1,P}\right)&\\
&~~~\underbrace{+h\left(Y^{n}_{2,(1,NP)},Y^{n}_{3,(1,NP)}\vert Y^{n}_{1,P},\widehat{Y}^{n}_{1,NP},{Y}^{n}_{1,NP},M^{[1]},{\bf{\Theta}}\right)}_{\leq no\left(\log(P)\right)}&\\
&= \underbrace{\sum_{i=1}^{n\Upsilon_3}{h\left(Y_{2,P} ,Y_{3,(1,P)} \right)}}_{\leq n\Upsilon_3\log{(P)}}-\underbrace{\sum_{i=1}^{n\frac{\Upsilon_3}{2}}{h\left(Y_{1,P} \right)}}_{= n\frac{\Upsilon_{3}}{2}\log{(P)}}+no\left(\log(P)\right)&\\
&\leq n\Upsilon_3 \log{(P)}-n\frac{\Upsilon_{3}}{2}\log{(P)}+no\left(\log(P)\right)&\\
&=n\frac{\Upsilon_{3}}{2}{\log{(P)}}+no\left(\log(P)\right)&\\
\end{aligned}
\end{equation}\\
where (a) comes from conditional entropy, (b) comes from chain rule and also if we have copies of values $Y^{n}_{1,P},\widehat{Y}^{n}_{1,NP},{Y}^{n}_{1,NP}$ and $M^{[1]},{\bf{\Theta}}$ we can estimate random variables $Y^{n}_{2,(1,NP)},Y^{n}_{3,(1,NP)}$ within noise power. Substituting result of \ref{entropy5} in \ref{entropy4} we have:
 \begin{equation}
\begin{aligned}
&n\left(\frac{3}{2} R_1+R_2+R_3\right)&\\
&~~~~~~\leq \frac{3}{2} n\log{(P)}+\frac{\Upsilon_{3}}{2}n{\log{(P)}}+no\left(\log(P)\right)+n{\varepsilon_n}'.&\\
\end{aligned}
\end{equation}
Therefore:
\begin{equation}
\left(\frac{3}{2} d_1+d_2+d_3\right)\leq \frac{3}{2}+\frac{\Upsilon_{3}}{2},
\end{equation}
and similarly we can bound 3-tuple $\left(d_1,d_2,d_3 \right)$ as follows:
\begin{equation}
\begin{aligned}
&\left( d_1+\frac{3}{2}d_2+d_3\right)\leq \frac{3}{2}+\frac{\Upsilon_{3}}{2},&\\
&\left( d_1+d_2+\frac{3}{2}d_3\right)\leq \frac{3}{2}+\frac{\Upsilon_{3}}{2},&
\end{aligned}
\end{equation}
summing up all the above bounds for 3-tuple $\left(d_1,d_2,d_3 \right)$ we get:
\begin{equation}
\frac{7}{2}\left(d_1+d_2+d_3\right) \leq \frac{9}{2}+\frac{3\Upsilon_{3}}{2}
\end{equation}
so:
\begin{equation}
d_1+d_2+d_3\leq \frac{9}{7}+\frac{3\Upsilon_3}{7}
\end{equation}
and finally for $d_1+d_2+d_3=\frac{3}{2}$ we should have $\Upsilon_3\geq \frac{1}{2}$, which shows the proof of this theorem.
\end{IEEEproof}
\begin{thm}
\label{theorem_4} \it{For even values of $K\geq 4$, $\frac{1}{2}$ is the minimum value for $\Upsilon_K$ to achieve maximum DoF of $\frac{K}{2}$.} 
\end{thm}
\begin{IEEEproof}
 Consider the channel output at first receiver is denoted as follow:
\begin{equation}
Y^{n}_1=\left(Y^{n}_{1,P},Y^{n}_{1,NP}\right),
\end{equation}
$Y^{n}_{i,p}$ is the channel outputs at $i^{th}$ receiver where the perfect channel state is presented and $Y^{n}_{i,NP}$ is the channel outputs where the perfect values of channel state is not presented.
Next, we add up $\frac{K}{2}-1$ artificial receivers that are statistically similar to the first receiver. The outputs of these artificial receivers can be denoted as follows:
\begin{equation}
\widehat{Y}^{n}_i=\left(Y^{n}_{1,P},\widehat{Y}^{n}_{i,NP}\right),~1\leq i \leq \frac{K}{2}-1
\end{equation}
where, $Y^{n}_{1,P}$ is exactly similar to what is received at first receiver and $\widehat{Y}^{n}_{i,NP}$ has the same distribution of $Y^{n}_{1,NP}$ but not equal. Let $\bf{\Theta}$ be the total channel state information of interference channel, we can upper bound $R_1$ as follows:
\begin{equation}
\begin{aligned}
nR_1&=H\left(M^{[1]}\right)&\\
&=H\left(M^{[1]}\vert{\bf{\Theta}}\right)&\\
&\leq I\left(M^{[1]};Y^{n}_{1}\vert{\bf{\Theta}}\right)+n\varepsilon_{n}&\\
&=I\left(M^{[1]};Y^{n}_{1,P},Y^{n}_{1,NP}\vert \bf{\Theta}\right)+n\varepsilon_{n}&\\
&=h\left(Y^{n}_{1,P},Y^{n}_{1,NP}\vert {\bf{\Theta}}\right)-h\left(Y^{n}_{1,P},Y^{n}_{1,NP}\vert M^{[1]},{\bf{\Theta}}\right)+n\varepsilon_{n}&\\
&\leq n\log{(P)}-{h\left(Y^{n}_{1,P}\vert M^{[1]},{\bf{\Theta}}\right)}-h\left(Y^{n}_{1,NP} \vert M^{[1]},{\bf{\Theta}},Y^{n}_{1,P}\right)+n\varepsilon_{n}&
\end{aligned}
\end{equation}
in the similar way for all other $1\leq i \leq \frac{K}{2}-1$ artificial receivers we get:
\begin{equation}
\begin{aligned}
nR_1&\leq n\log{(P)}-\underbrace{h\left(Y^{n}_{1,P}\vert M^{[1]},{\bf{\Theta}}\right)}_{\geq no\left(\log{(P)}\right)}-h\left(\widehat{Y}^{n}_{i,NP} \vert M^{[1]},{\bf{\Theta}},Y^{n}_{1,P}\right)+n\varepsilon_{n}&\\
&\leq n\log{(P)}-no\left(\log{(P)}\right)-h\left(\widehat{Y}^{n}_{i,NP} \vert M^{[1]},{\bf{\Theta}},Y^{n}_{1,P}\right)+n\varepsilon_{n}.&
\end{aligned}
\end{equation}
adding up all these $K/2$ bound we have:
\begin{equation}
\label{entropy6}
\begin{aligned}
n\frac{K}{2}R_{1}&\leq n\frac{K}{2}\log{(P)}-h(Y^{n}_{1,p}\vert M^{[1]},{\bf{\Theta}})-no\left(\log{(P)}\right)-\underbrace{\sum_{i=1}^{\frac{K}{2}-1}{h\left(\widehat{Y}^{n}_{i,NP}\vert M^{[1]},{\bf{\Theta}},Y^{n}_{1,P}\right)}}_{\geq h\left(\widehat{Y}^{n}_{1,NP},\dots\widehat{Y}^{n}_{\frac{K}{2}-1,NP}\vert M^{[1]},{\bf{\Theta}},Y^{n}_{1,P}\right)}+n\frac{K}{2}\varepsilon_{n}&\\
&\leq n\frac{K}{2}\log{(P)}-h(Y^{n}_{1,P}\vert M^{[1]},{\bf{\Theta}})-{ h\left(\widehat{Y}^{n}_{1,NP},\dots\widehat{Y}^{n}_{\frac{K}{2}-1,NP}\vert M^{[1]},{\bf{\Theta}},Y^{n}_{1,P}\right)}-no\left(\log{(P)}\right)+n\frac{K}{2}\varepsilon_{n}.&
\end{aligned}
\end{equation}
Then we can write the following bounds for the receivers $2,3\dots,K$ as follows:
\begin{equation}
\label{entropy7}
\begin{aligned}
&n\left(R_2+R_3+\dots+R_K\right)&\\
&\leq H\left(M^{[2]},M^{[3]},\dots,M^{[K]}\right)&\\
&=H\left(M^{[2]},M^{[3]},\dots,M^{[K]}\vert M^{[1]},{\bf{\Theta}}\right)&\\
&\leq I\left(M^{[2]},\dots,M^{[K]};Y^{n}_{1},Y^{n}_{2},\dots,Y^{n}_{K},\widehat{Y}^{n}_{1,NP},\dots,\widehat{Y}^{n}_{\frac{K}{2}-1,NP}\vert M^{[1]},{\bf{\Theta}}\right)+n\varepsilon_n&\\
&=h\left(Y^{n}_{1,P},Y^{n}_{2},\dots,Y^{n}_{K},Y^{n}_{1,NP},\widehat{Y}^{n}_{1,NP},\dots,\widehat{Y}^{n}_{\frac{K}{2}-1,NP}\vert M^{[1]},{\bf{\Theta}}\right)&\\
&~~~-h\left(Y^{n}_{1,P},Y^{n}_{2},\dots,Y^{n}_{K},Y^{n}_{1,NP},\widehat{Y}^{n}_{1,NP},\dots,\widehat{Y}^{n}_{\frac{K}{2}-1,NP}\vert M^{[1]},M^{[2]},\dots,M^{[K]},{\bf{\Theta}}\right)+n\varepsilon_n&\\
&\leq h\left(Y^{n}_{1,P},\dots,Y^{n}_{K},Y^{n}_{1,NP},\widehat{Y}^{n}_{1,NP},\dots,\widehat{Y}^{n}_{\frac{K}{2}-1,NP}\vert W_1,{\bf{\Theta}}\right)+n\varepsilon_n&\\
&= h\left(Y^{n}_{1,P},\widehat{Y}^{n}_{1,NP},\dots,\widehat{Y}^{n}_{\frac{K}{2}-1,NP}\vert M^{[1]},{\bf{\Theta}}\right)+h\left(Y^{n}_{2},\dots,Y^{n}_{K}\vert Y^{n}_{1,P},\widehat{Y}^{n}_{1,NP},\dots,\widehat{Y}^{n}_{\frac{K}{2}-1,NP},M^{[1]},{\bf{\Theta}}\right)+n\varepsilon_n&\\
&=h\left(Y^{n}_{1,P}\vert M^{[1]},{\bf{\Theta}}\right)+h\left(\widehat{Y}^{n}_{1,NP},\dots,\widehat{Y}^{n}_{\frac{K}{2}-1,NP}\vert M^{[1]},{\bf{\Theta}},Y^{n}_{1,P}\right)&\\
&~~~+h\left(Y^{n}_{2},\dots,Y^{n}_{K}\vert Y^{n}_{1,P},\widehat{Y}^{n}_{1,NP},\dots,\widehat{Y}^{n}_{\frac{K}{2}-1,NP},M^{[1]},{\bf{\Theta}}\right)+n\varepsilon_n&
\end{aligned}
\end{equation}
adding relations \eqref{entropy6} and \eqref{entropy7} we get:
\begin{equation}
\label{entropy8}
\begin{aligned}
&n\left(\frac{K}{2} R_1+R_2+\dots+R_K\right)&\\
&~~~~~~\leq n\frac{K}{2}\log{(P)}+h\left(Y^{n}_{2},\dots,Y^{n}_{K}\vert Y^{n}_{1,P},\widehat{Y}^{n}_{1,NP},\dots,\widehat{Y}^{n}_{\frac{K}{2}-1,NP},M^{[1]},{\bf{\Theta}}\right)-no\left(\log{(P)}\right)+n{\varepsilon_n}'.&
\end{aligned}
\end{equation}
Then $Y^{n}_2,Y^{n}_{3},\dots,Y^{n}_{K}$ can be divided into two parts as follows:
\begin{align}
&Y^{n}_2=\left(Y^{n}_{2,(1,P)},Y^{n}_{2,(1,NP)}\right)&\\
&Y^{n}_3=\left(Y^{n}_{3,(1,P)},Y^{n}_{3,(1,NP)}\right)&\\
&~~~~~~~~~~~~~~~~\vdots&\\
&Y^{n}_K=\left(Y^{n}_{K,(1,P)},Y^{n}_{K,(1,NP)}\right)&
\end{align}
where $Y^{n}_{k,(1,P)}$ is channel output at $k^{th}$ receiver for those instances in which perfect CSIT is presented. Also, $Y^{n}_{k,(1,NP)}$ is channel output at $k^{th}$ receiver for those instances in which perfect CSIT is not presented. 

The term $h\left(Y^{n}_{2},\dots,Y^{n}_{K}\vert Y^{n}_{1,P},\widehat{Y}^{n}_{1,NP},\dots,\widehat{Y}^{n}_{\frac{K}{2}-1,NP},M^{[1]},{\bf{\Theta}}\right)$ can be bounded as follows:
\begin{equation}
\label{entropy9}
\begin{aligned}
&h\left(Y^{n}_{2},\dots,Y^{n}_{K}\vert Y^{n}_{1,P},\widehat{Y}^{n}_{1,NP},\dots,\widehat{Y}^{n}_{\frac{K}{2}-1,NP},M^{[1]},{\bf{\Theta}}\right)&\\
&=h\left(Y^{n}_{2,(1,P)},Y^{n}_{3,(1,NP)},\dots,Y^{n}_{K,(1,P)},Y^{n}_{K,(1,NP)}\vert Y^{n}_{1,P},\widehat{Y}^{n}_{1,NP},\dots,\widehat{Y}^{n}_{\frac{K}{2}-1,NP},M^{[1]},{\bf{\Theta}}\right)&\\
&\leq h\left(Y^{n}_{2,(1,P)},\dots,Y^{n}_{K,(1,P)}\vert Y^{n}_{1,P},\widehat{Y}^{n}_{1,NP},\dots,\widehat{Y}^{n}_{\frac{K}{2}-1,NP},M^{[1]},{\bf{\Theta}}\right)&\\
&~~~+h\left(Y^{n}_{2,(1,NP)},\dots,Y^{n}_{K,(1,NP)}\vert Y^{n}_{1,P},\widehat{Y}^{n}_{1,NP},\dots,\widehat{Y}^{n}_{\frac{K}{2}-1,NP},M^{[1]},{\bf{\Theta}}\right)&\\
&\overset{(a)}{\leq} h\left(Y^{n}_{2,(1,P)},\dots,Y^{n}_{K,(1,P)}\vert Y^{n}_{1,P}\right)&\\
&~~~+h\left(Y^{n}_{2,(1,NP)},\dots,Y^{n}_{K,(1,NP)}\vert Y^{n}_{1,P},\widehat{Y}^{n}_{1,NP},\dots,\widehat{Y}^{n}_{\frac{K}{2}-1,NP},M^{[1]},{\bf{\Theta}}\right)&\\
&\overset{(b)}{=}h\left(Y^{n}_{1,P},Y^{n}_{2,(1,P)},\dots,Y^{n}_{K,(1,P)}\right)-h\left(Y^{n}_{1,P}\right)&\\
&~~~\underbrace{+h\left(Y^{n}_{2,(1,NP)},\dots,Y^{n}_{K,(1,NP)}\vert Y^{n}_{1,P},\widehat{Y}^{n}_{1,NP},\dots,\widehat{Y}^{n}_{\frac{K}{2}-1,NP},M^{[1]},{\bf{\Theta}}\right)}_{\leq no\left(\log(P)\right)}&\\
&= \underbrace{\sum_{i=1}^{n\Upsilon_K}{h\left(Y^{n}_{1,P} ,Y^{n}_{2,(1,P)} ,\dots,Y^{n}_{K,(1,P)} \right)}}_{\leq n\Upsilon_K \frac{K}{2}\log{(P)}}-\underbrace{\sum_{i=1}^{n\Upsilon_K}{h\left(Y^{n}_{1,P} \right)}}_{= n\Upsilon_{K}\log{(P)}}+no\left(\log(P)\right)&\\
&\leq n\Upsilon_K \frac{K}{2}\log{(P)}-n\Upsilon_{K}\log{(P)}+no\left(\log(P)\right)&\\
&=n\Upsilon_{K}\left(\frac{K}{2}-1\right){\log{(P)}}+no\left(\log(P)\right)&
\end{aligned}
\end{equation}
(a) comes from conditional entropy, (b) comes from chain rule, by substituting the result of \ref{entropy9} in \ref{entropy8} we have:
 \begin{equation}
\begin{aligned}
&n\left(\frac{K}{2} R_1+R_2+\dots+R_K\right)&\\
&~~~~~~\leq n\frac{K}{2}\log{(P)}+n\Upsilon_{K}\left(\frac{K}{2}-1\right){\log{(P)}}+no\left(\log(P)\right)+n{\varepsilon_n}'.&\\
\end{aligned}
\end{equation}
Therefore:
\begin{equation}
\left(\frac{K}{2} d_1+d_2+\dots+d_K\right)\leq n\frac{K}{2}+n\Upsilon_{K}\left(\frac{K}{2}-1\right),
\end{equation}
and similarly for every value of $1\leq i \leq K$ we have:
\begin{equation}
\left( d_1+\dots+\frac{K}{2}d_i+\dots+d_K\right)\leq \frac{K}{2}+\Upsilon_{K}\left(\frac{K}{2}-1\right),
\end{equation}
summing up all the above bounds for different values of $1\leq i \leq K$ we get:
\begin{equation}
\begin{aligned}
&\sum_{i=1}^{K}\left( d_1+\dots+\frac{K}{2}d_i+\dots+d_K\right)&\leq K\left(\frac{K}{2}+\Upsilon_{K}\left(\frac{K}{2}-1\right)\right)&\\
\end{aligned}
\end{equation}
so:
\begin{equation}
d_1+d_2+\dots+d_K\leq \frac{K\left(\frac{K}{2}+\Upsilon_{K}\left(\frac{K}{2}-1\right)\right)}{\left(\frac{K}{2}+\left(K-1\right)\right)}
\end{equation}
and finally for $d_1+d_2+\dots+d_K=\frac{K}{2}$ and $K\geq4$ we should have $\Upsilon_K\geq \frac{1}{2}$, which shows the proof of this theorem.
\end{IEEEproof}
\begin{thm}
\label{theorem_5} For odd values of $K>3$, $\frac{1}{2}$ is the minimum value for $\Upsilon_K$ to achieve maximum DoF of $\frac{K}{2}$. 
\end{thm}
\begin{IEEEproof}
Since every odd $K$ value is between two even numbers of $K_1$ and $K_2$ where $K_1\leq K \leq K_2$. Because of \textit{Theorem3}, we should have $\Upsilon_{K_1}\leq \Upsilon_{K} \leq \Upsilon_{K_2}$. Also, we have $\Upsilon_{K_2}=\Upsilon_{K_1}=\frac{1}{2}$ which shows $\Upsilon_K=\frac{1}{2}$. 
\end{IEEEproof}
\section{Interpretation of Leakage Rate}
In this paper, we have proposed different solutions for IA with imperfect CSI. Let us analyze the interpretation of leakage rate to the specific receiver from its interference paths. For the sake of simplicity, we consider a case in which all message sets of $\mathcal{M}=\{M^{[1]},\dots,M^{[k]},\dots,M^{[K]} \},~k\neq \{q_1,q_2\}$ are eliminated. Therefore, we can simplify the analysis of leakage rate of $K-$user interference channel to the $2-$user interference channel problem. In other words, we can assume $\bar{\bf{X}}^{[j]}=\bar{\bf{O}}, j\in \{1,\dots,K\}-\{q_1,q_2\}$, where, $\bar{\bf{O}}$ is an $n\times 1$ zero matrix. Therefore, at $\mathrm{RX}_{q_1}$ and $\mathrm{RX}_{q_2}$ we have:
\begin{equation}
\label{2user}
\begin{aligned}
\bar{\bf{Y}}^{[q_1]}&={\bar{\bf{H}}}^{[q_{1}q_{1}]} \bar{\bf{X}}^{[q_1]} +{\bar{\bf{H}}}^{[q_{1}q_{2}]} \bar{\bf{X}}^{[q_2]} +\bar{\bf{Z}}^{[q_1]} &\\
\bar{\bf{Y}}^{[q_2]}&={\bar{\bf{H}}}^{[q_{2}q_{2}]} \bar{\bf{X}}^{[q_2]} +{\bar{\bf{H}}}^{[q_{2}q_{1}]} \bar{\bf{X}}^{[q_1]} +\bar{\bf{Z}}^{[q_2]} .&\\
\end{aligned}
\end{equation}
 In all the proposed methods of this paper, we can express all the cross channels e.g. ${\bar{\bf{H}}}^{[q_{1}q_{2}]}=\mathrm{diag}\left(\left[ h^{[q_{1}q_{2}]}_{1},\dots,h^{[q_{1}q_{2}]}_{n}\right] \right) $ and ${\bar{\bf{H}}}^{[q_{2}q_{1}]}=\mathrm{diag}\left(\left[ h^{[q_{2}q_{1}]}_{1},\dots,h^{[q_{2}q_{1}]}_{n}\right] \right)$ with the summation of some basic diagonal matrices. Let the sets $U^{[{q_{1}}{q_{2}}]}$ and $U^{[{q_{2}}{q_{1}}]}$ represent our uncertainly about the exact value of the channel in signaling time duration. Similar to the previous sections these two sets are defined as follows:
\begin{align}
U^{[{q_{1}}{q_{2}}]}&=\left\lbrace {\forall j \in \{1,\dots,n\}\vert h^{[q_{1}q_{2}]}_{j}~\text{is unknown}}\right\rbrace&\\
U^{[{q_{2}}{q_{1}}]}&=\left\lbrace {\forall j \in \{1,\dots,n\}\vert h^{[q_{2}q_{1}]}_{j}~\text{is unknown}}\right\rbrace.&
\end{align}
We define the basic matrices $\bar{\bf{Q}}^{{[{q_1}{q_2}]}_j}=\mathrm{diag}\left(\left[ q^{{[{q_1}{q_2}]}_j}_1,\dots,q^{{[{q_1}{q_2}]}_j}_n\right]\right)$ and $\bar{\bf{Q}}^{{[{q_2}{q_1}]}_j}=\mathrm{diag} \left(\left[ q^{{[{q_2}{q_1}]}_j}_1,\dots,q^{{[{q_2}{q_1}]}_j}_n\right]\right)$ as follows:
\begin{equation}
\left\{\begin{matrix}
q^{{[{q_1}{q_2}]}_j}_r &={h}^{[{q_1}{q_2}]}_r$ if$~r\notin U^{[{q_{1}}{q_{2}}]}& \\ \\
q^{{[{q_1}{q_2}]}_j}_r &=\gamma^{{[{q_1}{q_2}]}_j}_{r}$ if$~r\in U^{[{q_{1}}{q_{2}}]}&
\end{matrix}\right.
,
\end{equation}
and similarly we have:
\begin{equation}
\left\{\begin{matrix}
q^{[{q_2}{q_1}]}_r&={h}^{[{q_2}{q_1}]}_r$ if$~r\notin U^{[{q_{2}}{q_{1}}]}& \\ \\
q^{[{q_2}{q_1}]}_r&=\gamma^{[{q_2}{q_1}]}_{r}$ if$~r\in U^{[{q_{2}}{q_{1}}]}&\\
\end{matrix}\right.
\end{equation} 
where, $\gamma^{[{q_1}{q_2}]}_{r}$ and $\gamma^{[{q_2}{q_1}]}_{r}$ are both random variables with desired distributions. Similar to the Lemma 2 all the channel matrices can be represented as follows:
\begin{align}
&{\bar{\bf{H}}}^{[{q_1}{q_2}]} =\sum_{j=1}^{\vert U^{[{q_1}{q_2}]} \vert+1}{\beta_j \bar{\bf{I}}\left({\bar{\bf{Q}}^{[{q_1}{q_2}]_{j}}}\right)}&\\
&{\bar{\bf{H}}}^{[{q_2}{q_1}]} =\sum_{j=1}^{\vert U^{[{q_2}{q_1}]} \vert+1}{\beta_j \bar{\bf{I}}\left({\bar{\bf{Q}}^{[{q_2}{q_1}]_{j}}}\right)}&.
\end{align}
Therefore, the relation \eqref{2user} can be represented as follows:
\begin{align}
&\bar{\bf{Y}}^{[q_1]} ={\bar{\bf{H}}}^{[q_{1}q_{1}]} \bar{\bf{X}}^{[q_1]} +\left(\sum_{j=1}^{\vert U^{[{q_1}{q_2}]} \vert+1}{\beta_j \bar{\bf{I}}{\bar{\bf{Q}}^{[{q_1}{q_2}]_{j}}}}\right)\bar{\bf{X}}^{[q_2]} +\bar{\bf{Z}}^{[q_1]} &\\
&\bar{\bf{Y}}^{[q_2]} ={\bar{\bf{H}}}^{[q_{2}q_{2}]} \bar{\bf{X}}^{[q_2]} +\left(\sum_{j=1}^{\vert U^{[{q_2}{q_1}]} \vert+1}{\beta_j \bar{\bf{I}}{\bar{\bf{Q}}^{[{q_2}{q_1}]_{j}}}}\right)\bar{\bf{X}}^{[q_1]} +\bar{\bf{Z}}^{[q_2]} .&
\end{align}
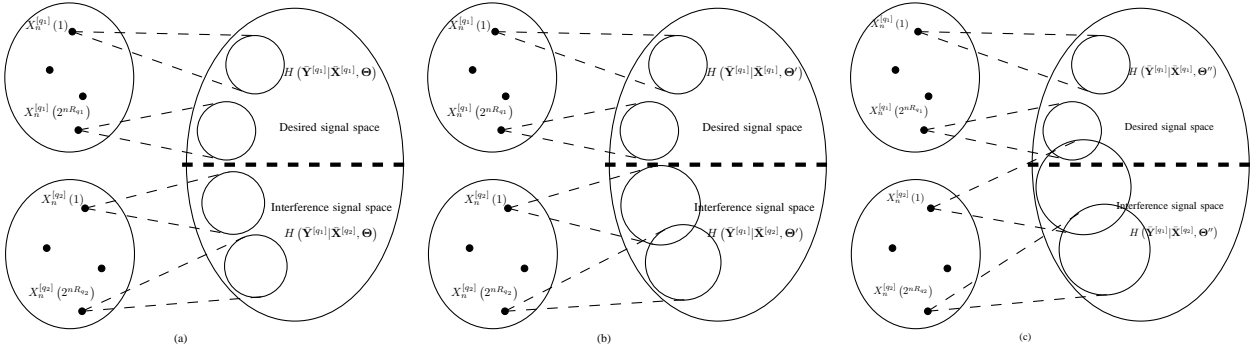
\begin{figure}
\begin{tikzpicture}[scale=0.5, every node/.style={scale=0.45}]
\clip(-5.,-5.) rectangle (6.,5.);
\draw [rotate around={90.:(3.,0.)}] (3.,0.) ellipse (4.162277660168379cm and 2.8852305489053665cm);
\draw [line width=1.6pt,dash pattern=on 4pt off 4pt] (0.,0.)-- (6.,0.);
\draw(1.94,2.66) circle (0.7762087348130015cm);
\draw(1.18,0.9) circle (0.7762087348130011cm);
\draw(1.36,-1.02) circle (0.8335466393669883cm);
\draw (2.48,1.22) node[anchor=north west] {Desired signal space};
\draw (2.26,-0.9) node[anchor=north west] {Interference signal space};
\draw [rotate around={90.:(-3.,2.32)}] (-3.,2.32) ellipse (1.984324964736322cm and 1.7139269429225403cm);
\draw [rotate around={90.:(-2.98,-2.38)}] (-2.98,-2.38) ellipse (1.9843249647363215cm and 1.7139269429225399cm);
\draw (2.58,2.74) node[anchor=north west] {$H\left( \bar{\bf{Y}}^{[q_1]} \vert  \bar{\bf{X}}^{[q_1]}, \bf{\Theta} \right)$};
\draw (2.6,-1.54) node[anchor=north west] {$H\left( \bar{\bf{Y}}^{[q_1]} \vert  \bar{\bf{X}}^{[q_2]}, \bf{\Theta} \right)$};
\draw [dash pattern=on 4pt off 4pt] (-2.92,3.54)-- (1.7169580612933557,1.9165268709778531);
\draw [dash pattern=on 4pt off 4pt] (-2.76,0.92)-- (1.088242122381309,1.6707661719970028);
\draw [dash pattern=on 4pt off 4pt] (-2.76,0.92)-- (0.9722524951101321,0.1521089823964762);
\draw [dash pattern=on 4pt off 4pt] (-2.58,-1.16)-- (1.2590955322788127,-0.19258336468626414);
\draw [dash pattern=on 4pt off 4pt] (-2.58,-1.16)-- (1.2009012370982184,-1.8382222092091625);
\draw [dash pattern=on 4pt off 4pt] (-2.92,3.54)-- (1.94,3.4362087348130017);
\draw (-4.32,1.74) node[anchor=north west] {$X_{n}^{[{q_1}]}\left({2^{n{R_{q_1}}}}\right)$};
\draw (-4.28,4.06) node[anchor=north west] {$X_{n}^{[{q_1}]}\left(1 \right)$};
\draw (-3.86,-0.58) node[anchor=north west] {$X_{n}^{[{q_2}]}\left(1 \right)$};
\draw (-4.2,-3.06) node[anchor=north west] {$X_{n}^{[{q_2}]}\left({2^{n{R_{q_2}}}}\right)$};
\draw (-0.32,-4.38) node[anchor=north west] {(a)};
\draw(1.96,-2.7) circle (0.8318653737234165cm);
\draw [dash pattern=on 4pt off 4pt] (-2.66,-3.9)-- (1.6624856229027747,-1.9231569042461345);
\draw [dash pattern=on 4pt off 4pt] (-2.66,-3.9)-- (2.1331175434374594,-3.51365245415606);
\begin{scriptsize}
\draw [fill=black] (-2.92,3.54) circle (2.5pt);
\draw [fill=black] (-3.52,2.52) circle (2.5pt);
\draw [fill=black] (-2.64,1.82) circle (2.5pt);
\draw [fill=black] (-2.76,0.92) circle (2.5pt);
\draw [fill=black] (-2.58,-1.16) circle (2.5pt);
\draw [fill=black] (-3.6,-2.22) circle (2.5pt);
\draw [fill=black] (-2.14,-2.76) circle (2.5pt);
\draw [fill=black] (-2.66,-3.9) circle (2.5pt);
\end{scriptsize}
\end{tikzpicture}
\begin{tikzpicture}[scale=0.5, every node/.style={scale=0.45}]
\clip(-5.,-5.) rectangle (6.,5.);
\draw [rotate around={90.:(3.,0.)}] (3.,0.) ellipse (4.162277660168379cm and 2.8852305489053665cm);
\draw [line width=1.6pt,dash pattern=on 4pt off 4pt] (0.,0.)-- (6.,0.);
\draw(1.94,2.66) circle (0.7762087348130015cm);
\draw(1.18,0.9) circle (0.7762087348130011cm);
\draw(1.48,-1.08) circle (1.0567875850898325cm);
\draw (2.48,1.22) node[anchor=north west] {Desired signal space};
\draw (2.26,-0.9) node[anchor=north west] {Interference signal space};
\draw [rotate around={90.:(-3.,2.32)}] (-3.,2.32) ellipse (1.984324964736322cm and 1.7139269429225403cm);
\draw [rotate around={90.:(-2.98,-2.38)}] (-2.98,-2.38) ellipse (1.9843249647363215cm and 1.7139269429225399cm);
\draw (2.58,2.74) node[anchor=north west] {$H\left( \bar{\bf{Y}}^{[q_1]} \vert  \bar{\bf{X}}^{[q_1]}, \bf{\Theta}' \right)$};
\draw (2.6,-1.54) node[anchor=north west] {$H\left( \bar{\bf{Y}}^{[q_1]} \vert  \bar{\bf{X}}^{[q_2]}, \bf{\Theta} '\right)$};
\draw [dash pattern=on 4pt off 4pt] (-2.92,3.54)-- (1.7169580612933557,1.9165268709778531);
\draw [dash pattern=on 4pt off 4pt] (-2.76,0.92)-- (1.088242122381309,1.6707661719970028);
\draw [dash pattern=on 4pt off 4pt] (-2.76,0.92)-- (0.9722524951101321,0.1521089823964762);
\draw [dash pattern=on 4pt off 4pt] (-2.58,-1.16)-- (1.3520712390504879,-0.030984160214002188);
\draw [dash pattern=on 4pt off 4pt] (-2.58,-1.16)-- (1.2782913138904404,-2.1173589571348748);
\draw [dash pattern=on 4pt off 4pt] (-2.92,3.54)-- (1.94,3.4362087348130017);
\draw (-4.32,1.74) node[anchor=north west] {$X_{n}^{[{q_1}]}\left({2^{n{R_{q_1}}}}\right)$};
\draw (-4.28,4.06) node[anchor=north west] {$X_{n}^{[{q_1}]}\left(1 \right)$};
\draw (-3.86,-0.58) node[anchor=north west] {$X_{n}^{[{q_2}]}\left(1 \right)$};
\draw (-4.2,-3.06) node[anchor=north west] {$X_{n}^{[{q_2}]}\left({2^{n{R_{q_2}}}}\right)$};
\draw (-0.32,-4.38) node[anchor=north west] {(b)};
\draw(2.08,-2.6) circle (1.0043903623591774cm);
\draw [dash pattern=on 4pt off 4pt] (-2.66,-3.9)-- (1.7207825322957837,-1.6620432787723254);
\draw [dash pattern=on 4pt off 4pt] (-2.66,-3.9)-- (2.2890213124337744,-3.5824001684387388);
\begin{scriptsize}
\draw [fill=black] (-2.92,3.54) circle (2.5pt);
\draw [fill=black] (-3.52,2.52) circle (2.5pt);
\draw [fill=black] (-2.64,1.82) circle (2.5pt);
\draw [fill=black] (-2.76,0.92) circle (2.5pt);
\draw [fill=black] (-2.58,-1.16) circle (2.5pt);
\draw [fill=black] (-3.6,-2.22) circle (2.5pt);
\draw [fill=black] (-2.14,-2.76) circle (2.5pt);
\draw [fill=black] (-2.66,-3.9) circle (2.5pt);
\end{scriptsize}
\end{tikzpicture}
\begin{tikzpicture}[scale=0.5, every node/.style={scale=0.4}]
\clip(-5.,-5.) rectangle (6.,5.);
\draw [rotate around={90.:(3.,0.)}] (3.,0.) ellipse (4.162277660168379cm and 2.8852305489053665cm);
\draw [line width=1.6pt,dash pattern=on 4pt off 4pt] (0.,0.)-- (6.,0.);
\draw(1.94,2.66) circle (0.7762087348130015cm);
\draw(1.18,0.9) circle (0.7762087348130011cm);
\draw(1.48,-0.6) circle (1.2631706139710503cm);
\draw (2.48,1.22) node[anchor=north west] {Desired signal space};
\draw (2.26,-0.9) node[anchor=north west] {Interference signal space};
\draw [rotate around={90.:(-3.,2.32)}] (-3.,2.32) ellipse (1.984324964736322cm and 1.7139269429225403cm);
\draw [rotate around={90.:(-2.98,-2.38)}] (-2.98,-2.38) ellipse (1.9843249647363215cm and 1.7139269429225399cm);
\draw (2.58,2.74) node[anchor=north west] {$H\left( \bar{\bf{Y}}^{[q_1]} \vert  \bar{\bf{X}}^{[q_1]}, \bf{\Theta}'' \right)$};
\draw (2.6,-1.54) node[anchor=north west] {$H\left( \bar{\bf{Y}}^{[q_1]} \vert  \bar{\bf{X}}^{[q_2]}, \bf{\Theta} ''\right)$};
\draw [dash pattern=on 4pt off 4pt] (-2.92,3.54)-- (1.7169580612933557,1.9165268709778531);
\draw [dash pattern=on 4pt off 4pt] (-2.76,0.92)-- (1.088242122381309,1.6707661719970028);
\draw [dash pattern=on 4pt off 4pt] (-2.76,0.92)-- (0.9722524951101321,0.1521089823964762);
\draw [dash pattern=on 4pt off 4pt] (-2.58,-1.16)-- (1.3270876704144716,0.6538811026013301);
\draw [dash pattern=on 4pt off 4pt] (-2.58,-1.16)-- (1.2388990555233979,-1.8399477144510934);
\draw [dash pattern=on 4pt off 4pt] (-2.92,3.54)-- (1.94,3.4362087348130017);
\draw (-4.32,1.74) node[anchor=north west] {$X_{n}^{[{q_1}]}\left({2^{n{R_{q_1}}}}\right)$};
\draw (-4.28,4.06) node[anchor=north west] {$X_{n}^{[{q_1}]}\left(1 \right)$};
\draw (-3.86,-0.58) node[anchor=north west] {$X_{n}^{[{q_2}]}\left(1 \right)$};
\draw (-4.2,-3.06) node[anchor=north west] {$X_{n}^{[{q_2}]}\left({2^{n{R_{q_2}}}}\right)$};
\draw (-0.32,-4.38) node[anchor=north west] {(c)};
\draw(2.04,-2.26) circle (1.208304597359457cm);
\draw [dash pattern=on 4pt off 4pt] (-2.66,-3.9)-- (1.6078531635256677,-1.1316165936503557);
\draw [dash pattern=on 4pt off 4pt] (-2.66,-3.9)-- (2.2914574235525365,-3.4418498906969193);
\begin{scriptsize}
\draw [fill=black] (-2.92,3.54) circle (2.5pt);
\draw [fill=black] (-3.52,2.52) circle (2.5pt);
\draw [fill=black] (-2.64,1.82) circle (2.5pt);
\draw [fill=black] (-2.76,0.92) circle (2.5pt);
\draw [fill=black] (-2.58,-1.16) circle (2.5pt);
\draw [fill=black] (-3.6,-2.22) circle (2.5pt);
\draw [fill=black] (-2.14,-2.76) circle (2.5pt);
\draw [fill=black] (-2.66,-3.9) circle (2.5pt);
\end{scriptsize}
\end{tikzpicture}
  \caption{This figure shows the effects of CSI on the decodable information of interference path at $\mathrm{RX}_{q_1}$. The left hand side of each figure shows the selectable code-words at each transmitter e.g. the $X_{n}^{[{q_1}]}\left({1}\right)$ at $\mathrm{TX}_{q_1}$. At $\mathrm{RX}_q$, we have two types of signal spaces which are separated by using horizontal dashed line (desired and interference signal spaces). With these assumptions, the figure with subtitle (a) shows that, if we have the perfect CSI, the number of jointly typical sequences with transmitted code-words is limited and the receiver can distinguish among different transmitted code-words of both desired and undesired transmitters. In this case, our knowledge about the perfect CSI is depicted by random variable of $\bf{\Theta}$. The figure with subtitle (b) shows that at interference signal space in the case of imperfect CSI (${\bf{\Theta}}^{'}$), the number of jointly typical sequences with a specific transmitted code-word increases and overlaps with other transmitted sequences. When our uncertainty about CSI is not larger than a specific value, we can decode the desired signal but our uncertainty about the transmitted code-word of interference signal increases. Therefore, the leakage rate from interference path is reduced while we can accommodate interference signal in interference subspace. In a specific case when our uncertainty about CSI ($\bf{\Theta}^{''}$) increases from a specific value, not only the leakage rate is reduced but also the desired signal space is polluted by the interference signal. This fact can be figured out by the figure with the subtitle (c). In this case, we can assume $\bf{\Theta}^{'}, \bf{\Theta}^{''}$ are two degraded versions of the random variable $\bf{\Theta}$, in other words, $I\left({\bf{\Theta} ;\bf{\Theta}^{'} }\right) < I\left({\bf{\Theta} ;\bf{\Theta}^{''} }\right)$.}
\end{figure} 
From this definition, in the present of imperfect CSI the leakage rate can be analyzed as follows:
\begin{equation}
\label{up1}
\begin{aligned}
R^{[q_1]}_{L}&= I\left( M^{[q_2]}; \bar{\bf{Y}}^{[q_1]} \vert {\bf{\Theta'}} \right)&\\
&= I\left( M^{[q_2]}; {\bar{\bf{H}}}^{[{q_1}{q_1}]} \bar{\bf{X}}^{[q_1]} +{\bar{\bf{H}}}^{[{q_1}{q_2}]} \bar{\bf{X}}^{[q_2]} +\bar{\bf{Z}}^{[q_1]} \vert {\bar{\bf{H}}}^{[{q_1}{q_1}]} , {\bar{{\bf{Q}}}^{[{q_1}{q_2}]_j}} \right)&\\
&\overset{(a)}{\leq}  I\left( M^{[q_2]}; {\bar{\bf{H}}}^{[{q_1}{q_1}]} \bar{\bf{X}}^{[q_1]} +{\bar{\bf{H}}}^{[{q_1}{q_2}]} \bar{\bf{X}}^{[q_2]} +\bar{\bf{Z}}^{[q_1]}  \vert M^{[q_1]}, {\bar{\bf{H}}}^{[{q_1}{q_1}]} , {\bar{{\bf{Q}}}^{[{q_1}{q_2}]_j}}\right)&\\
&\overset{(b)}{=}  I\left( M^{[q_2]}; {\bar{\bf{H}}}^{[{q_1}{q_1}]} \bar{\bf{X}}^{[q_1]} +{\bar{\bf{H}}}^{[{q_1}{q_2}]} \bar{\bf{X}}^{[q_2]} +\bar{\bf{Z}}^{[q_1]}  \vert M^{[q_1]}, \bar{\bf{X}}^{[q_1]} ,{\bar{\bf{H}}}^{[{q_1}{q_1}]} , {\bar{{\bf{Q}}}^{[{q_1}{q_2}]_j}}\right)&\\
&\overset{(c)}{=} I\left( M^{[q_2]};{\bar{\bf{H}}}^{[{q_1}{q_2}]} \bar{\bf{X}}^{[q_2]} +\bar{\bf{Z}}^{[q_1]} \vert {\bar{{\bf{Q}}}^{[{q_1}{q_2}]_j}}\right)&\\
&\overset{(d)}{\leq} I\left({\bar{\bf{X}}}^{[q_2]} ;{\bar{\bf{H}}}^{[{q_1}{q_2}]} \bar{\bf{X}}^{[q_2]} +\bar{\bf{Z}}^{[q_1]}  \vert {\bar{{\bf{Q}}}^{[{q_1}{q_2}]_j}} \right)&\\
&= H\left( {\bar{\bf{X}}}^{[q_2]}  \vert {\bar{{\bf{Q}}}^{[{q_1}{q_2}]}_j} \right)-H\left( {\bar{\bf{X}}}^{[q_2]}  \vert {\bar{\bf{H}}}^{[{q_1}{q_2}]} \bar{\bf{X}}^{[q_2]} +\bar{\bf{Z}}^{[q_1]}, {\bar{{\bf{Q}}}^{[{q_1}{q_2}]_j}} \right)&\\
&= H\left( {\bar{\bf{X}}}^{[q_2]} \right)-H\left( {\bar{\bf{X}}}^{[q_2]}  \vert {\bar{\bf{H}}}^{[{q_1}{q_2}]} \bar{\bf{X}}^{[q_2]} +\bar{\bf{Z}}^{[q_1]} , {\bar{{\bf{Q}}}^{[{q_1}{q_2}]_j}} \right)&
\end{aligned}
\end{equation}
where (a) follows from conditional mutual information, (b) follows from that $\bar{\bf{X}}^{[q_1]} $ is a function of $ M^{[q_1]}$ and ${\bar{{\bf{Q}}}^{[{q_1}{q_2}]} }$, (c) follows from functionality of ${\bar{\bf{H}}}^{[{q_1}{q_1}]} \bar{\bf{X}}^{[q_1]} $ from conditions of mutual information relation, (d) follows from Markov chain of $M^{[q_2]}-{\bar{\bf{X}}}^{[q_2]} -{\bar{\bf{Y}}}^{[q_1]} $. Now, let us we analyze the information leakage in the present of perfect CSI, similarity, we get:
\begin{equation}
\label{up2}
\begin{aligned}
R^{[q_1]}_{L} &\leq  H\left( {\bar{\bf{X}}}^{[q_2]} \right)-H\left( {\bar{\bf{X}}}^{[q_2]}  \vert {\bar{\bf{H}}}^{[{q_1}{q_2}]} \bar{\bf{X}}^{[q_2]} +\bar{\bf{Z}}^{[q_1]} , {\bar{\bf{H}}}^{[{q_1}{q_2}]}  \right)&\\
&= H\left( {\bar{\bf{X}}}^{[q_2]} \right)-o\left({\log {\left(\mathrm{SNR}\right)}}\right).&
\end{aligned}
\end{equation}

Comparing two upper bounds of equations \eqref{up1} and \eqref{up2}, in the case of imperfect and perfect CSI, we can conclude that the entropy term of $H\left( {\bar{\bf{X}}}^{[q_2]}  \vert {\bar{\bf{H}}}^{[{q_1}{q_2}]} \bar{\bf{X}}^{[q_2]} +\bar{\bf{Z}}^{[q_1]} , {\bar{{\bf{Q}}}^{[{q_1}{q_2}]} } \right)$ shows the trade off between available CSI and leakage rate. Figure 7 shows that how the channel state information effects on the number of jointly typical sequences at $\mathrm{RX}_q$.   
\section{Conclusion}
In this paper, we investigate IA problem for the $K$-user interference channel with imperfect CSI. While CJ method \cite{jafar} changes our viewpoint to interference channel, several challenges must be solved to transform it to a practical method. One of the most challenges in this method is the assumption of global channel knowledge. In this method a transmitter should have knowledge of channel state information for its own precoder design. In the practical situation, it is very hard to know the cross channels especially for fast fading channel cases.
  We explore two separate IA problem models, in the first one we use channel coherence time and changing pattern to align interferences. Through one example and a theorem we show that by the use of different changing pattern of direct and cross channels, we can achieve more than one DoF. In other words, we show that time variant characteristics of the channels can help us to accomplish completely or partially align interference in receivers. Also, we show that in a case in which all the channels ended to the same destination have similar changing pattern, the IA still can be applicable. The optimality of this method is also proved by a theorem. In the second one, the results of this paper extended to the more complicated problem of IA in the fast fading channel and we show that direct channel memory and permutation characteristics can help us to find a solution for this problem. Using converse proof, we show that our solution is optimum to achieve maximum DoF of $\frac{K}{2}$ and half of the channel knowledge is the minimum requirement of channel to achieve maximum DoF. The key insight has been explored at the last section of this paper which we show that there exists a trade off between leakage rate from interference paths and the CSI.   

%

\appendices
\section{Proof of Alignment Condition}
From Lemma 1, ${\bar{\bf{H}}}^{[pk]} =\sum_{j=1}^{\sigma'+1}\beta^{[pq]_j}{\bf{\bar{I}}}({\bar{{\bf{Q}}} })^j$ therefore,
\begin{equation}
\label{Hrep}
{\bar{\bf{H}}}^{[pk]} \bar{\bf{V}}^{[k]} =\left(\sum_{j=1}^{\sigma'+1}\beta^{[pq]}_j{\bf{\bar{I}}}({\bar{{\bf{Q}}} })^j\right)\bar{\bf{V}}^{[k]}.
\end{equation}
Since ${\bar{{\bf{Q}}}}^j,~j>\sigma'+1$ has similar changing pattern to ${\bar{{\bf{Q}}} }$, for every $l>\sigma'+1$ the matrix ${\bar{{\bf{Q}}}}^l$ can be represented as follows:
\begin{equation}
{\bar{{\bf{Q}}}}^l=\sum_{i=1}^{\sigma'+1}{\lambda_{i}'\bar{\bf{I}}\bar{{\bf{Q}}}^i},~\lambda_{i}\in \mathbb{R}
\end{equation}
Also from the definition of the span of the matrix, if $I_l\in I$ and $v_{l}^{[k]}\in I$ we have:
\begin{align}
{\mathrm{span}}\left({\bar{\bf{I}}}^{[p]}\right) &=\left\{\sum_{l=1}^{f}\lambda_{l}I_{l}\vert f\in\mathbb{N},~\lambda_{l}\in \mathbb{R} \right\}&\\
\mathrm{span}\left({\bar{\bf{V}}^{[k]}}\right)&=\left\{\sum_{l=1}^{f}\lambda_{l}v^{[k]}_{l}\vert f\in\mathbb{N},~\lambda_{l}\in \mathbb{R}\right\}.&
\end{align}
Since basic vectors of ${\bar{\bf{V}}^{[k]}}$ and ${\bf{I}}^{[k]}$ are chosen from similar set we have:
\begin{align}
\label{Vrep}
\mathrm{span}\left({\bar{\bf{V}}^{[k]}}\right)&=\left\{\sum_{l=1}^{f}\lambda_{l}I_{l}\vert f\in\mathbb{N},~\lambda_{l}\in \mathbb{R} \right\}&\\
&=\left\{\sum_{l=1}^{f}\lambda_{l}{\bar{{\bf{Q}}}}^{\alpha} \bar{\bf{\Gamma}}^{j}\vert \alpha \in \{1,\dots,\sigma'\},~j\in \{1,\dots,\varrho\},~f\in\mathbb{N},~\lambda_{l}\in \mathbb{R} \right\}&
\end{align}  
Therefore, from equations \eqref{Hrep} and \eqref{Vrep}, the $\mathrm{span}\left(\bar{\bf{H}}^{[pk]} \bar{\bf{V}}^{[k]}\right)$ can be given as follows:
\begin{align}
\mathrm{span}\left({{\bar{\bf{H}}}^{[pk]} \bar{\bf{V}}^{[k]}}\right)&=\left\{\left(\sum_{i=1}^{\sigma'+1}\beta^{[pk]}_j{\bf{\bar{I}}}({\bar{{\bf{Q}}} })^i\right)\sum_{l=1}^{f}\lambda_{l}I_{l}\vert f\in\mathbb{N},~\lambda_{l}\in \mathbb{R}\right\}&\\
&=\left\{\sum_{i=1}^{\sigma'+1}{\sum_{l=1}^{f}{\beta^{[pk]}_i}\lambda_{l}{\bar{{\bf{Q}}}}^{\alpha+i} \bar{\bf{\Gamma}}^{j}} \vert \alpha \in \{1,\dots,\sigma'\},~j\in \{1,\dots,\varrho\},~f\in\mathbb{N},~\lambda_{l}\in \mathbb{R}\right\}&\\
&=\left\{{\sum_{l=1}^{f}{\lambda'_l}{\bar{{\bf{Q}}}}^{\alpha} \bar{\bf{\Gamma}}^{j}} \vert \alpha \in \{1,\dots,\sigma'\},~j\in \{1,\dots,\varrho\},~f\in\mathbb{N},~\lambda'_{l}\in \mathbb{R}\right\}&\\
&=\mathrm{span}\left({\bf{I}}^{[p]}\right),&
\end{align}
thus we have $\mathrm{span}\left({{\bar{\bf{H}}}^{[pk]} \bar{\bf{V}}^{[k]}}\right)=\mathrm{span}\left({\bf{I}}^{[p]}\right)$, and the proof is complete.
\section{Proof of Theorem 1}
\textit{Step1) Finding the dimension of the space spanned by the interference subspace} $\bar{\textbf{I}} $:\\ 
First we should show that $\mathrm{rank}\left(\bar{\textbf{I}} \right)=\mathrm{rank}\left(\bar{\bf{V}} \right)=\varrho\left(|\xi|+1\right)$. Let $\bar{\bf{P}}$ be a matrix which its columns are the member of the following set:
\begin{equation}
P=\left\lbrace\left(\bar{{\bf{Q}}} \right)^{\alpha}{\bf{W}}:\forall {\alpha} \in \{1,\dots,\sigma'+1\}\right\rbrace.
\end{equation}
Assume $j^{th}$ row of the matrix $\bf{P}$ is expressed by ${\bf{P}}_j$. By choosing the rows of ${\bf{P}}_j,~j \in C'\cup \{1\}$, we can generate a new matrix of $\bar{\bf{P}}'$ which can be represented as follows:
\begin{equation}
\bar{\bf{P}}'=
\begin{bmatrix}
 \gamma_1 &  \gamma_1^2&  \dots& \gamma_1^{\sigma'+1}\\ 
 \gamma_2&  \gamma_2^2&  \dots& \gamma_2^{\sigma'+1}\\ 
 \vdots &  \vdots &  \dots& \vdots\\ 
 \gamma_{\sigma'+1}&  \gamma_{\sigma'+1}^2&  \dots&\gamma_{\sigma'+1}^{\sigma'+1} 
\end{bmatrix}.
\end{equation}
All the columns of the matrix $\bar{\bf{P}}'$ are the columns of the Vandermonde matrix multiplied by $\mathrm{diag}\left(\left[\gamma_{1},\gamma_{2},\dots,\gamma_{r}\right]\right)$, also the $\bar{\bf{P}}'$ is a full rank matrix with the rank of $\sigma'+1$. Since, all the rows of the $\bar{\bf{P}}$ are repetitive rows of the matrix $\bar{\bf{P}}'$, the rank of the matrix $\bar{\bf{P}}$ equals to the rank of the matrix $\bar{\bf{P}}'$ and it equals to $\sigma'+1$. On the other hand the matrix $\bar{\textbf{I}}$ can be represented as follows:
 \begin{equation}
 \bar{\textbf{I}} =\left[\bar{\bf{\Gamma}}\bar{\bf{P}}~~\bar{\bf{\Gamma}}^{2}\bar{\bf{P}}\dots~~\bar{\bf{\Gamma}}^{\varrho}\bar{\bf{P}}\right].
 \end{equation}
Because the matrix $\bar{\bf{\Gamma}}$ has random elements, we can easily show that all the columns of $\bar{\textbf{I}}$ are linearly independent. Therefore, the rank of the matrix $\bar{\textbf{I}} $ equals to its columns number $\varrho \left( \sigma'+1\right)$.

\textit{Step2) Finding the dimension of the space spanned by the free interference subspace at $\mathrm{RX}_k$:}\\
In this case all the transmitters use the similar precoder vectors. Therefore, in order to find the dimension of free interference subspace at $\mathrm{RX}_k$ we should find $\mathrm{rank}\left( \left[ \bar{\textbf{I}} ~~{\bar{\bf{H}}}^{[kk]} \bar{\textbf{I}} \right]\right)$.
Since ${\bar{\bf{H}}}^{[kk]} $ and $\bar{\bf{\Gamma}}^{\varrho}$ are diagonal matrices, we have:
\begin{equation}
\begin{aligned}
{\bar{\bf{H}}}^{[kk]} \bar{\textbf{I}} &={\bar{\bf{H}}}^{[kk]}\left[ \bar{\bf{\Gamma}}\bar{\bf{P}}~~\bar{\bf{\Gamma}}^{2}\bar{\bf{P}}\dots~~\bar{\bf{\Gamma}}^{\varrho}\bar{\bf{P}}\right]&\\
&=\left[\bar{\bf{\Gamma}}{\bar{\bf{H}}}^{[kk]} \bar{\bf{P}}~~\bar{\bf{\Gamma}}^{2}{\bar{\bf{H}}}^{[kk]} \bar{\bf{P}}\dots~~\bar{\bf{\Gamma}}^{\varrho}{\bar{\bf{H}}}^{[pp]} \bar{\bf{P}}\right].
\end{aligned}
\end{equation}
To find the $\mathrm{rank}\left( \left[ \bar{\bf{P}}~~{\bar{\bf{H}}}^{[kk]} \bar{\bf{P}} \right] \right)$, the matrix $\bar{\bf{P}}$ has the following structure: 
\begin{equation}
\bar{\bf{P}}=
\begin{bmatrix}
 \gamma_1 &  \gamma_1^2&  \dots& \gamma_1^{\sigma'+1}\\
 \vdots &  \vdots&  \dots& \vdots\\
 \gamma_1 &  \gamma_1^2&  \dots& \gamma_1^{\sigma'+1}\\
 \vdots &  \vdots &  \dots& \vdots\\ 
 \gamma_m&  \gamma_m^2&  \dots& \gamma_m^{\sigma'+1}\\ 
 \vdots &  \vdots &  \dots& \vdots\\
 \gamma_m&  \gamma_m^2&  \dots& \gamma_m^{\sigma'+1}\\
 \vdots &  \vdots &  \dots& \vdots\\
 \gamma_{\sigma'+1}&  \gamma_{\sigma'+1}^2&  \dots&\gamma_{\sigma'+1}^{\sigma'+1} \\
 \vdots &  \vdots &  \dots& \vdots\\ 
 \gamma_{\sigma'+1}&  \gamma_{\sigma'+1}^2&  \dots&\gamma_{\sigma'+1}^{\sigma'+1} 
\end{bmatrix}.
\end{equation}
Let, we analysis the pattern of the matrix $\bar{\bf{P}}$ when it multiply by the direct channels. In the following two cases multiplying the matrix $\bar{\bf{P}}$ with the direct channel matrix ${\bar{\bf{H}}}^{[kk]}=\mathrm{diag} \left(\left[{h_{1}^{[kk]},\dots,h_{1}^{[kk]},\dots,h_{R(k,k)}^{[kk]},\dots,h_{R(k,k)}^{[kk]}}\right]\right)$  can not change the pattern of the matrix $\bar{\bf{P}}$.
\begin{itemize}
  \item The changing points of the direct channel is the same with the changing points of the matrix $\bar{\bf{P}}$, in other words, $\left({C^{[kk]}-\cup_{p,q}{C^{[pq]}}=\varnothing,~p\neq q}\right)$.
  \item There is no changing points between or simultaneous with the changing points of the matrix $\bar{\bf{P}}$, in other words, \\$\left({C^{[kk]}-\cup_{p,q}{C^{[pq]}}=\varnothing,~p\neq q}\right)$.
\end{itemize}
If multiplying the matrix ${\bar{\bf{H}}}^{[kk]}$ by the $\bar{\bf{P}}$ do not change the pattern of the matrix ${\bar{\bf{H}}}^{[kk]}$, all the columns of the ${\bar{\bf{H}}}^{[kk]} \bar{\bf{P}}$ can be generated by linear combination of the columns of the matrix $\bar{\bf{P}}$.\\
In other words, $\mathrm{span} \left( {\bar{\bf{P}}} \right)=\mathrm{span} \left({{\bar{\bf{H}}}^{[kk]}\bar{\bf{P}}}\right)$ therefore, $\mathrm{rank}\left( \left[ \bar{\bf{P}}~{\bar{\bf{H}}}^{[kk]} \bar{\bf{P}}\right]\right)=\mathrm{rank} \left( \bar{\bf{P}} \right)=\sigma^{'}+1$. Now, consider for the channel matrix $\bar{{\bf H}}^{[kk]}$ the changing pattern set of $C^{[kk]}$ has an element of $c^{[kk]}_m$ in which $c^{[kk]}_m \in C'_{\mathcal{B}_{m}}$. In this case for the matrix $\left[\bar{\bf{P}}~{\bar{\bf{H}}}^{[kk]} \bar{\bf{P}}\right]$ we can find $\sigma^{'}+2$ rows which are linearly independent from each other almost surely. Therefore, the rank of the matrix $\left[\bar{\bf{P}}~{\bar{\bf{H}}}^{[kk]} \bar{\bf{P}}\right]$ is equal to $\sigma^{'}+2$. Let, we add a new $c^{[kk]}_{m'}$ to the changing pattern set of the $C^{[kk]}$, in this case we have two different states:
\begin{itemize}
\item {The $c^{[kk]}_{m'}$ is the member of the set $C'_{\mathcal{B}_{m}}$. In this case all the new generated rows of the matrix $\left[\bar{\bf{P}}~{\bar{\bf{H}}}^{[kk]}\bar{\bf{P}}\right]$ can be expressed by linear combination of the rows in the row number of the set $C'_{\mathcal{B}_{m}}$.}
\item {The $c^{[kk]}_{m'}$ is not the member of the set $C'_{\mathcal{B}_{m}}$. In this case all the new generated rows in the matrix $\left[\bar{\bf{P}}~{\bar{\bf{H}}}^{[kk]}\bar{\bf{P}}\right]$ can not be generated by any linear combination of the row numbers of the set $C'_{\mathcal{B}_{m}}$.}
\end{itemize}
Therefore, we can conclude that:
\begin{equation}
\mathrm{rank} \left( \left[ \bar{\bf{P}}~{\bar{\bf{H}}}^{[kk]}\bar{\bf{P}}\right] \right)=\sigma'+1+\sum_{m} {\mathbbm{1}}{\left(\left\vert {C}^{[kk]}_{\mathfrak{B}_{m}}-\bigcup_{p,q} {C}^{[pq]}_{\mathfrak{B}_{m}}\right\vert\right)}.
\end{equation} 
Finally, since $\bar{\bf{\Gamma}}$ is the random diagonal matrix we have:
\begin{equation}
\mathrm{rank}\left(\left[\bar{\bf{\Gamma}}\left[\bar{\bf{P}}~{\bar{\bf{H}}}^{[kk]} \bar{\bf{P}}\right]~~\bar{\bf{\Gamma}}^{2}\left[\bar{\bf{P}}~{\bar{\bf{H}}}^{[kk]} \bar{\bf{P}}\right]\dots~\bar{\bf{\Gamma}}^{\varrho}\left[\bar{\bf{P}}~{\bar{\bf{H}}}^{[kk]} \bar{\bf{P}}\right]\right]\right)\\
=\varrho \left( \sigma'+1+\sum_{m}{\mathbbm{1}}{\left(\lvert {C}^{[kk]}_{\mathfrak{B}_{m}}-\bigcup_{p,q} {C}^{[pq]}_{\mathfrak{B}_{m}}\rvert \right)}\right),
\end{equation}
because the rank of the matrix $\left[\bar{\bf{\Gamma}}{\bar{\bf{H}}}^{[kk]} \bar{\bf{P}}~~\bar{\bf{\Gamma}}^{2}{\bar{\bf{H}}}^{[kk]} \bar{\bf{P}}\dots~~\bar{\bf{\Gamma}}^{\varrho}{\bar{\bf{H}}}^{[pp]} \bar{\bf{P}}\right]$ is not larger than the row numbers of it. Therefore, we have:
\begin{equation}
\mathrm{rank}\left( {\left[ {\bar{\bf{\Gamma}} \left[{\bar{\bf{P}}~{\bar{\bf{H}}}^{[kk]} \bar{\bf{P}}}\right]~~\dots~\bar{\bf{\Gamma}}^{\varrho}\left[\bar{\bf{P}}~{\bar{\bf{H}}}^{[kk]} \bar{\bf{P}}\right]}\right]}\right)\\
= \min{\left( 2N, \varrho {\left( \sigma'+1+\sum_{m} {\mathbbm{1}{\left( \lvert {{C}^{[kk]}_{\mathfrak{B}_{m}}-\bigcup_{p,q} {C}^{[pq]}_{\mathfrak{B}_{m}}}\rvert \right)}}\right)}\right)},
\end{equation}
where $N=\varrho(\sigma'+1)$.
Finally the desired signal space rank can be calculated by subtracting interference rank from the above equation:
\begin{equation}
D_{k}= \min \left(N,\varrho \sum_{m} {\mathbbm{1}}{\left(\left\vert {C}^{[kk]}_{\mathcal{B}_{m}}-\bigcup_{p,q} {C}^{[pq]}_{\mathcal{B}_{m}}\right\vert > 0\right)}\right),~p\neq q
\end{equation}
where it completes the proof of this theorem.
\section{Achievable method for the $K-$user interference channel}
We show that when we don't know half of the channel state information (CSI), we can achieve $\frac{K}{2}$ DoF asymptotically. Referring section two, during signaling time for every $\left(p\neq q \in \lbrace 1,2,\dots,K\rbrace\right)$ the set $U^{[pq]}=\lbrace u^{[pq]}_1,\dots,u^{[pq]}_{\lvert U^{[pq]} \rvert}\rbrace$ shows that during $n$ transmission the exact value of the channel $h^{[pq]}_{r^{[pq]}} ,~r^{[pq]}\in U^{[pq]}$ between $\mathrm{TX}_q$ and $\mathrm{RX}_p$ is unknown. From the assumption of this paper we assume $\mathrm{RX}_p$  knows the channel matrix ${{\bar{\bf{H}}}}^{[pp]}$ between $\mathrm{TX}_p$ and $\mathrm{RX}_p$. Now, our objective is to find proper encoding $e_{q}\left(M^{[q]},{\bar{X}}^{[q]}\vert {{\bar{\bf{H}}}}^{[pq]} \right)$ and $d_{q}\left({\bf{\bar{Y}}}^{[q]}\vert {{\bar{\bf{H}}}}^{[pq]} \right)$ functions at transmitters and receivers respectively that satisfy IA conditions. Referring Lemma1, every channel matrix such as ${{\bar{\bf{H}}}}^{[pq]} $ can be represented as follows:
\begin{equation}
\label{a1}
{\bar{\bf{H}}}^{[pq]} =\sum_{j=1}^{\lvert U^{[pq]} \rvert+1}{\beta^{[pq]}_j{\bf{I}}\bar{{\bf{Q}}}^{[pq]_j} },
\end{equation}
from Lemma 4 and IA conditions we should have:
\begin{equation}
\label{a2}
{\mathrm{span}}\left(\bar{{\bf{Q}}}^{[1q]_j}\bar{\bf{V}}^{[q]}\right)={\mathrm{span}}\left(\bar{{\bf{Q}}}^{[1q']_{j'}}\bar{\bf{V}}^{[q']}\right),~(q,q'\neq 1)
\end{equation}
Also, interference received at second receiver should satisfy the following relations:
\begin{equation}
\label{a3}
\begin{aligned}
&{\mathrm{span}}\left(\bar{{\bf{Q}}}^{[23]_{j_{1}}}\bar{\bf{V}}^{[3]}\right)\subseteq{\mathrm{span}}\left(\bar{{\bf{Q}}}^{[21]_{j'_{1}}}\bar{\bf{V}}^{[1]}\right)&\\
&{\mathrm{span}}\left(\bar{{\bf{Q}}}^{[24]_{j_{2}}}\bar{\bf{V}}^{[4]}\right)\subseteq{\mathrm{span}}\left(\bar{{\bf{Q}}}^{[21]_{j'_{2}}}\bar{\bf{V}}^{[1]}\right)&\\
&~~~~~~~~~~~~~~~~~~~~~~~\vdots&\\
&{\mathrm{span}}\left(\bar{{\bf{Q}}}^{[2K]_{j_{K-2}}}\bar{\bf{V}}^{[K]}\right)\subseteq{\mathrm{span}}\left(\bar{{\bf{Q}}}^{[21]_{j'_{K-2}}}\bar{\bf{V}}^{[1]}\right)&\\
\end{aligned}
\end{equation}
above relations align the interference from $K-2$ transmitters within the interference from the first transmitter at the second receiver. Similarly at remain receivers we should have:
\begin{equation}
\label{a4}
{\mathrm{span}}\left(\bar{{\bf{Q}}}^{[pq]_{j_{1}}}\bar{\bf{V}}^{[q]}\right)\subseteq{\mathrm{span}}\left(\bar{{\bf{Q}}}^{[p1]_{j'_{1}}}\bar{\bf{V}}^{[1]}\right),~~q \notin \{1,p\}
\end{equation}
Relations \eqref{a1}, \eqref{a2} and \eqref{a3} can be equivalently expressed as
\begin{equation}
\mathrm{span}\left({\bar{\bf{V}}}^{[q]}\right)=\mathrm{span}\left({\bf S}^{[q]}\textbf{B}\right),~q=2,\dots,K~\text{at receiver 1}
\end{equation}  
and\\
\begin{equation}
\left.\begin{matrix}
\mathrm{span}\left({\textbf{T}}^{[2]}_{3}\textbf{B}\right)&=\mathrm{span}\left(\textbf{B}\right)&\prec&\mathrm{span}\left(\bar{\bf{V}}^{[1]}\right)\\
&\mathrm{span}\left({\textbf{T}}^{[2]}_{4}\textbf{B}\right)&\prec&\mathrm{span}\left(\bar{\bf{V}}^{[1]}\right)\\
&~&\vdots&\\ 
&\mathrm{span}\left({\textbf{T}}^{[2]}_{K}\textbf{B}\right)&\prec&\mathrm{span}\left(\bar{\bf{V}}^{[1]}\right)\\
\end{matrix}\right\}~\text{at receiver 2}
\end{equation}
also, we can generalize the above relation at other receivers as follows:
\begin{equation}
\left.\begin{matrix}
&\mathrm{span}\left({\textbf{T}}^{[i]}_{2}\textbf{B}\right)&\prec&\mathrm{span}\left(\bar{\bf{V}}^{[1]}\right)\\
&~&\vdots&\\
&\mathrm{span}\left({\textbf{T}}^{[i]}_{i-1}\textbf{B}\right)&\prec&\mathrm{span}\left(\bar{\bf{V}}^{[1]}\right)\\
&\mathrm{span}\left({\textbf{T}}^{[i]}_{i+1}\textbf{B}\right)&\prec&\mathrm{span}\left(\bar{\bf{V}}^{[1]}\right)\\
&~&\vdots&\\
&\mathrm{span}\left({\textbf{T}}^{[i]}_{K}\textbf{B}\right)&\prec&\mathrm{span}\left(\bar{\bf{V}}^{[1]}\right)\\
\end{matrix}\right\}~\text{at receiver i where i=3,\dots,\textit{K}}
\end{equation}
where,
\begin{equation}
\begin{aligned}
\textbf{B}&=&{\left({\bar{{\bf{Q}}} }^{[21]_{j'_1}}\right)^{-1}}{{\bar{{\bf{Q}}} }}^{[23]_{j_1}}\bar{\bf{V}}^{[3]}~~~~~~~~~~~~~~~~~~~~~~~~~~~~~~~~~~~~~~&\\
{\bf S}^{[q]}&=&{\left({\bar{{\bf{Q}}} }^{[1q]_{j}}\right)^{-1}}{{\bar{{\bf{Q}}} }}^{[13]_{j_1}}{\left({\bar{{\bf{Q}}} }^{[23]_{j_1}}\right)^{-1}}{{\bar{{\bf{Q}}} }}^{[21]_{j_1}},~q=2,\dots,K&\\
\textbf{T}^{[p]}_{q}&=&{\left({\bar{{\bf{Q}}} }^{[p1]_{j_1}}\right)^{-1}}{{\bar{{\bf{Q}}} }}^{[pq]_{j_1}}{\textbf{S}}^{[q]},~p,q=2,\dots,K.~~~~~~~~~~~~~~~~~~~~~&
\end{aligned}
\end{equation}
Now, we set $\mathrm{span}\left(\textbf{B}\right)$ and $\mathrm{span}\left({\bar{\bf{V}}}^{[1]}\right)$ as follows:
\begin{equation}
\mathrm{span}\left(\textbf{B}\right)=\mathrm{span}\left \{\left(\prod_{m,k \in \{2,\dots,K\}, m \neq k,(m,k) \neq (2,3)}{{\bar{\bf{\Gamma}}^j}{\left(\textbf{T}^{[m]}_k\right)^{\alpha_{mk}}}}\right){\bf{W}}: \alpha_{mk} \leq n^{*}, 1 \leq  j \leq  L \right \}
\end{equation}
and
\begin{equation}
\mathrm{span}\left({\bar{\bf{V}}}^{[1]}\right)=\mathrm{span}\left \{\left(\prod_{m,k\in\{2,\dots,K\},m \neq k,(m,k)\neq(2,3)}{{\bar{\bf{\Gamma}}^j}{\left(\textbf{T}^{[m]}_k\right)^{\alpha_{mk}}}}\right){\bf{W}}: \alpha_{mk} \leq n^{*}+1, 1 \leq  j \leq  L \right \}
\end{equation}
where, ${\bf{W}}=[1~1\dots~1]^{\mathrm{T}}$ is an $n \times 1$ column matrix, similar to Theorem 4 the matrix ${\bar{\bf{\Gamma}}}=\mathrm{diag}\left(\left[{\Gamma_{1},\Gamma_{2},\dots,\Gamma_{n}}\right]\right)$ is defined as follows:
\begin{equation}
\left\{\begin{matrix}
\Gamma_{r}=1,~~~~\text{if}~~~~r\notin \bigcup_{p,q}{U^{[pq]}}, p\neq q\\ 
\Gamma_{r}=\gamma_{r},~~~\text{if}~~~~r\in \bigcup_{p,q}{U^{[pq]}}, p\neq q
\end{matrix}\right.
\end{equation}
where, $\gamma_{r}$ is a random variable with an arbitrarily distribution, also $n=2L+\left({n^{*}}\right)^N+{\left( {n^{*}+1} \right)}^N$, $L=\lvert {\bigcup_{p,q}{U^{[pq]}}}\rvert$ and $N=\left({K-1}\right)\left({K-2}\right)-1$. From Lemma 6 we can easily show that $\mathrm{dim}\left({\bf{B}}\right)=L+{\left({n^{*}}\right)}^{N}$ and $\mathrm{dim}\left({\bar{\bf{V}}}^{[1]}\right)=L+{\left({n^{*}+1}\right)}^{N}$. Similar to Lemma 5 we can show that all the above IA conditions are satisfied.
\section*{Acknowledgment}
The authors would like to thank the anonymous reviewers for their valuable comments and suggestions to improve the quality of the paper.

\ifCLASSOPTIONcaptionsoff
  \newpage
\fi



%

\end{document}